\documentclass[amsfonts,floatfix,superscriptaddress,nofootinbib,reprint,prc,aps]{revtex4-2}

\usepackage{graphicx,amssymb,color,amsmath,bm,soul}
\usepackage[normalem]{ulem}
\usepackage{braket}
\usepackage{multirow}

\usepackage[colorinlistoftodos]{todonotes}
\usepackage[colorlinks,citecolor=blue]{hyperref}

\newcommand{\pp}{\mathfrak{p}}
\newcommand{\hh}{\mathfrak{h}}

\newcommand{\n}{\mathtt{n}}
\newcommand{\p}{\mathtt{p}}

\definecolor{orange}{rgb}{1,0.5,0}

\newcommand{\orcid}[1]{\href{https://orcid.org/#1}{\hskip2pt\includegraphics[width=9pt]{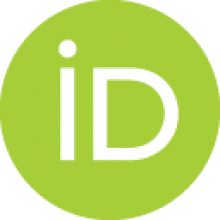}}}

\begin{document}

\title{Competition between allowed and first-forbidden $\beta$ decay in $r$-process waiting-point nuclei within a relativistic beyond-mean-field approach }

\author{Caroline E. P. Robin\orcid{0000-0001-5487-270X}}
\email{crobin@physik.uni-bielefeld.de}
\affiliation{Fakult\"at f\"ur Physik, Universit\"at Bielefeld, D-33615, Bielefeld, Germany}
\affiliation{GSI Helmholtzzentrum f\"ur Schwerionenforschung, Planckstra{\ss}e 1, 64291 Darmstadt, Germany}

\author{Gabriel Mart\'inez-Pinedo\orcid{0000-0002-3825-0131}}
\affiliation{GSI Helmholtzzentrum f\"ur Schwerionenforschung, Planckstra{\ss}e 1, 64291 Darmstadt, Germany}
\affiliation{Institut f\"ur Kernphysik (Theoriezentrum), Fachbereich Physik, Technische Universit\"at Darmstadt, Schlossgartenstra{\ss}e 2, 64289 Darmstadt, Germany}
\affiliation{Helmholtz Forschungsakademie Hessen f\"ur FAIR, GSI
    Helmholtzzentrum f\"ur Schwerionenforschung,
  Planckstra{\ss}e~1,
    64291 Darmstadt, Germany}

\date{\today}


\begin{abstract}
\begin{description}
\item[Background] $\beta$-decay rates of neutron-rich nuclei are a crucial ingredient to the simulations of the $r$-process nucleosynthesis. Up to now global calculations of these rates have been performed within the quasiparticle random-phase approximation (QRPA) which is known to have limited accuracy and predictive power due to a limited treatment of nuclear correlations. Although extensions of this approach have been developed to include a more precise description of correlations, and have been applied to the study of allowed $\beta$-decay of selected nuclei, no systematic calculations including first-forbidden transitions have so far been performed.
\item[Purpose] The goal of this work is to compute $\beta$-decay half-lives of isotonic nuclear chains located at neutron shell closures $N=50$, $82$, $126$ and $184$, which are of particular importance for the $r$-process nucleosynthesis, and study the role of first-forbidden transitions in a framework that includes complex nucleonic correlations beyond the QRPA.
\item[Method] The many-body approach based on the
  relativistic QRPA extended to
  account for the coupling between single-particle and collective degrees of freedom is applied. Both Gamow-Teller and first-forbidden contributions to the decay rates are considered
\item[Results] Overall, the fragmentation of the transition strength
  distributions due to the coupling between single particles and
  collective vibrations systematically increases the transition strength at low excitation energies and yields a decrease of the $\beta$-decay half-lives
  which is particularly important for low Q-values.
  Such effects are crucial to bring Gamow-Teller transitions within the decay energy window, and to reproduce the measured decay rates. 
  Generally, quasiparticle-vibration coupling tends to decrease the probability of decay via FF transition near stability, due to the appearance of Gamow-Teller transitions within the decay $Q$ value.
  While in the lighter systems allowed transitions dominate,
  the decay of $N=126$ and $N=184$ nuclei is found to occur to a large extent
  via first-forbidden transitions, and in particular those
  induced by $1^-$ and $0^-$ operators. 
\item[Conclusions] The relativistic approach based on the coupling
  between nucleons and vibrations has the ability to capture relevant
  correlations that are essential for an accurate predictions of
  $\beta$-decay and provides an ideal framework for future large-scale
  calculations. Upcoming experimental measurements of $\beta$-decay
  rates in the $N=126$ region by radioactive-beam facilities will be
  crucial in order to validate the
  approach. 
\end{description}
\end{abstract}

\maketitle

\section{Introduction}
The synthesis of heavy elements via the $r$ process
\cite{RevModPhys.29.547,Cameron:1957} is an extremely complex
phenomenon which involves several thousands of nuclei and a delicate
interplay between various reaction and decay mechanisms such as
neutron capture, $\beta^-$ decay and fission.  Although the
multi-messenger neutron star merger event GW170817
\cite{LIGOScientific:2017vwq} has provided tremendous information, a
full comprehension of the $r$-process nucleosynthesis requires a more
precise understanding of the nuclear physics involved.  In particular,
an accurate knowledge of masses, $\beta$-decay half-lives,
neutron-capture rates and fission yields, to name a few, are needed
\cite{Cowan.Sneden.ea:2021,Mendoza-Temis.Wu.ea:2015,Eichler_2015,MUMPOWER201686,KAJINO2019109}.
Out of these many ingredients, $\beta$-decay half-lives are of great
importance as they set the time scale of the nucleosynthesis and also
have a crucial impact on the final abundance pattern
\cite{Pfeiffer.Kratz.ea:2001,Arcones.Martinez-Pinedo:2011,Mumpower2014,Eichler_2015,Shafer2016,Marketin2016}.
Because most of the nuclei of interest to the $r$ process are
extremely short lived, they cannot be produced in laboratories and
thus, astrophysical simulations heavily rely on theoretical
predictions.  Providing consistent, precise and predictive nuclear
physics input, for the whole range of nuclei involved, is a tremendous
challenge for nuclear theory.  Microscopic methods based on the
self-consistent mean field or density functional theory (DFT) in
principle represent good candidates to reach this goal as they can
provide a consistent description of most of the nuclear
chart. However, in their present forms, most of these approaches
suffer important limitations and do not meet the degree of precision
needed for reliable astrophysical simulations.

The method currently used to provide global sets of $\beta$-decay rates is known as the quasiparticle random phase approximation (QRPA) \cite{ring2004nuclear}, applied in the charge-exchange (proton-neutron) channel, and corresponds to the small amplitude limit of the time-dependent mean field or time-dependent DFT. 
Several relativistic and non-relativistic versions have been developed and applied to the calculations of $\beta^-$ decay (see \textit{e.g.} Refs.~\cite{Engel1999,Borzov2003,Moeller2003,Niksic2005,NIU2013172,PhysRevC.88.024314,PhysRevC.88.034304,PhysRevC.89.044306,PhysRevC.91.044304,Marketin2016,Ney2020}).
Overall it is known that QRPA suffers from major shortcomings as it only includes a very limited amount of correlations in the description of nuclei due to the neglect of retardation effects in the one-nucleon self-energy. In the charge-exchange channel, this basically restricts the daughter configurations to one-particle-one-hole (1p-1h), or two-quasiparticle (2qp), proton-neutron excitations of the parent ground state. Such approximation typically leads to an imprecise description of transition strength distributions, in particular the very low-energy states, that are essential for an accurate prediction of decay rates. To compensate, QRPA usually relies on extra empirical parameters (proton-neutron pairing) which have to be fitted to the $\beta$-decay half-lives, sometimes separately for each mass region \cite{Engel1999,Niksic2005}. Such a procedure drastically limits the reliability and predictive power of this approach, in particular in unexplored regions of the nuclear chart.
As of today, the three existing global sets of $\beta$-decay rates
have been calculated within different versions of the charge-exchange
QRPA.  The earlier set is based on the finite range droplet model
(FRDM) and combines a microscopic description of Gamow-Teller (GT)
transitions using a schematic separable interaction with a
description of the first-forbidden (FF) modes based on the gross
  theory~\cite{Moeller2003}. More recently, two self-consistent and
fully microscopic QRPA calculations have become available:
Ref. \cite{Marketin2016} based on a covariant framework and
Ref. \cite{Ney2020} based on a non-relativistic Skyrme
functional. While the former allowed a fully relativistic treatment of
$\beta$-decay operators, it considered all nuclei as spherical and
treated odd systems in the same way as paired even-even nuclei with an
odd number of nucleons on average.  The latter, on the other hand,
relied on non-relativistic reductions of the transition operators but
incorporated a more correct treatment of deformation and odd systems.
Overall, the predictions of these three QRPA frameworks disagree on
several points, in particular on the relative importance of GT and FF
transitions in different mass regions and on the order of magnitude of
the total rates above $N=126$. These discrepancies constitute an
important problem as they can lead to large uncertainties in
subsequent astrophysical simulations and predictions of elemental
abundances \cite{Eichler_2015,Marketin2016,Shafer2016,PhysRevC.108.L062802,Chen:2023fpe}. For example, it was found in
Refs.~\cite{Marketin2016,Eichler_2015} that the shortenings of
half-lives in the heavy region yields a broadening of the third
$r$-process abundance peak towards lower masses.

While understanding the origin of the discrepancies between the QRPA
predictions constitute an interesting problem on its own and should be
investigated further, in order to ultimately limit the uncertainties
of $r$-process simulations related to the nuclear physics input, it is
desirable to develop extensions of the (Q)RPA to incorporate
higher-order dynamical processes in the nucleonic self-energy and a
more precise treatment of correlations in the nuclear response. Such
approaches include the Second RPA (SRPA) and methods based on the
coupling of single nucleons to collective nuclear vibrations
(phonons). SRPA introduces a second-order treatment of the nucleonic
self-energy leading to explicit 2p-2h configurations in the excited
states. Recently a charge-exchange version of this approach based on a
Skyrme functional was applied to GT transitions and $\beta^-$-decay of
a few nuclei \cite{Gambacurta:2020dhb,Gambacurta:2021zlv}. However,
due to the large number of 2p-2h configurations, and because the
available versions are currently limited to doubly magic nuclei, the
present range of applicability of this approach remains
limited. 
On the other hand, methods based on the particle-vibration coupling
(PVC), and its extension to open-shell nuclei ---quasiparticle
vibration-coupling (QVC)--- provide an efficient and relatively
low-cost way to include complex nucleonic configurations relevant to
mid- and heavy-mass nuclei.  In this approach, virtual particle-hole
excitations are included up to infinite order in the nucleonic
self-energy and excitations are built from 1p-1h $\otimes$ phonon
configurations on top of the initial state.  Microscopic versions of
the PVC scheme include the non-relativistic one based on Skyrme
interactions \cite{PhysRevC.82.064307} as well as the relativistic
framework based on the meson-nucleon Lagrangian
\cite{PhysRevC.73.044328}.  These two frameworks have been
successfully applied to GT and $\beta$ decay of selected closed and
open-shell nuclei
\cite{Niu:2012mi,PhysRevLett.114.142501,Niu:2016kfj,Niu:2018art,Robin:2016wuh,Robin:2018cjm,Litvinova:2018pmr,Robin:2019jzm}. 
Recently, the formalism of QVC based on a Skyrme functional has been successively extended and applied to the description of GT modes and $\beta$-decay of a few deformed nuclei within the finite amplitude method~\cite{Liu:2023xlv}.
Overall the coupling of nucleons to vibrations induces fragmentation and
spreading of the transition strength distribution which have been
found crucial for an accurate description of the width of the
Gamow-Teller resonances and $\beta$-decay half-lives.  So far,
however, no extensive study of $\beta$ decay including FF transitions
have been performed, even though these transitions are known to
importantly contribute to the decay in many nuclei relevant to the $r$
process, due to large neutron excess.

Although global calculations are not presently possible, one can
study selected regions along the $r$-process path. Among those, the
so-called waiting-point nuclei, located at neutron shell closures, are
of crucial importance. Because these nuclei present low $Q$-values
  for neutron capture, a long sequence of alternating $\beta$ decay
and neutron capture takes place until the $r$-process can move to
heavier neutron numbers. Matter thus accumulates at nuclei closer
  to stability with long beta-decay half-lives and leads to peaks in
the abundance patterns.  While experimental measurements of
neutron-rich $N=50$ and $N=82$ nuclei have been performed, almost no
data is available yet for the $N=126$ and $N=184$ chains (see
\textit{e.g.}  Ref.~\cite{Cowan.Sneden.ea:2021} for a recent review on the
experimental status).  The $N=126$ isotones are particularly
interesting because they have been calculated within several
theoretical frameworks which largely disagree on the contribution of
the forbidden transitions
\cite{Moeller2003,Borzov2003,PhysRevC.88.034304,Marketin2016,Ney2020,SM1,SM2}.

In the present work, we perform systematic calculations of
$\beta^-$-decay half-lives of even-even nuclei with magic neutron
numbers $N=50, 82, 126$ and $184$ within the relativistic QVC
framework and include both allowed GT and FF transitions.  The
manuscript is organized as follows. In section \ref{sec:formalism} we
review the formalism of our approach. We present both the theory of
$\beta^-$-decay that we apply and the formalism of our nuclear
many-body method. The numerical scheme is also explained in detail. In
section \ref{sec:isotones} we present the $\beta^-$-decay half-lives
of the $r$-process waiting-point nuclei and study the role of QVC and
its interplay with first-forbidden transitions.   
In section
\ref{sec:comparison} we compare our results to other existing
theoretical calculations.

\section{Formalism} \label{sec:formalism}

\subsection{Theory of allowed and first-forbidden $\beta$-decay} \label{sec:beta_decay_theory}
In this work we follow the theory of allowed and forbidden $\beta$-decay as was derived by Behrens and Buehring \cite{BEHRENS1971}. The equations have already been summarized in a number of papers, as for instance Ref.~\cite{Marketin2016}. Below we repeat them for completeness.

The rate $\lambda_{\text{tot}}$ for the $\beta^-$ decay from a parent nucleus $(Z,N)$, initially in its ground state, to states in the daughter nucleus $(Z+1,N-1)$ reads
\begin{equation}
\lambda_{\text{tot}} =  \frac{\mbox{ln 2} }{K}   \sum_{f , E_f < - m_e} \Psi(E_f, Z+1) \; .
\label{eq:rate}
\end{equation}
In Eq. \ref{eq:rate}, $K=6144 \pm 2 $ s \cite{K}, the index $f$
denotes the states of the daughter nucleus, $E_f$ is the decay energy of these states defined as $E_f = M_f -M_i$, where $M_i$ and $M_f$ are the initial and final nuclear masses, respectively. The summation is restricted to states that are energetically accessible (with energy $E_f$ up to the electron mass $m_e$).
Finally, $\Psi(E_f, Z+1)$ is the
so-called integrated shape function.  When both allowed and FF
transitions are considered, this function takes the form
\begin{eqnarray}
\Psi(E_f, Z+1) = && \int_1^{W_f} dW W \sqrt{W^2-1} (W_f -W)^2 \nonumber \\
&& \times F(Z+1, W)  C(W)  \; . \nonumber \\ 
\label{eq:shape_fct}
\end{eqnarray}
In Eq.~\ref{eq:shape_fct}, $W = E_e/ m_e$ denotes the electron energy in unit of the electron mass, $W_f \equiv (M_i - M_f)/m_e = - (E_f/m_e) $ is the maximal electron energy, and $F(Z+1, W) $ is the Fermi function \cite{BEHRENS1971} accounting for the Coulomb interaction between the emitted electron and the daughter nucleus. Finally, $C(W)=C_{\text{all.}}(W)$ or $C_{\text{FF}}(W)$ is the shape factor corresponding to the allowed or FF transition, as described below. 

In the case of allowed transitions, the shape factor
$C_{\text{all.}}(W)$ is in fact independent of the energy $W$ and
coincides with the GT reduced transition probabilities,
\begin{equation}
C_{\text{all.}} = B(\text{GT}) = g_A^2  \frac{| \langle f || \bm{O}_{GT} || i \rangle |^2 }{2J_i +1} \; ,
\end{equation}
where $g_A= -1.2764(13)$ \cite{Workman:2022ynf} is the weak axial coupling constant, $J_i$ is the angular momentum of the initial state, and $\bm{O}_{GT}$ is the relativistic GT transition operator
\begin{equation}
\bm{O}_{GT} = \sum_{k=1}^A \bm{\Sigma}^k\, \bm{t}_-^k =
\sum_{k=1}^A
\begin{pmatrix}
\bm{\sigma}^k & 0 \\
0 & \bm{\sigma}^k
\end{pmatrix}
\bm{t}_-^k \; ,
\end{equation}
with $\bm{\sigma}$ the Pauli spin operator and $\bm{t}_- = \bm{\tau}/2$ the operator converting a neutron into a proton.

In the case of first-forbidden decays, the shape factor $C_{\text{FF}}(W)$ depends on the energy $W$ and the nuclear transition matrix elements of forbidden operators: 
\begin{equation}
C_{\text{FF}}(W) = k + ka \, W + kb \, W^{-1} + kc \, W^2 \; .
\end{equation}
Within the formalism of Behrens and Buehring \cite{BEHRENS1971}, the coefficients $k$, $ka$, $kb$ and $kc$ read
\begin{eqnarray}
k &=& \bigl[ \zeta_0^2 + \frac{1}{9} w^2 \bigr]_{(0)}  \nonumber \\
&&+ \bigl[ \zeta_1^2 + \frac{1}{9} (x+u)^2 - \frac{4}{9} \mu_1 \gamma_1 u (x+u) \nonumber \\
&&+ \frac{1}{18} W_f^2 (2x+u)^2 - \frac{1}{18} \lambda_2 (2x-u)^2 \bigl]_{(1)} \nonumber \\
&&+ \bigl[ \frac{1}{12} z^2 ( W_f^2 - \lambda_2)\bigr]_{(2)} \; , \label{eq:k} \\
ka &=& \bigl[ -\frac{4}{3} u Y - \frac{1}{9} W_f (4x^2 + 5u^2) \bigr]_{(1)}  \nonumber \\ 
&& - \bigl[ \frac{1}{6} W_f z^2\bigr]_{(2)} \; , \label{eq:ka} \\
kb &=& \frac{2}{3} \mu_1 \gamma_1 \left( - \bigl[ \zeta_0 w \bigr]_{(0)} + \bigl[ \zeta_1 (x+u) \bigr]_{(1)} \right)  \; , \label{eq:kb} \\
kc &=& \frac{1}{18} \bigl[ 8u^2 + (2x+u)^2 + \lambda_2 (2x-u)^2 \bigr]_{(1)}  \nonumber \\
&& + \frac{1}{12} \bigl[ (1+\lambda_2) z^2\bigr]_{(2)} \; , \label{eq:kc}
\end{eqnarray}
where 
\begin{eqnarray}
\zeta_0 &= V +\frac{1}{3} w W_f &\mbox{ with } V = \xi 'v + \xi w' \label{eq:zeta0} \\
\zeta_1 &= Y + \frac{1}{3} (u-x) W_f &\mbox{ with } Y = \xi 'y - \xi (u' + x') \label{eq:zeta1} \\
\gamma_1 &= \sqrt{1-(\alpha Z)^2}  \; . \label{eq:gamma1}
\end{eqnarray}
In the above equations, $\xi = \alpha Z /2R$ where $\alpha$ is the fine structure constant and $R$ is the radius of a uniformly charged sphere 
approximating the nuclear charge distribution. Under this approximation, $R$ is determined from the root mean square charge radius $\sqrt{\langle r \rangle ^2}$ 
(calculated in the mean-field approximation, see section \ref{sec:num_scheme}) as 
$R = \sqrt{5 \langle r \rangle ^2 /3}$.
The parameters
$\mu_1$ and $\lambda_2$ are related to the emitted electron and are
approximated by $\mu_1 =1$ and $\lambda_2=1$
\cite{behrens1982electron,Warburton1991}.  
The number $(J)$ in subscript after the square brackets in Eqs.~(\ref{eq:k})-(\ref{eq:kc}) denotes the rank of the operators appearing in the bracket. 

The nuclear transition matrix elements are given by (in the Condon-Shortley phase convention)
\begin{eqnarray}
&& w = - g_A \sqrt{3} \hat{J}_i^{-1} \langle f || \sum_k r_k \bigl[  \boldsymbol{C}_1^k \otimes \boldsymbol{\Sigma}^k \bigr]_{(0)} \bm{t}_-^k || i \rangle  \; , \label{eq:FF_w} \\
&& x = - \hat{J}_i^{-1} \langle f || \sum_k  r_k \boldsymbol{C}_1^k  \bm{t}_-^k || i \rangle \; , \label{eq:FF_x} \\
&& u = - g_A \sqrt{2} \hat{J}_i^{-1} \langle f || \sum_k r_k \bigl[  \boldsymbol{C}_1^k \otimes \boldsymbol{\Sigma}^k \bigr]_{(1)} \bm{t}_-^k || i \rangle  \; , \label{eq:FF_u} \\
&& z = 2 g_A  \hat{J}_i^{-1} \langle f || \sum_k r_k \bigl[  \boldsymbol{C}_1^k \otimes \boldsymbol{\Sigma}^k \bigr]_{(2)} \bm{t}_-^k || i \rangle  \; ,  \label{eq:FF_z}\\
&& w' =- g_A \frac{2}{\sqrt{3}} \hat{J}_i^{-1}  \nonumber \\
&& \hspace{2mm} \times \langle f || \sum_k r_k I(1,1,1,1,r_k)\bigl[  \boldsymbol{C}_1^k \otimes \boldsymbol{\Sigma}^k \bigr]_{(0)} \bm{t}_-^k || i \rangle  \; , \label{eq:FF_wp} \\
&& x' = - \frac{2}{3} \hat{J}_i^{-1} \nonumber \\
&& \hspace{2mm} \times \langle f || \sum_k  r_k I(1,1,1,1,r_k) \boldsymbol{C}_1^k  \bm{t}_-^k || i \rangle \; , \label{eq:FF_xp} \\
&&u' = - g_A \frac{2\sqrt{2}}{3} \hat{J}_i^{-1} \nonumber \\ 
&& \hspace{2mm} \times \langle f || \sum_k  r_k I(1,1,1,1,r_k)  \bigl[ \boldsymbol{C}_1^k \otimes \boldsymbol{\Sigma}^k \bigr]_{(1)} \bm{t}_-^k || i \rangle  \; , \label{eq:FF_up}
\end{eqnarray}
where 
\begin{equation}
\boldsymbol{C}_{LM} = \sqrt{\frac{4\pi}{2L+1}} Y_{LM} \; ,
\end{equation}
and $\hat{J}_i \equiv \sqrt{2J_i +1} =1$ in the case of decay from even-even ground states.
The function $I(1,1,1,1,r_k)$ accounts for the nuclear charge distribution which is approximated by a uniform spherical charge distribution as \cite{BEHRENS1971}
\begin{eqnarray}
I(1,1,1,1,r_k) = \frac{3}{2}  \left\{
    \begin{array}{ll}
        \displaystyle{1- \frac{1}{5} \left( \frac{r}{R} \right)^2 \; , 0 \leq r \leq R \; ,} \\[4mm]
         \displaystyle{\frac{R}{r} - \frac{1}{5}  \left( \frac{R}{r} \right)^3 \; , r >R \;.}
    \end{array}
\right.
\end{eqnarray} 
Finally the transition matrix elements originating from relativistic contributions are
\begin{eqnarray}
\xi 'v &=& - g_A \hat{J}_i^{-1} \langle f || \sum_k  \gamma_5^k  \bm{t}_-^k || i \rangle \; , \label{eq:gamma5} \\
\xi 'y &=& - \hat{J}_i^{-1} \langle f || \sum_k  \boldsymbol{\alpha}^k
           \bm{t}_-^k || i \rangle \label{eq:alpha} \; ,
\end{eqnarray}
where $\boldsymbol{\alpha} = \gamma^0 \boldsymbol{\gamma}$.
These transitions
connect the small and large components of the Dirac spinors, and thus,
also change the parity of the nuclear state.  Since we will work in a
relativistic framework, we do not perform any non-relativistic
reductions of the matrix elements (\ref{eq:gamma5}) and
(\ref{eq:alpha}).

\subsection{Nuclear Many-Body Method: Proton-Neutron Relativistic Time Blocking Approximation (RQTBA)}
In this work the nuclear transition matrix elements are calculated within the relativistic charge-exchange (or proton-neutron) QVC approach which represents an extension of the charge-exchange relativistic QRPA (RQRPA) accounting for the coupling between single quasiparticles and collective nuclear vibrations, as first developed in Ref. \cite{Robin:2016wuh}.

This approach is based on the calculation of the transition strength distribution $S({\bm O},E)$ 
\begin{eqnarray}
S({\bm O},E) &=& \sum_f |\braket{f|\bm{O}|i}|^2 \delta (E-E_f) \, , 
\label{eq:strength1}
\end{eqnarray}
which characterizes the response of a nucleus to an external field
$\bm{O}$. Here $\bm{O}$ will coincide with the GT or FF operators from
section \ref{sec:beta_decay_theory}, which convert neutrons into
protons. 
In Eq.~(\ref{eq:strength1}), $E$ denotes the energy variable, and as in the previous section, $\ket{i}$ and $\ket{f}$ are the initial parent ground state and final daughter state, respectively.
The definition (\ref{eq:strength1}) for the strength distribution can easily be re-written in terms of the response function $R(\omega)$, or propagator of a correlated proton-neutron quasiparticle pair in the nuclear medium (see appendix~\ref{sec:appendix}), as
\begin{eqnarray}
S(\bm{O},E) &=& -\frac{1}{\pi} \lim_{\Delta\rightarrow 0^+} \mbox{Im} \sum_{1234} O_{12}^{*} R_{1423} (\omega) O_{34} \; ,
\label{eq:strength}
\end{eqnarray}
where $\omega = E+i\Delta$ is the complex energy variable.
In Eq.~(\ref{eq:strength}) we have introduced a single-quasiparticle basis
$i = \{ \eta_i, k_i \}$ where $k_i = \{ n_i, \pi_i, j_i\}$ denotes single-particle states and
$\eta_i=\pm$ denotes the upper and lower component in the Nambu-Gorkov space, for nuclei with superfluid pairing correlations. In the following we will use odd (resp. even) numbers to denote proton (resp. neutron) single-particle states.

In the present framework, the response function is obtained within the framework of the linear response theory, by solving the following Bethe-Salpeter equation which accounts for QVC effects in the time-blocking approximation \cite{TBA}:
\begin{eqnarray}
R_{1423} (\omega)  &=& \widetilde{R}^{(0)}_{1423}  (\omega) \nonumber \\
&&  + \sum_{5678} \widetilde{R}^{(0)}_{1625} (\omega) W_{5867} (\omega)  R_{7483} (\omega) \; .  \nonumber \\
\label{eq:BSE_pn}
\end{eqnarray}
The free two-quasiparticle proton-neutron propagator $\widetilde{R}^{(0)}$ is given by
\begin{eqnarray}
R^{(0)}_{1423}  (\omega) = \frac{  \delta_{\eta_1, -\eta_2} \delta_{\eta_3, -\eta_4} \delta_{13}  \delta_{24}  }{\eta_1 (\omega-\lambda_{pn}) - \mathcal{E}_{k_1} - \mathcal{E}_{k_2} } \; ,\nonumber \\
\end{eqnarray}
where $\mathcal{E}_{k_i}$ and $\mathcal{E}_{k_i}$ are neutron and proton quasiparticle energies, respectively, and 
$\lambda_{pn} = \lambda_p - \lambda_n$ is the proton-neutron chemical potential difference.

\begin{figure}[ht]
\centering{\includegraphics[width=\columnwidth]{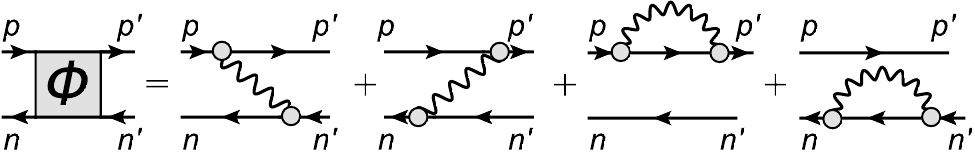} }
\caption{Energy-dependent proton-neutron QVC interaction $\Phi(\omega)_{pn',np'}$ from Eq.~(\ref{eq:Phi}). The gray blobs represent the vertices of Eq.~\ref{eq:vertex}.}
\label{f:Phi_QVC}
\end{figure}

The effective proton-neutron interaction $W(\omega)$ is given by
\begin{equation}
W(\omega) = \widetilde{V} +  \Phi(\omega) \; ,
\label{eq:int}
\end{equation}
which is the sum of the static interaction $\widetilde{V}$ and an energy-dependent term $\Phi(\omega)$ responsible for damping effects. 
The interaction $\Phi(\omega)$ originates from QVC and describes the virtual emission and re-absorption of a nuclear vibration by a quasiparticle, as well as the exchange of a vibration between a proton and a neutron quasiparticle, as represented in Fig.~\ref{f:Phi_QVC}. 

It is given by
\begin{eqnarray}
\Phi_{1423} (\omega) &\equiv& 
\Phi^{\eta_1 \eta_4 , \eta_2 \eta_3}_{p_1 n_4, n_2 p_3} (\omega)  \nonumber \\
&=&  \delta_{\eta_1,\eta_3} \delta_{\eta_1, -\eta_2} \delta_{\eta_3, -\eta_4} \times \sum_{\mu \xi} \delta_{\xi,\eta_1} \nonumber \\
\times \Bigl[ && \delta_{k_1 k_3} \sum_{k_6} \frac{g_{\mu ;k_6 k_2}^{\eta_1; -\xi -\xi} g_{\mu ;k_6 k_4}^{\eta_1; -\xi -\xi *}}{\eta_1 (\omega - \lambda_{pn}) - \mathcal{E}_{k_1} -\mathcal{E}_{k_6} - \Omega_\mu } \nonumber \\
&&+ \delta_{k_2 k_4} \sum_{k_5} \frac{g_{\mu ;k_1k_5}^{\eta_1; \xi \xi} g_{\mu ;k_3k_5}^{\eta_1; \xi \xi *} }{\eta_1 (\omega - \lambda_{pn}) - \mathcal{E}_{k_5} -\mathcal{E}_{k_2} - \Omega_\mu} \nonumber \\
&&- \frac{g_{\mu ;k_1k_3}^{\eta_1; \xi \xi} g_{\mu ;k_2k_4}^{\eta_1; -\xi -\xi *} }{\eta_1 (\omega - \lambda_{pn}) - \mathcal{E}_{k_3} -\mathcal{E}_{k_2} - \Omega_\mu} \nonumber \\
&&- \frac{g_{\mu ;k_3k_1}^{\eta_1; \xi \xi *} g_{\mu ;k_4 k_2}^{\eta_1; -\xi -\xi } }{\eta_1 (\omega - \lambda_{pn}) - \mathcal{E}_{k_1} -\mathcal{E}_{k_4} - \Omega_\mu} \Bigr] \;, \nonumber \\
\label{eq:Phi}
\end{eqnarray}
where the index $\mu$ labels the collective phonons of frequency $\Omega_\mu$ that couple to the quasiparticle states through the vertices 
\begin{equation}
g_{\mu ;k_1 k_2}^{\eta_\mu ; \eta_1 \eta_2} = \delta_{\eta_\mu,+1} g_{\mu ;k_1 k_2}^{\; \eta_1 \eta_2} + \delta_{\eta_\mu,-1} g_{\mu ;k_2 k_1}^{\; \eta_2 \eta_1*} \; ,
\label{eq:vertex}
\end{equation}
where $k$ denotes generic nucleon states.
These vertices are obtained from the solutions of the RQRPA as
\begin{equation}
g_{\mu ;k_1 k_2}^{\; \eta_1 \eta_2} = \sum_{k_3 k_4} \sum_{\eta_3 \eta_4} \widetilde{V}_{k_1 k_4 , k_2 k_3}^{\eta_1 \eta_4 \eta_2 \eta_3} \mathcal{R}_{\mu;k_3 k_4}^{\; \eta_3 \eta_4} \; ,
\end{equation}
where $\mathcal{R}_{\mu;k_3 k_4}^{\; \eta_3 \eta_4}$ denotes the RQRPA transition densities associated with each phonon $\mu$.
\\
\\
Due to the time-blocking approximation~\cite{TBA} that was applied in order to derive Eq.~(\ref{eq:BSE_pn}) the approach described above has usually been referred to as proton-neutron Relativistic Quasiparticle Time Blocking Approximation (RQTBA)~\cite{Robin:2016wuh}. We will use this term throughout the paper. Note that when $\Phi(\omega)$ is neglected, Eq.~(\ref{eq:BSE_pn}) reduces to the usual proton-neutron RQRPA.

\subsection{Numerical scheme}
\label{sec:num_scheme}


The starting point of the present approach is the effective Lagrangian
which describes the interaction between Dirac nucleons in terms of
meson ($\pi$, $\sigma$, $\omega$ and $\rho$) and photon exchange.
Because retardation effects are neglected, the resulting
meson-exchange interaction $\widetilde{V}$ is static
\cite{RING1996193}.

\paragraph*{Relativistic mean field:} In order to determine the basis
of quasiparticle states $(k_i,\eta_i)$, we start from a relativistic
mean-field (RMF) calculation.  In this work we consider the NL3
parametrization of the meson-nucleon Lagrangian \cite{NL3}, which
includes non-linear $\sigma$ terms, and whose coupling constants were
fitted to reproduce masses and radii of a few doubly-magic nuclei in
the RMF approximation. This parametrization was found to give a good
description of GT low-energy modes and giant resonances in previous
studies \cite{Robin:2016wuh,Robin:2018cjm}.  Since the RMF does not include the exchange interaction
(Fock term), the pion does not contribute at this level (as it would
break parity) \cite{RING1996193} and thus, is not included in the NL3
Lagrangian.

In order to treat open-shell nuclei, we also include neutron-neutron
and proton-proton isovector ($T=1$) pairing correlations.  It is well
known that in calculations based on the RMF, the pairing part of the
interaction is usually treated on a different footing than the
particle-hole (meson-exchange) component. Indeed, even though attempts
were made to describe pairing correlations from the effective
meson-exchange force, such studies so far did not succeed in
reproducing the empirical pairing gaps
\cite{Kucharek1991,Sierra2001}. Thus typically one uses a different
(non-relativistic) ansatz to describe the pairing component.  In this
work we use a simple monopole-monopole force for the $T=1$ pairing
(particle-particle) channel \cite{BONCHE1994185}:
\begin{eqnarray}
\widetilde{V}_{(k_1k_2k_3k_4)}^{(\pp \pp)(J,T=1)} = && - \frac{G}{2} \delta_{J0}  \delta_{(k_1k_2)} \delta_{(k_3k_4)} \sqrt{\frac{2 j_{k_1}+1}{1 + e^{(\varepsilon_{k_1}-w)/d}}} \nonumber \\
&& \times \sqrt{\frac{2 j_{k_3}+1}{1 + e^{(\varepsilon_{k_3}-w)/d}}}  \; ,
\label{eq:T1_pp_int}
\end{eqnarray}
for which the relativistic Hartree-Bogoliubov equations reduce to the relativistic Hartree-BCS problem.
These $T=1$ pairing correlations are included in a smooth window of value $w=20$ MeV and diffuseness $d=1$ MeV around the Fermi level. The parameter $G$ is typically adjusted to reproduce the pairing gaps of nuclei under study. Since no experimental data is available for most of the neutron-rich nuclei that we will consider, we adjust this parameter to the 
empirical gap formula $12/\sqrt{A}$.  

\paragraph*{Phonons:} As a second step, the spectrum of phonons that
couple to the single quasiparticles are calculated within the
RQRPA. Here we select neutral (non charge-exchange) natural-parity
phonons with multi-polarities $J^\pi=2^+,3^-,4^+,5^-,6^+$ with
excitation energy up to $\Omega_\mu^{\text{max}} = 20$ MeV.  The
collectivity of the phonons is ensured by selecting those realizing at
least $5\%$ of the highest transition probability for a given
multipole.  These criteria were found to provide good convergence of
the transition strength distributions in previous studies
\cite{Robin:2016wuh}. In Ref. \cite{Robin:2018cjm} we also
investigated the effect of charge-exchange phonons. Although such
phonons can be important for the description of the overall transition
strength, in particular the giant resonance, the effects at low
energies relevant to $\beta$ decay remained small in comparison to the
effect of neutral phonons. Thus, in order to keep the size of the
model space tractable for the heavy nuclei studied in this paper, we
do not include the coupling of quasiparticles to charge-exchange
phonons.

\paragraph*{Proton-neutron response in the RQTBA:} We then solve the
Bethe-Salpeter equation (\ref{eq:BSE_pn}) for the proton-neutron
response function with given spin and parity $J^\pi$ ($1^+$ for GT and
$0^-,1^-, 2^-$ for the FF modes).  

In the proton-neutron channel, the particle-hole component of the static interaction $\widetilde{V}$ in Eq.~(\ref{eq:int}) is described by exchange of isovector mesons
\begin{equation}
 \widetilde V^{(\pp \hh)}_{pn}= \widetilde V_\rho  + \widetilde V_\pi + \widetilde V_{\delta_{\pi}}
\end{equation}
where $\widetilde V_\rho$ and $\widetilde V_\pi$ are the rho-meson and pion exchange interactions, and $V_{\delta_\pi}$ is the zero-range Landau-Migdal term which accounts for short-range correlations \cite{PhysRevC.36.380}. In this work the strength of $V_{\delta_\pi}$ is taken to be $g'=0.6$, as the Fock term is not treated explicitly \cite{PhysRevLett.101.122502}.

The proton-neutron particle-particle component $\widetilde{V}_{pn}^{\pp \pp}$ of $\widetilde{V}$ can be coupled to $T=0$ in the case of unnatural-parity modes (GT, $0^-$ and $2^-$), or to $T=1$ in the case of natural parity modes ($1^-$) \cite{Ravlic:2021uvo,PhysRevC.103.064307}.
Due to isospin symmetry, the isovector component is given by Eq.~(\ref{eq:T1_pp_int}), and because of its monopole nature, does not contribute to $J=1$ transitions.
The isoscalar proton-neutron pairing is not constrained by isospin symmetry, and in general is difficult to determine \cite{Engel1999,Niksic2005}.
In most of our applications, we will actually not include the $T=0$ static pairing interaction. This is justified and investigated in more details below in section \ref{sec:pn_pairing}. 
Proton-neutron pairing is however naturally generated by an interplay between T=1 like-particle pairing and the dynamical proton-neutron interaction $\Phi_{pn}(\omega)$. Indeed, since each nucleon is described as a quasiparticle (superposition of particle and hole), $\Phi_{pn}(\omega)$ contains both particle-hole and particle-particle channels. Thus, such “dynamical proton-neutron pairing” arises naturally, without introducing extra parameters.

In summary the effective proton-neutron interaction in
Eq.~\eqref{eq:int} has the following components 
\begin{eqnarray}
\left\{
    \begin{array}{lll}
W_{pn}^{(\pp \hh)(J)} (\omega) &=&  \widetilde V^{(\pp \hh)(J)}_{pn} +  \Phi^{(\pp \hh) (J)}_{pn} (\omega)  \nonumber \\
W_{pn}^{(\pp \pp)(J,T=1)} (\omega) &=&  \widetilde V^{(\pp \pp)(J,T=1)}_{pn} \delta_{J0} +  \Phi^{(\pp \pp)(J,T=1)}_{pn} (\omega) \\
W_{pn}^{(\pp \pp) (J,T=0)} (\omega) &=&   [ \widetilde V^{(\pp \pp)(J,T=0)}_{pn} ] +  \Phi^{(\pp \pp) (J,T=0)}_{pn} (\omega) \; ,
    \end{array}
\right.
\end{eqnarray}
where $V^{(\pp \pp)(J,T=0)}_{pn}$ will be usually taken to be zero, except
in section \ref{sec:pn_pairing} (see explanations in that section). 

The quasiparticle-quasihole pairs entering the quasiparticle-phonon coupling are selected according to their excitation energy. 
In this work, we include QVC in an energy window of $30$ MeV for $N=50$ and $N=82$ nuclei. This value was determined based upon previous studies (for example, Ref.~\cite{Robin:2016wuh}) where we found that a $20$ MeV window provided overall convergence of the strength distribution and a $30$ MeV window ensured complete convergence of the detailed low-lying GT strength distribution and hence of the GT contribution to the half-lives in the Ni isotopic chain. For the heavy $N=126$ and $N=184$ chains, we used a $20$ MeV window in order to keep the size of the model space manageable. More detailed convergence studies of the GT and FF strength for the different mass regions would require improvement in the computational code and we leave it for a future work.

\paragraph*{Note on the subtraction procedure:} When applied in the
neutral (non charge-exchange) channel, the RQTBA originally applied a
so-called ``subtraction procedure'' (initially introduced in
\cite{Tselyaev2013}) in order to avoid double counting of correlation
effects that are implicitly contained in the parameters of the
meson-nucleon functional. In practice this procedure amounts to
subtracting from $\Phi(\omega)$ its value at zero frequency
$\Phi(0)$. In previous applications of the RQTBA to Gamow-Teller modes
\cite{Robin:2016wuh,Robin:2018cjm} we argued that since, in the case
of unnatural-parity modes, the pion provides almost all the
contribution to the transition, while the $\rho$ meson is negligible,
the subtraction procedure should not be applied. Indeed, as the pion
does not contribute at the RMF level, it is considered here with
free-space coupling constant ($f_\pi^2 / 4\pi = 0.08$). Thus in the
case of unnatural-parity charge-exchange transitions, such as GT
operators, as well as $0^-$ and $2^-$ FF operators, the
double-counting of correlations is expected to be negligible and,
therefore, no subtraction procedure is employed. For the case of $1^-$
modes, however, only the $\rho$ meson contributes and thus the
subtraction technique is applied.  In that case, the subtraction
energy is taken to be the proton-neutron chemical potential difference
$\lambda_{pn}$, which is the analogous of zero energy in the
charge-exchange channel, and ensures that we do not subtract at a pole
value. That is, we perform the substitution
\begin{equation}
\Phi(\omega) \rightarrow \Phi(\omega)  - \Phi(\lambda_{pn}) \; . 
\end{equation}
%

\paragraph*{Calculation of GT and FF rates:}
The GT contribution to the $\beta$-decay rates can then easily be obtained from the strength distribution $S(\bm{O}_{GT},E)$, as in Eq.~(\ref{eq:strength}).
The contributions to the FF transitions $k$, $ka$, $kb$ and $kc$ in Eqs.~(\ref{eq:k})-(\ref{eq:kc}) require the calculation of squared amplitudes $X^2$, as well as interference terms $X X'$, where $X,X' \in \{ w, x, u, z, w', x',u', \xi'v ,\xi'y \}$ (Eqs.(\ref{eq:FF_w})-(\ref{eq:FF_up}), (\ref{eq:gamma5}), (\ref{eq:alpha})). 
Similarly to the GT part, the FF squared amplitudes can be obtained from the strength distributions corresponding to the different FF multipoles operators $\bm{O}$. As mentioned in Ref.~\cite{Mustonen:2014bya}, the interference terms between two operators $\bm{O}$ and $\bm{O'}$ can be obtained from the calculation of a "mixed" strength distribution
where the response function is folded with both operators as
\begin{eqnarray}
    S(\bm{O},\bm{O'},E) =
 - \frac{1}{\pi} \lim_{\Delta \rightarrow 0^+} \mbox{Im}  \sum_{1423} \sum_{1423} 
            \bm{O}^{*}_{12}  R_{1423} (\omega) \bm{O'}_{34} \; . \nonumber \\
\end{eqnarray}
More details are provided in appendix~\ref{sec:appendix}.
In practice, we use a small value $\Delta = 20$ keV, 
to ensure that states that are close to the integration upper limit are fully included in the integral.
More precisely, this small value of the width allows to identify each individual peak in the integration window, from which we can straightforwardly deduce the discrete transition matrix elements. This can be done for both usual strength as well as “mixed strength” for the interference terms, and avoids any divergence issue related to increase of the phase space for decreasing energies.

Finally, we emphasize that, in this work, we do not apply any
quenching factor to the nuclear transitions and consider the bare
value of the axial constant $g_A$.

%
\section{$\beta$-decay half-lives of $r$-process waiting-point nuclei} \label{sec:isotones}
\subsection{$N=50$ chain} \label{sec:N50}
\subsubsection{$\beta$-decay half-lives and contribution of the FF transitions}
\begin{figure}[ht]
  \centering
  \includegraphics[width=\linewidth]{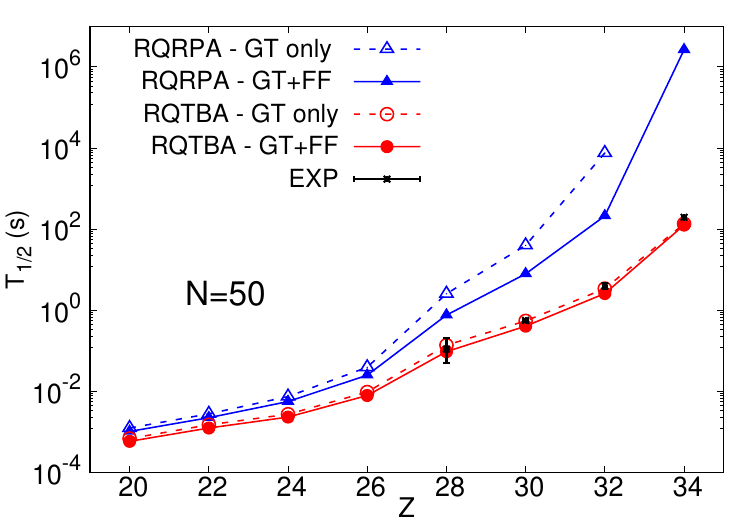}
  \caption{$\beta$-decay half-lives of $N=50$ isotones obtained within
    RQRPA (blue triangles) and RQTBA (red circles), in the allowed GT
    approximation (empty symbols) and including the FF transitions
    (full symbols). When no symbol is shown, the nucleus is
      predicted to be stable. The experimental data is taken from
    Ref.~\cite{data}.} 
\label{f:hlives_N50}
\end{figure}
We show in Fig. \ref{f:hlives_N50} the $\beta$-decay half-lives of
$N=50$ isotones obtained within RQRPA and RQTBA, {\it i.e.} without
and with the QVC interaction $\Phi(\omega)$ in Eq.~(\ref{eq:BSE_pn}),
respectively.  The empty symbols show the half-lives obtained in the
allowed Gamow-Teller approximation, while the full symbols show the
half-lives including the contribution of the first-forbidden
transitions.
We note that the effect of QVC becomes less and less
important near the dripline since the decay $Q$ value is
large. However the QVC is crucial closer to stability as the $Q$ value
becomes small and thus the half-lives are very sensitive to the
details of the transition strength distributions. In that case the
correlations are able to reproduce the trend of the data to a much
better extent. The half-life of $^{78}$Ni is within the experimental
error bar while those of $Z>28$ nuclei are slightly underestimated, up
to $\simeq 34 \%$ in the case of $^{84}$Se.

\begin{figure}[ht]
\centering{\includegraphics[width=\columnwidth]{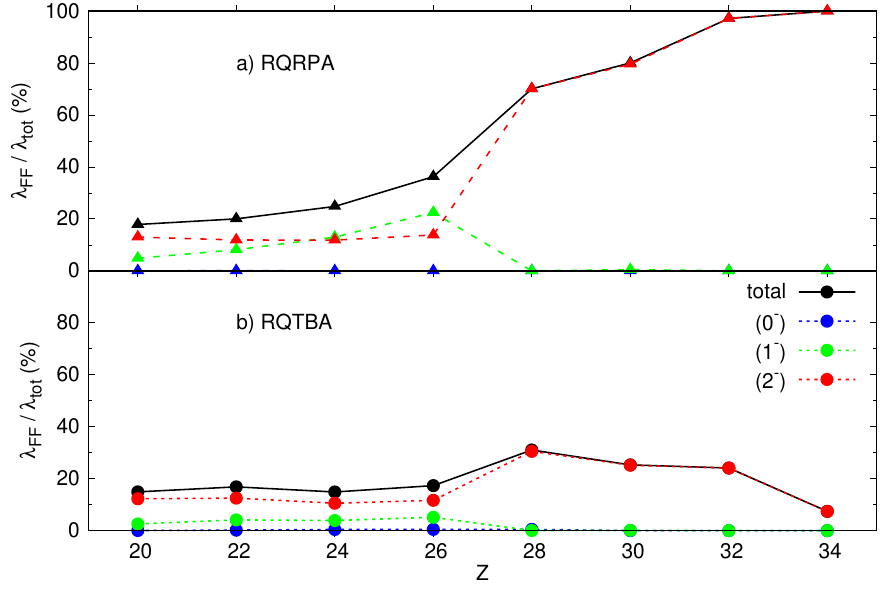} }
\caption{Contribution of first-forbidden transitions to the $\beta$-decay rate of $N=50$ nuclei (in $\%$) within a) RQRPA and b) RQTBA. The full contribution is shown with a plain black line while the dashed lines show the individual contribution of the different multipoles.}
\label{f:FF_N50}
\end{figure}

We show in Fig.~\ref{f:FF_N50} the contribution $\lambda_{\text{FF}}/\lambda_{\text{tot}}$ (in percent) of the FF transitions to the total $\beta$-decay rate, in both RQRPA and RQTBA. We also show in this figure the individual contributions of the different multipoles $0^-$, $1^-$ and $2^-$. 

\begin{figure}[ht]
\centering{\includegraphics[width=\columnwidth]{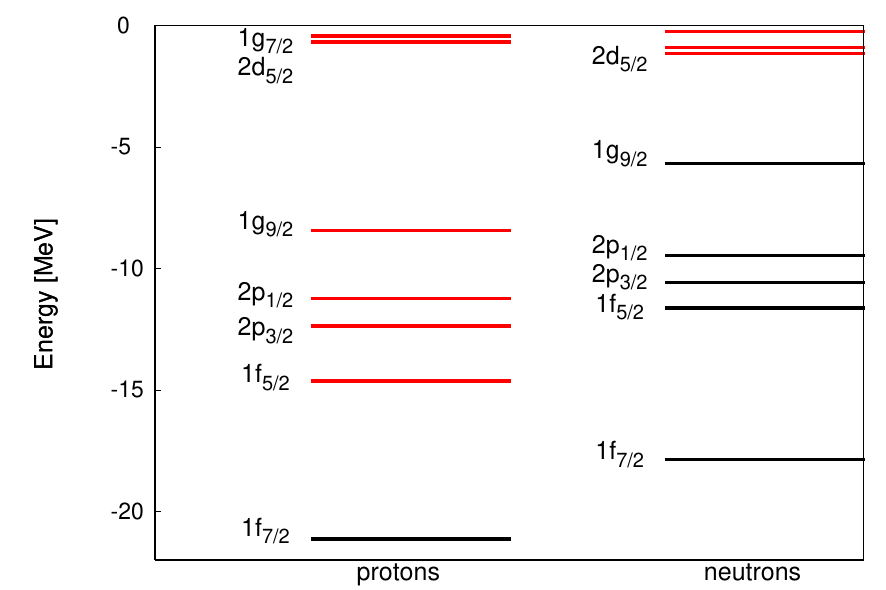} }
\caption{Single-particle spectrum in $^{78}$Ni. The black levels are fully occupied, while the red ones are empty.} 
\label{f:SPE_78Ni}
\end{figure}

In order to understand the results, we analyze the transition densities of the different modes. 
In the RQTBA framework, such densities can be obtained (up to a phase) using a formula from Ref.~\cite{Litvinova:2007gg}:
\begin{equation}
    \rho(\boldsymbol{O},\Omega^\nu)_{pn} = \lim_{\Delta \rightarrow 0} \sqrt{\frac{\Delta}{\pi S(\boldsymbol{O},\Omega^\nu)}} \; \text{Im} \, \delta \rho_{pn} (\Omega^\nu + i \Delta) \; ,
\end{equation}
where the index $\Omega^\nu$ denotes the energy of a particular peak $\nu$ in the strength distribution, and $\delta \rho_{pn} (\omega) = \sum_{p'n'} R_{pn'np'} \boldsymbol{O}_{p'n'}$. Interference terms of the first forbidden transitions and their signs can be analyzed using the techniques detailed in App.~\ref{sec:appendix}.
To guide the reader in the following
discussion, we show in Fig.~\ref{f:SPE_78Ni} the single-particle
spectrum of $^{78}$Ni obtained in the RMF approximation with the NL3
functional. 
The RMF occupations of the neutron levels remain the same
throughout the $N=50$ chain, while the occupations of the proton
states change. In particular, the proton $1f_{7/2}$ is partially empty
in isotones with $Z<28$, while it is basically fully filled in heavier isotones within the RMF.
\\ \\
In nuclei with $Z < 28$, we find in both RQRPA and RQTBA a dominant
contribution from Gamow-Teller modes to the total
rate, 
mainly originating from the low-energy $\n 1f_{5/2} \rightarrow \p 1f_{7/2}$
transition, although the weight of FF modes increases to
$\simeq 36 \%$ at $Z=26$ in RQRPA.  
In both cases the contribution of
$0^-$ modes is negligible (below $0.5 \%$), as these transitions occur
between states of same total angular momentum $j$ and different parity
(e.g. $\n 1d_{5/2} \rightarrow \p 1f_{5/2}$, $\n 2p_{1/2} \rightarrow \p 2s_{1/2}$
and $\n 2p_{3/2} \rightarrow \p 1d_{3/2}$) and thus involve proton states
that are almost fully occupied.  
RQTBA predicts a contribution from
$1^-$ and $2^-$ modes of roughly the same order (3--12\%), with the
major transitions being $\n 1g_{9/2} \rightarrow \p 1f_{7/2}$ for the $1^-$
modes and $\n 1g_{9/2} \rightarrow \p 1f_{7/2,5/2}$ for the $2^-$
operator. To a lesser extent, but with a similar order of magnitude,
the transition $\n 2d_{5/2} \rightarrow \p 1f_{7/2}$ also contributes to the
$2^-$ modes. Such a transition is due to the presence of ground-state
correlations which induce slight occupation of the neutron $2d_{5/2}$
sub-shell.

Since we do not include ground-state correlations (GSC) induced by QVC, these are
caused by the GSC of the RQRPA (the usual ``B'' interaction matrix
\cite{ring2004nuclear}).  In RQRPA, the contribution of the rank-1
operators is slightly larger and increases with $Z$ (up to $22 \%$ in
$^{76}$Fe) even though the occupation of the proton $1f_{7/2}$ becomes
larger (this is because the decrease in GT contribution is stronger
than the $1^-$ decrease, thus the relative contribution of $1^-$ goes
up).

In nuclei with $Z \geq 28$, the proton $1f_{7/2}$ sub-shell becomes
fully occupied, and transitions via rank-1 operators are blocked so
that only $2^-$ modes generate FF transitions. The main low-energy
Gamow-Teller transition $1f_{5/2} \rightarrow 1f_{7/2}$ is also
blocked. In RQRPA, transitions between other sub-shells are located
near the $Q$ value, which explains the small or absent contribution of
GT transitions in nuclei with large $Z$. 
The correlations due to QVC
lower down such transitions (in particular,
$\n 2p_{1/2,3/2} \rightarrow \p 2p_{1/2,3/2}$,
 $\n 1f_{5/2} \rightarrow \p 2p_{3/2}$, $ \n 2p_{3/2} \rightarrow \p 1f_{5/2}$, $\n 1f_{5/2} \rightarrow \p 1f_{5/2}$, or
$\n 1g_{9/2} \rightarrow \p 1g_{9/2}$) so that within RQTBA, GT transitions 
contribute to $\gtrsim 70 \%$ and thus remain dominant.
The trend found in Fig.~\ref{f:FF_N50} in RQTBA is similar to the shell-model calculations of Ref.~\cite{SM2} for $Z<28$.
For $Z\geq 28$, that reference finds a drop of the FF due to smaller contribution of the $2^-$ modes, which, are typically found at lower energies than in the present calculations (by $\simeq 2$ MeV). This is discussed further in the following paragraph.

\begin{figure}[ht]
\centering{\includegraphics[width=\columnwidth]{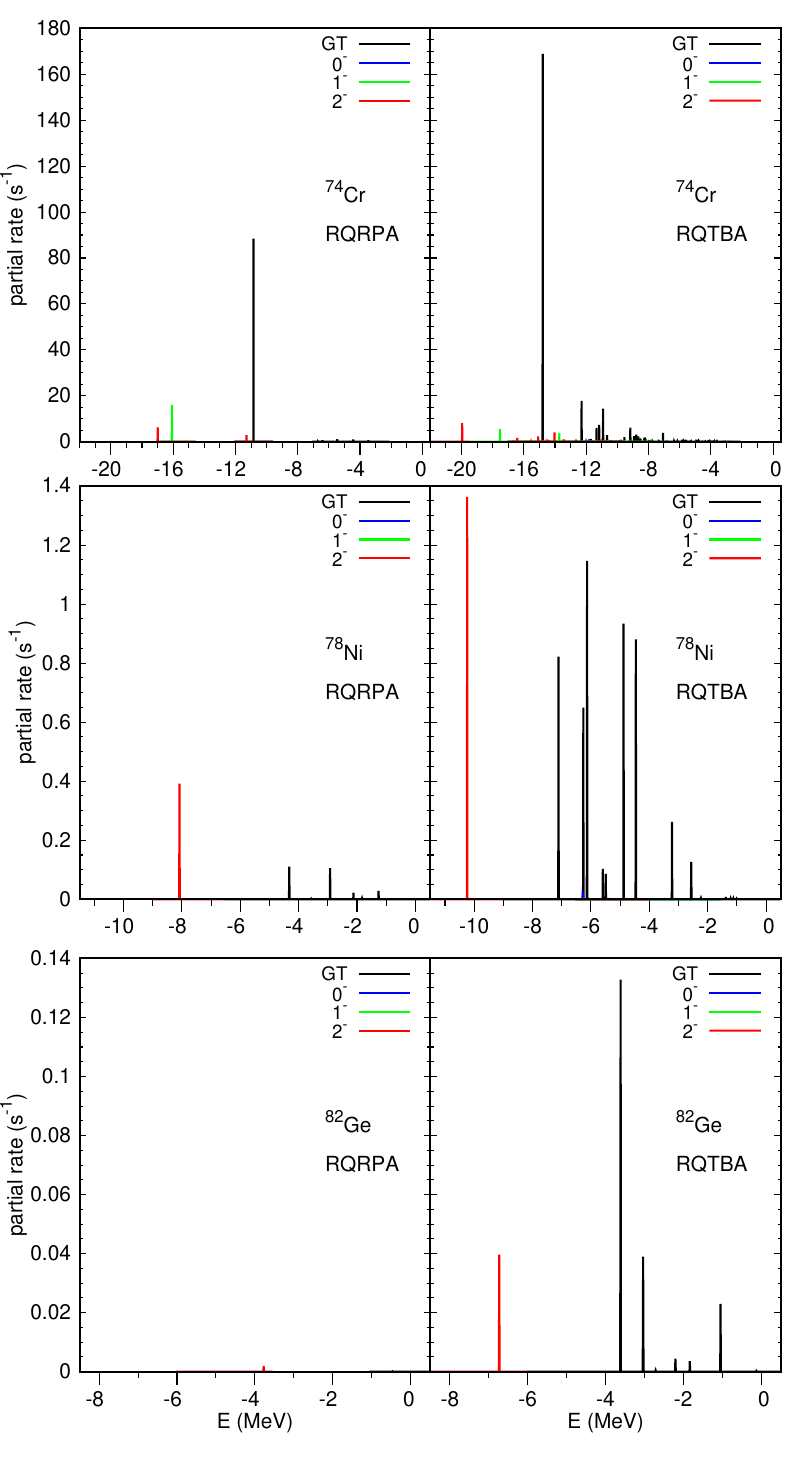} }
\caption{Distribution of partial rates (in $s^{-1}$) in $^{74}$Cr, $^{78}$Ni and $^{82}$Ge in the decay window, within RQRPA (left panels) and RQTBA (right panels). The excitation energy (in MeV) is $E=E_f = M_i - M_f$ as defined in section~\ref{sec:formalism}.}
\label{f:prates_N50}
\end{figure}

Finally we show in Fig.~\ref{f:prates_N50} distributions of partial
decay rates obtained within RQRPA (left panels) and RQTBA (right
panels) in $^{74}$Cr, $^{78}$Ni and $^{82}$Ge. 
Clearly we see that GT transitions are almost absent from the decay
window in RQRPA for large values of $Z$, while they are brought down
in energy due to the fragmentation process induced by QVC.  Overall we
note that FF transitions appear at lower energy than the GT
modes. This is in accordance with the shell model calculations of
Ref.~\cite{SM2}. The details of the RQTBA partial rate distributions
however present some discrepancies with those provided in that
reference.  In $^{74}$Cr we find a large GT transition at
$E_f \simeq -15.5$ MeV which appears to be split into two components
in the shell model case. The rest of the distribution is fragmented
over several states in both approaches.  For $Z \geq 28$, we observe
more dissimilarities between the two methods, in particular in
$^{78}$Ni. For this nucleus RQTBA shows a GT contribution which is
quite fragmented as well as very large contribution from the low-lying
$2^-$ state at $E_f \simeq -10.75$ MeV. The shell model of
Ref.~\cite{SM2}, on the other hand, predicts a GT strength that is
concentrated in only a few peaks and a very small contribution of the
$2^-$ mode.  In $^{82}$Ge, while the GT distributions are more
similar, we also predict an important low-lying $2^-$ peak which is
found to be almost negligible in the shell model.  Overall, we note
that our $2^-$ states seem to be predicted too low in excitation
energy.  For instance, the experimental $Q_\beta$ value for
  $^{82}$Ge is $4.69$ MeV and the estimated value for $^{78}$Ni is
  9.91~MeV~\cite{Wang.Huang.ea:2021}, while from
  Fig.~\ref{f:prates_N50} we see that our $Q_\beta$ values would be at
  least $\simeq 7.25$~MeV and $10.75$~MeV, respectively. Nevertheless, we
  find similar transition strengths to the low lying $2^-$ states
  that predicted by the shell-model. For $^{82}$Ge RQTBA predicts a
  $\log f_1 t = 8.0$ while the shell-model calculations give
  $\log f_1 t = 8.3$ to the low lying $2^-$ state. Both compare well
  with the measured vaue of
  $\log f_1 t = 8.6(3)$~\cite{Tuli.Browne:2019}. For $^{78}$Ni the low
  lying $2^-$ state has a $\log f_1 t = 4.5$ both in the RQTBA and
  shell-model approaches. For $^{74}$Cr we have $\log f_1 t=8.6$
  (RQTBA) and $\log f_1 t=8.9$ (Shell-Model). The large $Q_\beta$
  values are potentially caused by the fact that, in the present
study, the QVC correlations are only included in the description of
the daughter nucleus, and not in the parent ground state.  Indeed it
was observed in Ref.~\cite{Robin:2019jzm} that accounting for these
ground-state correlations could potentially bring the binding-energy
differences, and thus the $Q_\beta$ values, in better agreement with
the data.  Understanding in more details the differences between RQTBA
and shell model calculations, for different truncations of the model
spaces, and the role played by the ground-state correlations, will be
the subject of a future study.

\subsubsection{Isoscalar pairing interaction}
\label{sec:pn_pairing}
%
\begin{figure*}
\centering{\includegraphics[width=0.6\textwidth]{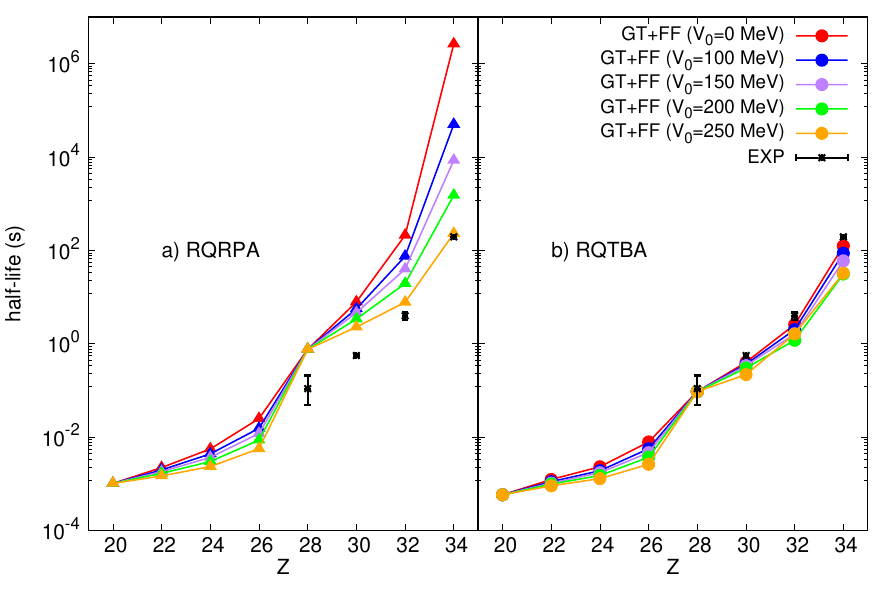} }
\caption{$\beta$-decay half-lives of $N=50$ isotones obtained within a) RQRPA and b) RQTBA, for different values of the static isoscalar pairing interaction $V_0$.}
\label{f:hlives_N50_pnpp}
\end{figure*}
In the calculations that we have discussed so far, we have not
included a $T=0$ proton-neutron ($pn$) particle-particle ($\pp \pp$)
component (usually referred to as "pairing" component)
$\widetilde{V}_{pn}^{(\pp \pp)(T=0)}$ of the static interaction. This
is so because it does not contribute to the structure of the parent
state in the RMF approximation. Thus, in standard QRPA calculations,
this component is often adjusted \emph{a posteriori}, when
investigating the nuclear response \cite{Engel1999,Niksic2005}. In
RQRPA, a Gogny-type ansatz is often considered:
\begin{equation}
\widetilde{\bm{V}}_{pn}^{(\pp \pp)(T=0)} = - V_0 \sum_{j=1,2} g_j e^{-r_{pn}^2/ \mu _j^2} \bm{\Pi}_{S=1,T=0} \; ,
\label{eq:V_isoscalar}
\end{equation}
where $\bm{\Pi}_{S=1,T=0} $ projects onto states with $S=1$ and $T=0$,
$\mu_1=1.2$ fm and $\mu_2 = 0.7$ fm are the range of the Gogny D1S
force \cite{D1S}, and the relative strengths are chosen to be
$g_1=1.0$ and $g_2=-2.0$ so that the interaction is repulsive at short
distance. Such interaction then usually increases the beta
strength at low-energy and yields a decrease of the half-lives. The
remaining free parameter $V_0$ is then typically adjusted in order to
reproduce the $\beta$-decay half-lives. However it was shown that the
resulting value of this parameter can vary strongly according to the
mass region and often has to be pushed close to the point of collapse
of the theory, when adjusted at the RQRPA level \cite{Niksic2005}.
These observations overall point out to the fact that by tuning $V_0$
to large values, one may be trying to correct for deficiencies of the model.


As a test, we include the interaction of the
form~\eqref{eq:V_isoscalar} in our calculations, and vary the
parameter $V_0$ to investigate the sensitivity of our half-lives to
such particle-particle component of the static $T=0$ interaction, when
QVC correlations are included. The results are shown in
Fig.~\ref{f:hlives_N50_pnpp} in both RQRPA and RQTBA.
First of all, we note that this interaction has no effect in $^{78}$Ni and $^{70}$Ca, and more generally in doubly-magic nuclei.
This is because the transition matrix elements in the particle-particle channel come with factors  
$\sim v_p^2 v_n^2$ and $ \sim u_p^2 u_n^2$ (where $v_k^2$ is the occupation of single-particle state $k$ and $u_k^2 + v_k^2 =1$) which are always zero in such nuclei.

In $Z > 28$, $\widetilde{V}_{pn}^{(\pp \pp)(T=0)}$ has much less
effect in RQTBA than in RQRPA, in particular close to stability.  Let
us analyze the case of $^{84}$Se. At the RQRPA level the main
transition responsible for the decay is the
$\n 1g_{9/2} \rightarrow \p 1f_{5/2}$ caused by the rank-2
operator. The corresponding particle-particle matrix element comes
with a factor $v_p^2 v_n^2 = v_p^2$, which is large ($\simeq 0.88$)
for $^{84}$Se as the proton $1f_{5/2}$ is largely occupied. Thus, the
T=0 particle-particle interaction has a large effect for this
nucleus. In RQTBA, the decay in this nucleus is caused by GT
transitions by more than $95 \%$ in agreement with experimental
  data, with the main contributions caused by transitions between the
neutron and proton $2p_{1/2,3/2}$ sub-shells. Since the occupation of
these proton levels are small ($v_p^2 \simeq 0.1$ and $0.05$) the
$T=0$ particle-particle static interaction is much less effective,
which is why the half-lives of $^{84}$Se, and other $Z>28$ nuclei, are
more stable against variations of the strength $V_0$ of this
interaction than within RQRPA.

For $Z<28$ isotones, however, the effect of $\widetilde{V}_{pn}^{(\pp \pp)(T=0)}$ appears
similar with and without QVC. For these nuclei, both GT and FF ($1^-$
and $2^-$) modes contribute in RQRPA and RQTBA, and the most important
transitions involve the proton $1f_{7/2}$ sub-shell which has $v_p^2$
varying from $0.25$ in $^{72}$Ti to $0.74$ in $^{76}$Fe. Therefore the
effect of $\widetilde{V}_{pn}^{(\pp \pp)(T=0)}$ increases with $Z$ and is found similar at
both levels of approximation. 

Clearly it seems difficult to find a single value of $V_0$ for which
RQRPA would reproduce the experimental half-lives of the whole
chain. In particular it is not possible for $^{78}$Ni which is
doubly-magic, and thus insensitive to the proton-neutron pairing
interaction. In RQTBA, the data is best reproduced for $V_0$=0, but as
noted above, varying this parameter has little effect for $Z \geq 28$,
and we cannot conclude for $Z<28$ since no data is available.

For these reasons, and because we find similar results as
Fig.~\ref{f:hlives_N50_pnpp} for other isotonic chains, we will
consider $V_0=0$ in the rest of this manuscript. We point, however, that the proton-neutron particle-particle channel 
is still naturally included in the
dynamical QVC interaction $\Phi(\omega)$ (see Eq. \ref{eq:Phi})
due to an interplay with the $T=1$ like-particle pairing, as explained in Sec.~\ref{sec:num_scheme}. This
may explains why the QVC results agree well with the data, without having to
resort to static $T=0$ particle-particle interaction.

\subsection{$N=82$ chain} \label{sec:N82}
\begin{figure}[ht]
\centering{\includegraphics[width=\columnwidth]{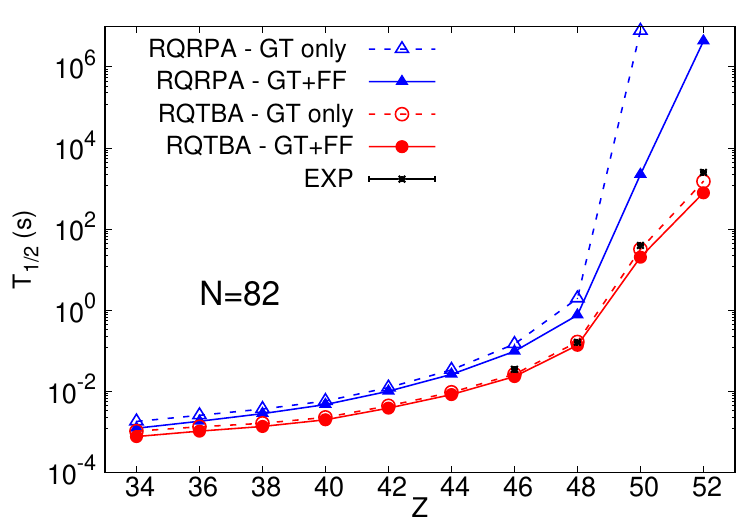} }
\caption{$\beta$-decay half-lives of $N=82$ isotones obtained within
  RQRPA (blue triangles) and RQTBA (red circles), in the allowed GT
  approximation (empty symbols) and including FF transitions (full
  symbols). When no symbol is shown, the nucleus is predicted to
    be stable. The experimental data is from
  \cite{data}.} 
\label{f:hlives_N82}
\end{figure}
We show in Fig. \ref{f:hlives_N82} the half-lives of $N=82$
isotones. As in the $N=50$ chain, the effect of QVC correlations is
very important closer to stability, as it reduces the half-lives by
several orders of magnitudes and leads to a much better agreement with
the available data, although the reduction tends to be slightly too
strong in some cases. As one goes towards the neutron dripline
(towards small $Z$), the effect of QVC on the half-lives fades out and
the predictions of RQTBA become similar to the ones of
RQRPA.

\begin{figure}[ht]
\centering{\includegraphics[width=\columnwidth]{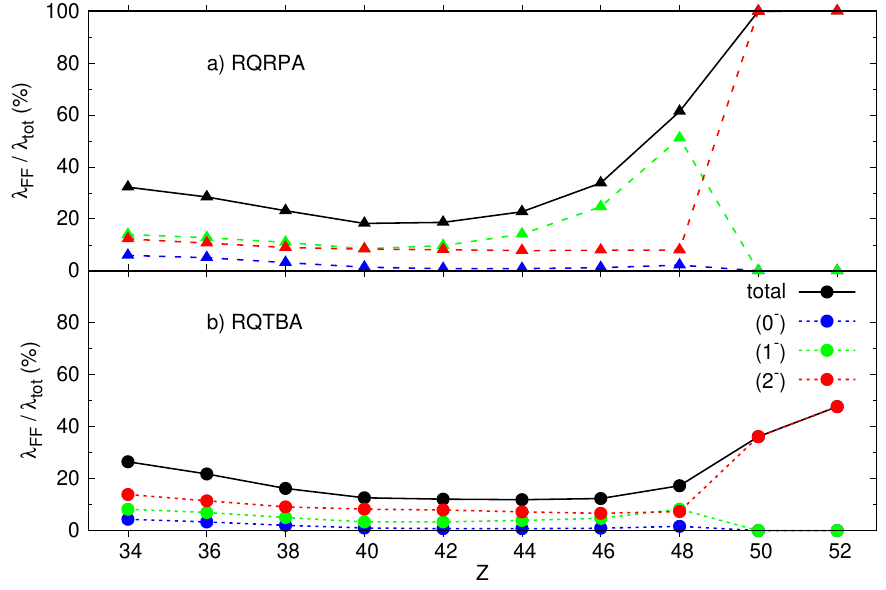} }
\caption{Contribution of first-forbidden transitions to the $\beta$-decay rate of $N=82$ nuclei (in $\%$) within a) RQRPA and b) RQTBA. The full contribution is shown with a plain black line while the dashed lines show the individual contribution of the different multipoles.}
\label{f:FF_N82}
\end{figure}

We show in Fig. \ref{f:FF_N82} the contribution of first forbidden
transitions to the $\beta$-decay rates.  In nuclei with $Z \geq 40$,
$pf$-shell orbits for protons are, to first approximation, fully
occupied.  This suppresses $0^-$ transitions (e.g.
$\n 2d_{3/2} \rightarrow \p 2p_{3/2}$ or
$\n 2d_{5/2} \rightarrow \p 1f_{5/2}$).  In lighter isotones, while
the transition $\n 2d_{3/2} \rightarrow \p 2p_{3/2}$ is Pauli
un-blocked, we find that the contribution of $0^-$ modes to the rate
remains small (below $6 \% $) due to a strong cancellation between the
spin-dipole component, Eq.~(\ref{eq:FF_w}), and the relativistic one,
Eq.~(\ref{eq:gamma5}). 
In nuclei with $Z \leq 42$, the decay proceeds essentially via GT 
(mostly $\n 1g_{7/2} \rightarrow \p 1g_{9/2}$), 
with a comparable contribution of $1^-$ and $2^-$ modes below $\simeq 14 \%$ at both levels of approximations.

In heavier isotones with $Z \geq 44$, the $1^-$ component increases in RQRPA, 
via transition $\n 1h_{11/2} \rightarrow \p 1g_{9/2}$,
until $Z=50$ when the proton $1g_{9/2}$ becomes fully filled. The main
low-energy GT and $1^-$ transitions are then completely blocked and the
decay proceeds solely via rank-$2$ FF transitions (mostly
$\n 1h_{11/2} \rightarrow \p 1g_{7/2}$).

\begin{figure}[ht]
\centering{\includegraphics[width=\columnwidth]{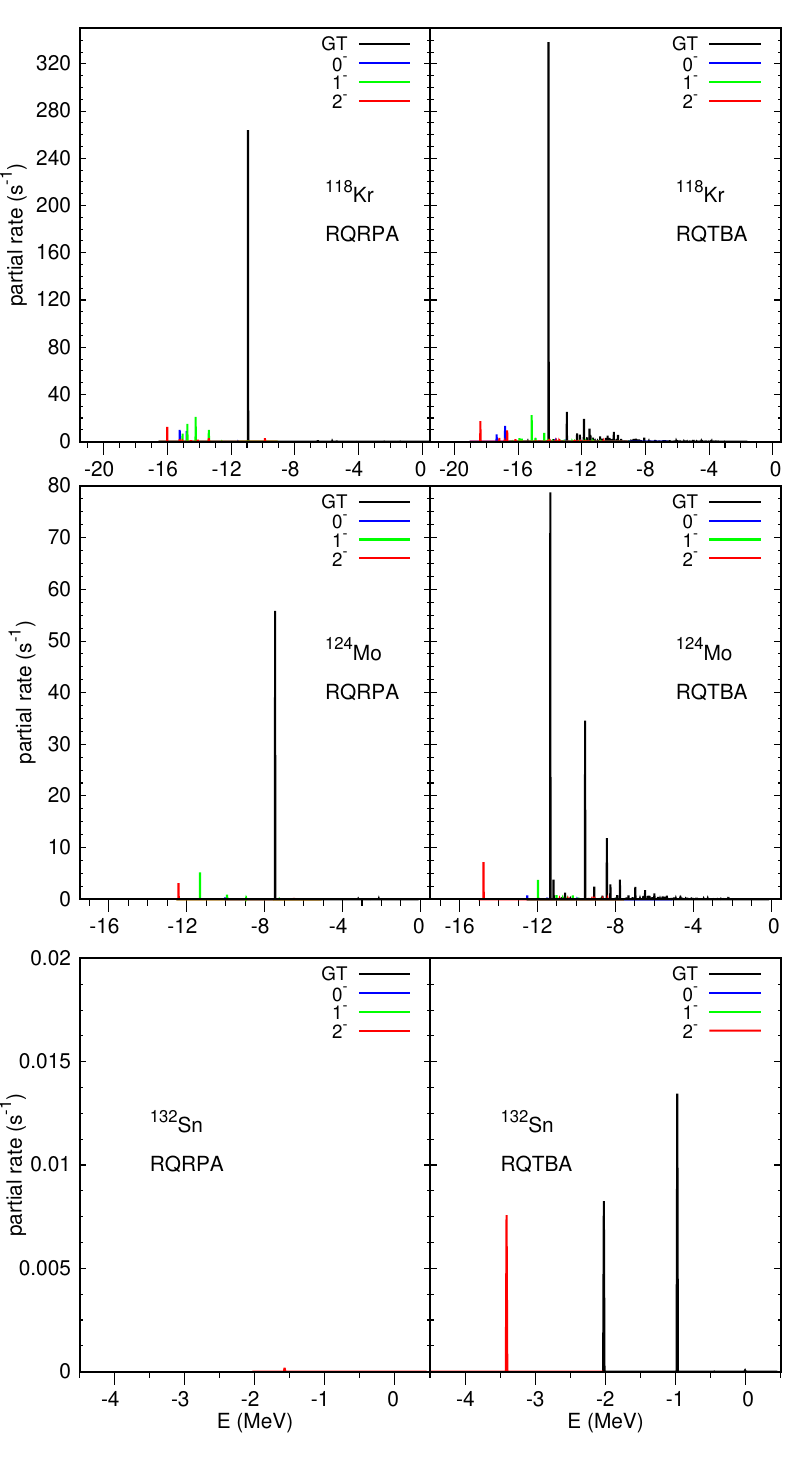} }
\caption{Distribution of partial rates (in $s^{-1}$) for $^{118}$Kr, $^{124}$Mo and $^{132}$Sn in the decay window, within RQRPA (left panels) and RQTBA (right panels). The excitation energy (in MeV) is $E=E_f = M_i - M_f$ as defined in section~\ref{sec:formalism}. }
\label{f:prates_N82}
\end{figure}

Within RQTBA, the trend is changed in nuclei with large $Z$, where the
GT remains the largest contribution ($> 50 \%$), even above the $Z=50$ shell
closure. In such nuclei, even though the proton $1g_{9/2}$ is 
filled, GT modes caused by transitions such as
$\n 1h_{11/2} \rightarrow \p 1h_{11/2}$ or $2d_{3/2} \rightarrow 2d_{5/2}$
are lowered in energy due to QVC, and contribute significantly to the
decay because of the phase space.
\\ \\
The RQRPA and RQTBA partial decay-rate distributions of $^{118}$Kr,
$^{124}$Mo and $^{132}$Sn are shown in Fig.~\ref{f:prates_N82}.  
In $^{118}$Kr the transitions of RQRPA and RQTBA actually show a similar
profile (with a bit more fragmentation in RQTBA), however the shift
due to QVC towards lower energy enhances the contributions to the
rates due to the phase space. In $^{124}$Mo and $^{132}$Sn, the effect
of QVC grows greatly.  As in the $N=50$ chain, FF transitions overall
occur at lower energy than GT modes. We note that the distribution in
$^{124}$Mo has a similar profile than the shell model calculation of
Ref.~\cite{SM2}, however the first FF transition occurs $~2$ MeV lower
in our calculation.

\subsection{$N=126$ chain} \label{sec:N126}
\begin{figure}[h!]
\centering{\includegraphics[width=\columnwidth]{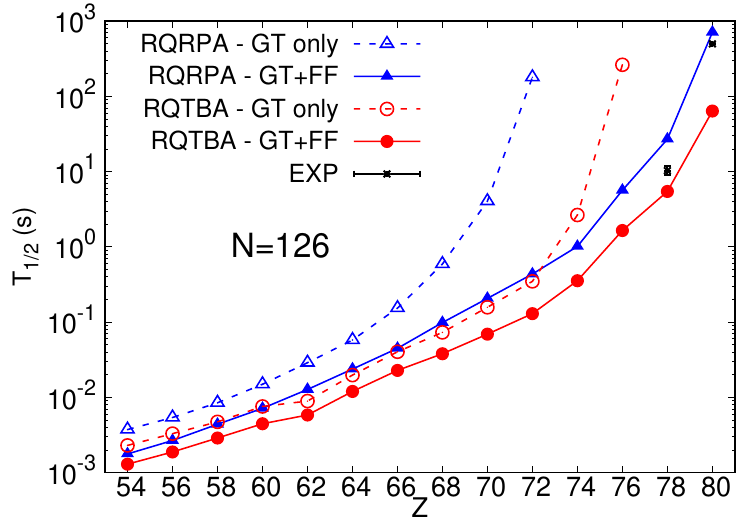} }
\caption{$\beta$-decay half-lives of $N=126$ isotones obtained within RQRPA (blue triangles) and RQTBA (red circles), in the allowed GT approximation (empty symbols) and including FF transitions (full symbols). When no symbol is shown, the half-life is predicted to be infinite. The experimental data is from \cite{data}.}
\label{f:hlives_N126}
\end{figure}

\begin{figure}
\centering{\includegraphics[width=\columnwidth]{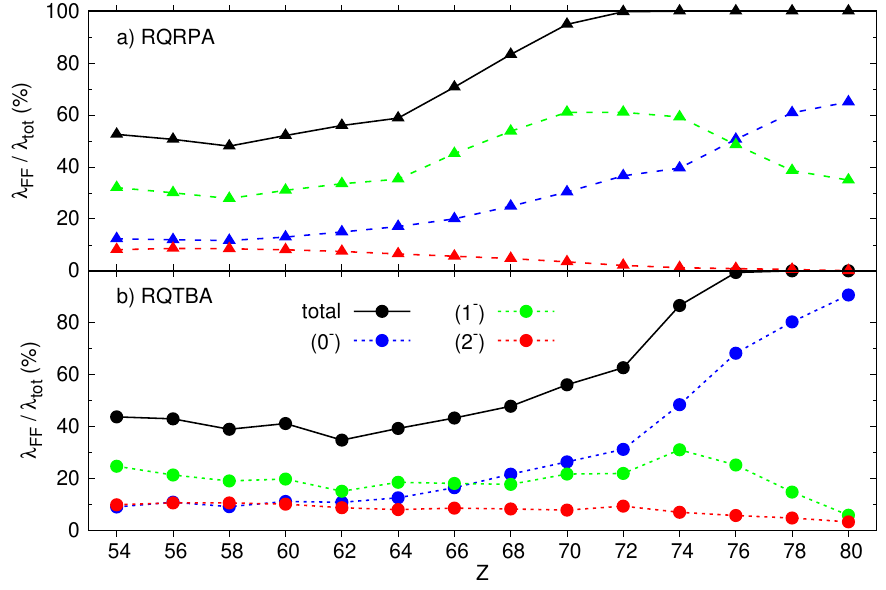} }
\caption{Contribution of first-forbidden transitions to the total $\beta$-decay rate of $N=126$ nuclei (in $\%$) within a) RQRPA and b) RQTBA.}
\label{f:FF_N126}
\end{figure}

Fig.~\ref{f:hlives_N126} shows the $\beta$-decay half-lives of $N=126$
isotones. 
In this chain, while QVC is important near stability in the
allowed GT approximation (empty symbols), its effect is reduced to
about one order of magnitude at most when FF transitions are included.
Again, the too strong decrease of the half-lives due to QVC observed
in $^{204}$Pt and $^{206}$Hg could potentially be explained by the
fact that, in the present study, we do not include ground-state
correlations induced by QVC. These correlations, that were implemented
and investigated in doubly-magic nuclei \cite{Robin:2019jzm}, can lead
to a small shift of the strength back to higher energy, re-increasing
slightly the half-lives. The effect of such correlations in the parent
ground state of open-shell nuclei will be investigated in the future.
We also note that, for these nuclei, FF transitions contribute to
the decay rate by a large fraction (see below) and we have not applied
any quenching to FF modes either.
\\ \\
In Fig.~\ref{f:FF_N126} is shown the contribution of FF transitions to
the rates.
The probability for the decay to occur via FF transitions is very large throughout the chain, in both RQRPA and RQTBA. 
The total contribution in the region far from stability is about $\sim 50-60 \%$ in RQRPA, and $\sim 35-45 \%$ in RQTBA. Above $Z \geq 68$ the proton $1h_{11/2}$ becomes more occupied and the main low-energy GT transition $\n 1h_{9/2} \rightarrow \p 1h_{11/2}$ is blocked by the Pauli principle. This explains the increase of the FF contribution towards stability. 
As will be discussed in section \ref{sec:comparison}, such an increase has also been observed in other theoretical methods.
However, the steepness of the increase and the overall contribution of the FF transitions in the $N=126$ isotonic chain, especially towards the neutron dripline, strongly varies between the different models.

\begin{figure}[ht]
\centering{\includegraphics[width=\columnwidth]{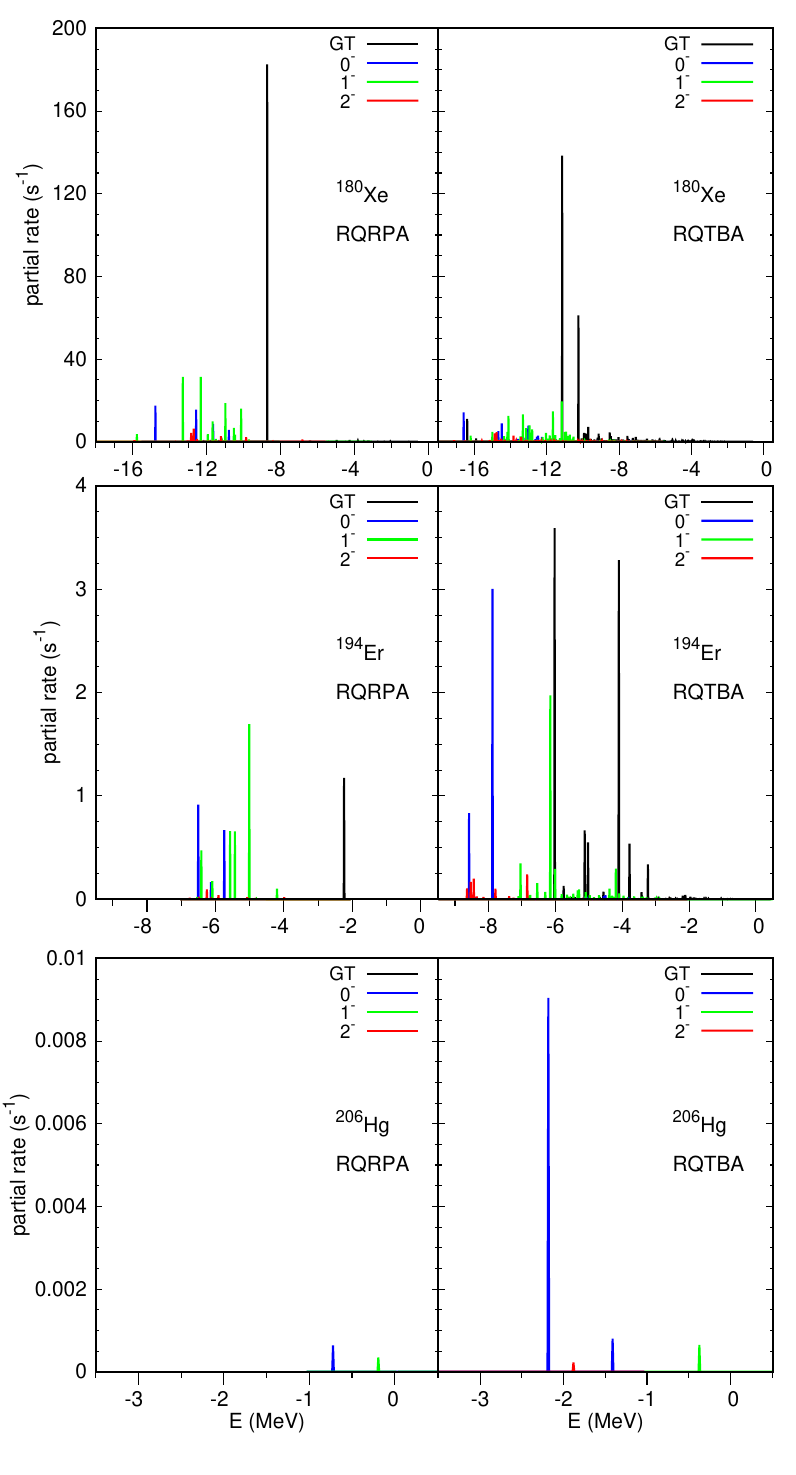} }
\caption{Distribution of partial rates (in $s^{-1}$) for $^{180}$Xe, $^{194}$Er and $^{206}$Hg in the decay window, within RQRPA (left panels) and RQTBA (right panels). The excitation energy (in MeV) is $E=E_f = M_i - M_f $ as defined in section~\ref{sec:formalism}. }
\label{f:prates_N126}
\end{figure}

Similarly to Ref.~\cite{SM2}, we find that FF transitions are dominated in these $N=126$ isotones by $0^-$ and $1^-$ modes, while contributions of $2^-$ transitions are very small throughout the chain. 
The main single-particle transitions contributing to FF modes involve (in order of increasing energy) the proton $2d_{5/2}$, $1h_{11/2}$, $2d_{3/2}$ and $3s_{1/2}$. For low values of $Z$, these sub-shells are mostly empty and leading to more available $1^-$ than $0^-$ transitions, based on selection-rules arguments. As the proton number $Z$ increases, these sub-shells are gradually filled, until the $\p 2d_{3/2}$. In that case the main remaining transition is $\n 3p_{1/2} \rightarrow \p 3s_{1/2}$ which contributes to both $0^-$ and $1^-$. The $0^-$ decay becomes more probable as the corresponding daughter states appear lower in energy, as can be seen from, e.g., Fig.\ref{f:prates_N126}.
In contrast, the shell-model calculations of Ref.~\cite{SM2} found a dominance of the rank-1 operators ($\simeq 70\%$ of the FF) over the rank-0 transitions in the range $66 \leq Z \leq 72$. 
\\
\\
As mentioned in section \ref{sec:beta_decay_theory}, $0^-$
transitions have two components: one originates from the spin-dipole
operator, $w$ and $w'$ terms in Eqs. (\ref{eq:FF_w},\ref{eq:FF_wp}),
while the other is due to the relativistic $\gamma_5$ operator, $\xi' v$ term in Eq. (\ref{eq:gamma5}), which originates from the time-like part of the axial current.
Because of interference between them, it is difficult to disentangle the effects of these two contributions.
This topic has been intensely studied by Warburton within a shell-model framework \cite{Warburton1991}. 
He found that there is typically a strong cancellation between the spin-dipole and $\xi' v$ terms, and thus,
that an enhancement of $\xi' v$ is necessary to avoid such cancellation.
In particular, an enhancement factor of $\sim 2$ was needed in the lead region.
Generally, even though the enhancement factor varies between different approaches, 
it is agreed that the predictions of methods based on impulse approximation within a
non-relativistic formalism typically largely underestimate the relative contribution the $\xi' v$ term.
While an important part of this enhancement in principle originates from two-body, or meson-exchange
currents, as originally demonstrated by Kubodera et al. based on
chiral-symmetry arguments~\cite{PhysRevLett.40.755}, a substantial
portion of it may also originate from other effects, such as
relativistic corrections~\cite{Warburton1991}. 

\begin{figure}[ht]
\centering{\includegraphics[width=\columnwidth]{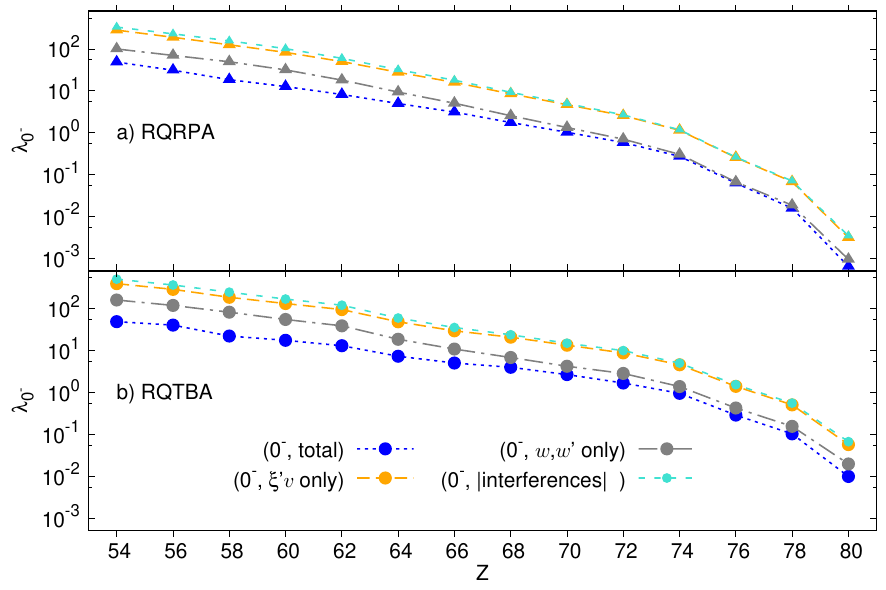} }
\caption{$\beta$-decay rates from the $0^-$ transitions of $N=126$ nuclei (in $\%$) within a) RQRPA and b) RQTBA. The blue symbols show the full $\lambda_{0^-}$ rates, the orange and grey symbols show the rates obtained when including only the relativistic ($\xi' v$) and spin-dipole ($w$ and $w'$) components, respectively. The turquoise symbols show the contribution of the absolute value of the destructive interference term. }
\label{f:FF0_N126}
\end{figure}

In our calculation, we also find that both spin-dipole and $\xi' v$ terms interfere destructively.
To investigate this aspect further, we try to evaluate the importance of both spin-dipole and $\xi' v$ components and show in Fig.~\ref{f:FF0_N126}, the $\beta$-decay rate $\lambda_{0^-}$ originating from the $0^-$ transitions, obtained when suppressing one of the two components. 
Specifically, the orange and grey symbols show the rates obtained when
including only the relativistic ($\xi' v$) and spin-dipole ($w$ and
$w'$) components, respectively. The turquoise symbols show the
absolute value of the destructive interference term.  
It appears that both components contribute by a similar
order of magnitude, with a somewhat larger contribution of the relativistic
operator (factor $\simeq 2-3$ in the rate). The interference term is negative and also of the same
order (with a value comparable to the relativistic term). 
Consequently, $\lambda_{0^-}$ also takes a value of the same order of magnitude. 
As a reminder, no quenching or enhancement factors have been applied in the
present study.  It would be interesting to investigate further the
contribution of the different components and their interference in a
shell-model framework. We leave such study for a future work.

\subsection{$N=184$ chain} \label{sec:N184}
\begin{figure}[ht]
\centering{\includegraphics[width=\columnwidth]{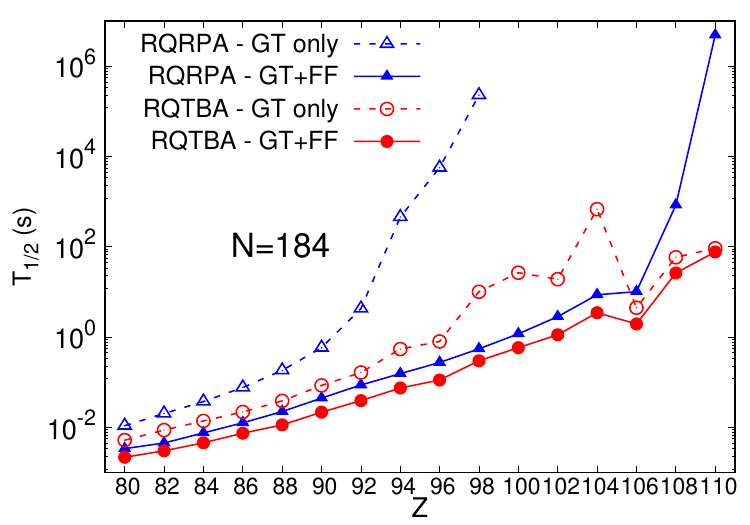} }
\caption{$\beta$-decay half-lives of $N=184$ isotones obtained within RQRPA (blue triangles) and RQTBA (red circles). The empty symbols show the half-lives obtained in the allowed Gamow-Teller approximation, while the full symbols show the half-lives with the contribution of the first-forbidden transitions.}
\label{f:hlives_N184}
\end{figure}

\begin{figure}
\centering{\includegraphics[width=\columnwidth]{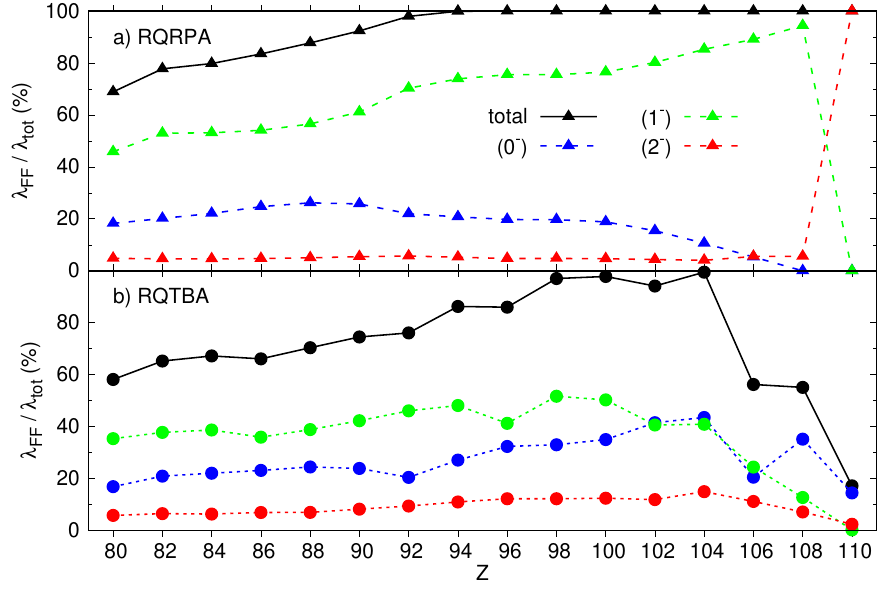} }
\caption{Contribution of first-forbidden transitions to the total $\beta$-decay rate of $N=184$ nuclei (in $\%$) within a) RQRPA and b) RQTBA.}
\label{f:FF_N184}
\end{figure}

We show in Figs.~\ref{f:hlives_N184} and \ref{f:FF_N184} the $\beta$-decay half-lives and contribution of FF transitions of $N=184$ isotones, respectively. We note that $^{290}$Sg ($Z=106$) has been treated here without pairing correlations, due to a failure (collapse) of the numerical procedure for that nucleus. This explain the non-smooth behaviour of some curves around $Z=106$.

Overall, the effect of QVC is of several orders of magnitude near stability, and the FF transitions largely dominate the decay, except close to stability in RQTBA. 

\begin{figure}
\centering{\includegraphics[width=\columnwidth]{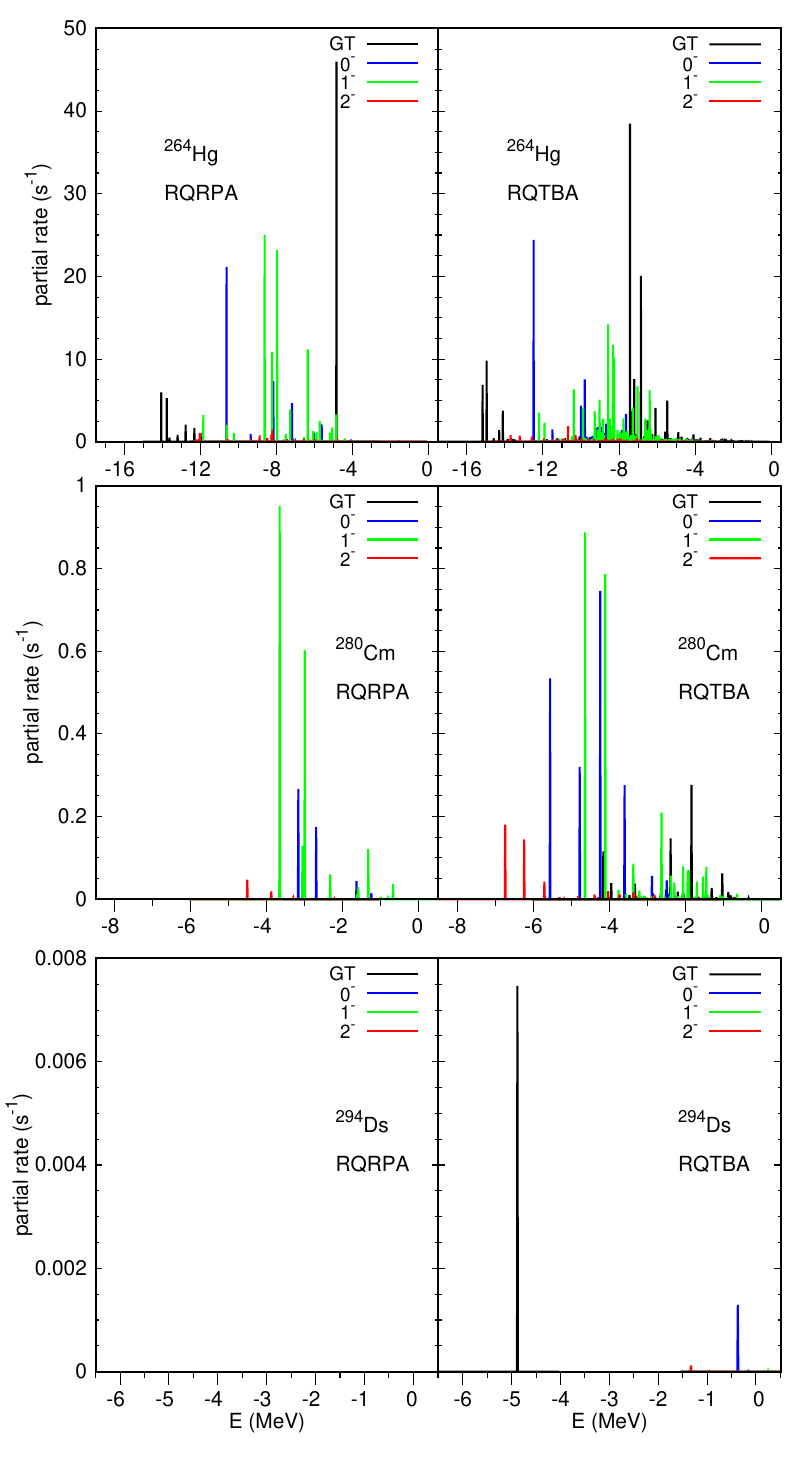} }
\caption{Distribution of partial rates (in $s^{-1}$) for $^{264}$Hg, $^{280}$Cm and $^{294}$Ds in the decay window, within RQRPA (left panels) and RQTBA (right panels). The excitation energy (in MeV) is $E=E_f = M_i - M_f $ as defined in section~\ref{sec:formalism}. }
\label{f:prates_N184}
\end{figure}

At the RQRPA level, the FF contribution remains above $\gtrsim 70 \%$
throughout the chain, and increases with $Z$ as the main GT transition
$\n 4s_{1/2} \rightarrow \p 3s_{1/2}$ becomes blocked due to the filling of
the $\p 3s_{1/2}$ level. Note that such transitions between states
with different main quantum numbers $n$ is only allowed due to
isospin-breaking terms of the interaction, in particular, the Coulomb
force. For nuclei with $Z \geq 92$, the decay is basically of purely
forbidden nature. 
Throughout the chain the decay proceeds mostly via rank-1 operators ($\n 3d_{3/2,5/2} \rightarrow \p 2f_{5/2,7/2}$) which are responsible for about $45-95 \%$ of the total rate, with an increasing contribution towards stability until $Z=108$. Rank-0 operators contribute to about $20 \%$ far from stability (via $\n 2g_{9/2} \rightarrow \p 1h_{9/2}$) and become unimportant near stability.
In that region, only few transitions occur below the decay $Q$ value. In particular, in $Z=110$, almost all transitions appear above that value, except for a very small $2^-$ state which corresponding partial rate is negligible ($\sim 10^8$) and not visible on Fig.~\ref{f:prates_N184}.

At the RQTBA level, the total contribution of FF modes is of similar
magnitude up to $Z=104$ $\gtrsim 70 \%$, after which it decreases to a value of
$\simeq 17 \%$ at $Z=110$. 
In the three isotones closest to stability
some GT transitions are lowered down in energy due to the correlations introduced by the QVC, in particular,
$\n 1i_{11/2} \rightarrow \p 1i_{13/2}$, as well as $\n 2h_{9/2} \rightarrow
\p 2f_{7/2}$. Note that the latter transition is due to GSC included in RQRPA, combined with QVC effects.
Overall the contribution of $0^-$ (and $2^-$) modes is enhanced compared to RQRPA,
especially at large $Z$, while the contribution of $1^-$
modes is decreased.  This is similar to the behaviour observed in the $N=126$ chain.

Finally we show in Fig. \ref{f:prates_N184} the distribution of partial decay rates within RQRPA and RQTBA in $^{264}$Hg, $^{280}$Hg and $^{294}$Ds, where one can again appreciate the importance of FF transitions. In $^{294}$Ds we also note a strong GT transitions which appears at lower energy than the FF one.

%

\section{Comparison with other approaches}
\label{sec:comparison}

\begin{figure*}
\centering{\includegraphics[width=\textwidth]{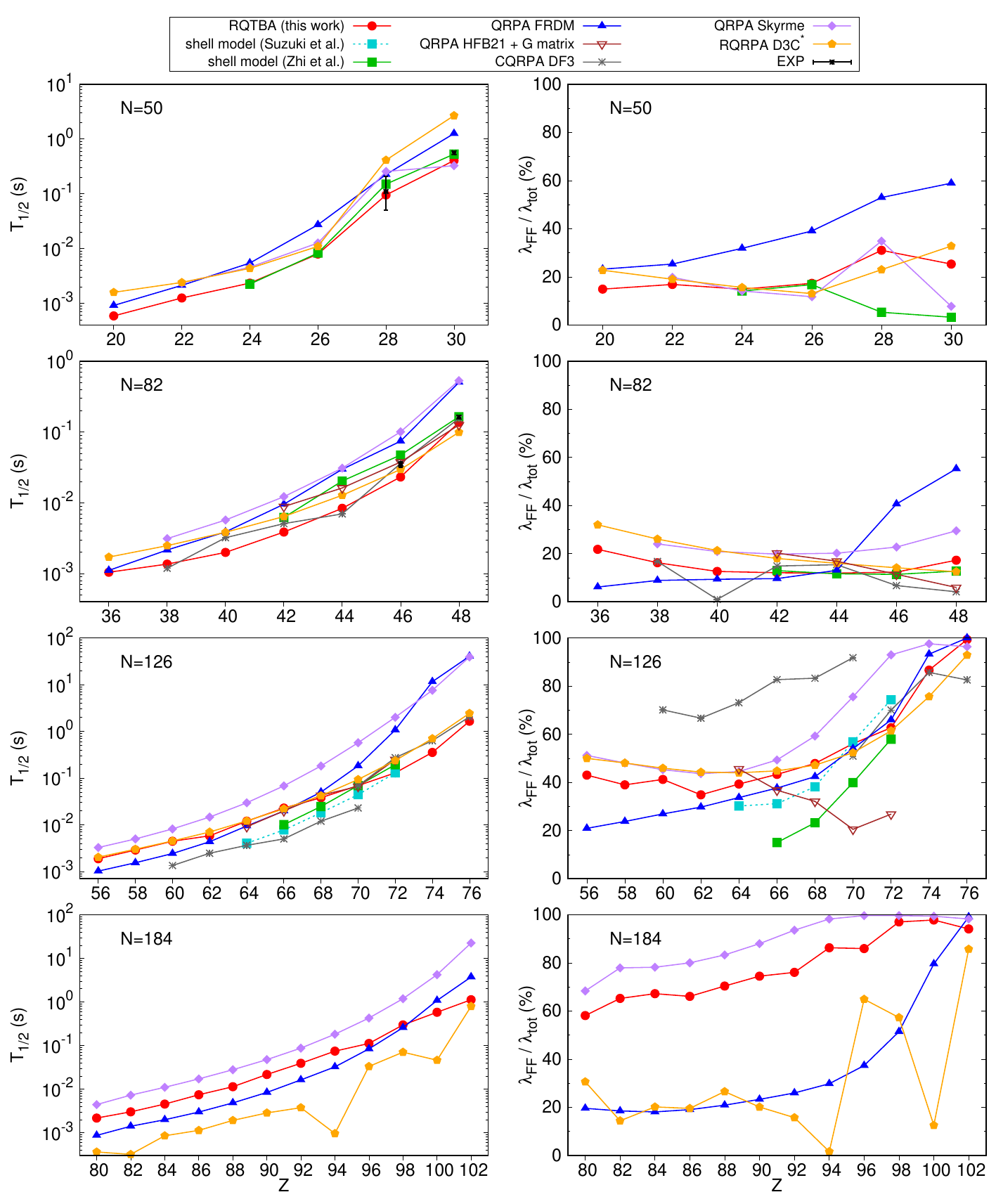} }
\caption{Comparison of our results (red circles) with other theoretical calculations from shell models and various QRPA approaches. The left panels show a comparison of the $\beta$-decay half-lives and the right panels show a comparison of the predicted contributions of FF transitions. See text for explanations of the different theoretical models. The experimental data is from \cite{data}.}
\label{f:comp_T_FF_all}
\end{figure*}

We now compare our results, obtained at the RQTBA GT+FF level, to other existing calculations.
The left panels of Fig.~\ref{f:comp_T_FF_all} show a comparison of the $\beta$-decay half-lives for the four isotonic chains under study, while the right panels compare the total contributions of the FF transitions to the rates. 
In these figures we selected nuclei with a half-life between $\sim 1$ ms and $1$ s, which are those of interest for the $r$ process.
We include results obtained within the shell model and various QRPA calculations, specifically:
\begin{itemize}
    \item the shell model calculations of Ref.~\cite{SM1} ('Suzuki et al.', turquoise squares),
    \item the shell model calculations of Ref.~\cite{SM2} ('Zhi et al.', green squares), which used a larger model space as well as a different effective interaction,
    \item the QRPA calculation of Ref.~\cite{Moeller2003} based on FRDM ('FRDM', blue triangles) using a schematic interaction and combining microscopic calculation of GT modes and macroscopic description of FF transitions,
    \item the microscopic (but non self-consistent) QRPA calculation of Ref.~\cite{PhysRevC.88.034304} ('HFB21 + G matrix', empty brown triangles) based on a G-matrix derived from the realistic CD-Bonn interaction \cite{Machleidt1989}, using single-particle energies obtained from a Skyrme functional and the HFB21 mass model \cite{HFB21} for the calculation of unknown Q values (similar results were also obtained with the FRDM mass model and are not shown here),
    \item the non-relativistic microscopic self-consistent continuum QRPA calculations of Refs.~\cite{Borzov2003,Borzov2011} using DF3 \cite{FAYANS200049} Fayans functional ('DF3', grey stars), 
    \item the non-relativistic microscopic self-consistent deformed QRPA approach using SkO' \cite{Sko} Skyrme functional of Ref.~\cite{Ney2020} ('Skyrme', purple diamonds),
    \item the relativistic microscopic self-consistent spherical QRPA approach using D3C$^*$ \cite{D3Cs} functional of Ref.~\cite{Marketin2016} ('D3C$^*$', orange pentagons). Note that the results shown here differ slightly from Ref.~\cite{Marketin2016} due to corrected phases in the computer code.
\end{itemize}
These QRPA calculations differ from the ones we have shown previously by the fact that some of them are not fully microscopic and/or self-consistent, they use different interactions or functionals (relativistic or not) and they all include residual (static) isoscalar proton-neutron pairing interaction which is typically fitted in different ways to reproduce known $\beta$-decay half-lives. Some of the non-relativistic QRPA's also treat further effects such as deformation or continuum.

 We also note that the methods we are comparing use different values for the quenching of the different multipole modes contributing to the decay.
 In particular, all the microscopic approaches apply a quenching $q$ to the Gamow-Teller transitions, ranging from $q=0.4$ in the QRPA of Ref.~\cite{PhysRevC.88.034304}, to $q\simeq 0.7$ in the shell model calculations ~\cite{SM1,SM2}, and $q \simeq 0.78$ (amounting to using $g_A \simeq -1.0$) in the remaining self-consistent QRPA calculations \cite{Borzov2003,Borzov2011,Ney2020,Marketin2016}.

 The quenching of FF transitions is however handled much differently. 
 In the shell model of Ref.~\cite{SM2}, the quenching (or enhancement for the $0^-$ operator) of the different multipoles contributing to FF modes are fixed to reproduce known $\beta$-decay half-lives. In the shell model of Ref.~\cite{SM1}, the FF contributions are quenched in the same way as the GT, except the $0^-$ modes which are enhanced (by a factor $2$ for the relativistic component). 
 The QRPA calculations of Ref.~\cite{Marketin2016} and Ref.~\cite{Borzov2003} use the same quenching for all GT and FF transitions, while Ref.~\cite{Ney2020} keeps the bare FF channel \cite{PhysRevC.93.014304}. Finally, Ref.~\cite{PhysRevC.88.034304} quenches the FF modes by a factor $0.5$ except the rank-0 operators that are enhanced.
As a reminder, the present RQTBA results are obtained without any quenching, neither in the GT nor in the FF cases, using the current bare value $g_A = -1.27641$ \cite{ga}.
\\ \\
In the $N=50$ isotonic chain the RQTBA half-lives are very similar to those predicted by the shell model. The contributions of FF transitions are also of the same order of magnitude although the shell model predicts a decrease of FF contribution above $Z=28$, while it increases by a few percents in RQTBA, due to low-lying $2^-$ transitions in this region, as discussed previously in section \ref{sec:N50}.
The QRPA calculations typically predict larger half-lives than RQTBA and shell model (except for the $Z=30$ case with Skyrme QRPA). 
\\ \\
In the $N=82$ chain, RQTBA predicts half-lives that are lower than other calculations in the middle of the chain, except for a few exceptions. In particular, they are lower than the shell-model predictions by up to a factor $\simeq 2.4$.
The contribution of FF transitions is predicted to be below $\simeq 32 \%$ by all the microscopic approaches (except the FRDM QRPA near stability which predicts a larger probability for $Z=46-48$), with a spread of the results which grows towards the dripline, 
where FF transitions can be unlocked by growing neutron excess, and close to stability where they can 
become relatively important due to blocking of GT transitions.
\\ \\
In the $N=126$ chain the spread in the predictions of the half-lives reaches $\sim$ one to two orders of magnitude.
The right panel of Fig.~\ref{f:comp_T_FF_all} shows a considerable spread in the predictions of FF modes contribution along the $N=126$ chain.
 In the range $Z=64-76$, almost all models (except Ref.~\cite{PhysRevC.88.034304}) predict an increase of the FF contribution due to the blocking of the $\n 1h_{9/2} \rightarrow \p 1h_{11/2}$ GT transition, as discussed previously in section \ref{sec:N126}, but the steepness of this increase varies depending on the models. Shell-model calculations predict smaller probabilities for the FF decay in the middle of the chain, compared to (R)QRPA and RQTBA caluclations.
\\  \\
Due to computational limits the $N=184$ nuclei cannot be calculated by the shell model. However they have been computed by several QRPA methods. Overall the spread in the predicted half-lives is of $\sim$ two orders of magnitude. 
The Skyrme QRPA and RQTBA calculations appear to agree on a large contribution of the FF modes and on the trend along the chain (with a somewhat larger contribution predicted by the Skyrme QRPA by up to $\simeq 10-15 \%$). 
The half-lives, however, differ by about one order of magnitude near stability. 
FRDM, on the other hand, predicts a low FF contribution of $20-30 \%$ in the $Z=80-94$ range which then grows towards stability. The relativistic QRPA calculation with D3C$^*$ provides FF contributions that are similar to the FRDM QRPA for most nuclei, but somehow shows strong variations, due to particularly strong GT contributions in $Z=94$ and $Z=100$ nuclei, which decrease the relative importance of the FF modes.

\section{Conclusion}

In this paper we have performed systematic calculations of $\beta$-decay half-lives of even-even $r$-process waiting-point nuclei with $N=50$, $82$, $126$ and $184$ in the approach based on relativistic QRPA extended to include quasiparticle-vibration coupling (RQTBA). The calculations are based on the NL3 relativistic functional, with both allowed Gamow-Teller and first-forbidden transitions included.
\\ \\
Overall we found that the coupling between nucleons and vibrations, which typically induces fragmentation and spreading of the transitions strength distributions, leads to a decrease of the $\beta$-decay half-lives which is particularly important when the neutron excess is not too large and the decay $Q$ value is small.
While FF transitions often constitute the leading mechanism for $\beta$-decay at the RQRPA level in nuclei close to stability, this is less true at the RQRPA level as QVC correlations unlock previously blocked GT transitions, thus reducing the relative effect of FF modes. 
\\ \\
In particular, in the $N=50$ and $N=82$ nuclei close to stability, the QVC is responsible for the appearance of GT transitions which then dominate the decay, in accordance with shell-model studies.
The $\beta$-decay half-lives predicted by the RQTBA approach agree well with available experimental data in these lighter nuclei, without introducing extra quenching of the transition matrix elements. This demonstrates the ability of the present approach to capture correlations that are essential for a precise description of charge-exchange modes at low energy and accurate predictions of $\beta$-decay. 
\\ \\
In the heavier nuclei, however, the decrease of the half-lives by QVC can be too strong and RQTBA tends to underestimate the available shell-model and (very few) experimental half-lives. This could be due to a too strong shift of the low-lying transition strength distributions towards lower energies due to missing ground-state correlations induced by QVC. Such correlations have been developed and implemented for doubly-magic nuclei and their impact on GT modes in $^{90}$Zr was investigated in Ref.~\cite{Robin:2019jzm}. While we found that these GSC are mostly crucial for the description of the GT$^+$ transitions, it was seen that they can also induce modifications of the GT$^-$ strength, in particular, a small shift of the low-lying states back to higher energy. Such shift could thus potentially correct the too strong decrease of the half-lives observed in some cases of the present study.
The effect of such correlations on the description of $\beta$-decay of selected nuclei should be investigated in a future paper.
\\ \\
The competition between GT and FF transitions in the present approach was also analyzed in detail. 
Overall the fragmentation induced by QVC was found to decrease the contribution of FF modes near stability, compared to RQRPA alone, which typically predicts very few GT states contributing to the decay, due to small decay $Q$ values. 
The contribution of FF modes was found to be below $\simeq 20-30 \%$ in most $N=50$ and $N=82$ nuclei when QVC is included, with an increase towards the dripline and near stability in $N=82$, up to a value of $\simeq 50 \%$.
In the heavier nuclei, the probability of decay via FF transition was found to be greater, with a contribution of more than $\simeq 40-50 \%$ up to $100 \%$ in the $N=126$ chain, and a contribution of more than $\simeq 60 \%$ in the $N=184$ chain (except near stability due to appearance of GT strength from QVC).

While in the light systems, $N=50$ and $N=82$, FF modes largely originate from rank-2 operators, we found that $1^-$ and $0^-$ modes contribute the most in heavy nuclei. Investigation of the rank-0 operator in $N=126$ showed a similar contribution from the spin-dipole component, relativistic component, and their destructive interference term (in absolute value). More detailed comparisons of the FF transitions components with shell-model calculations are planned for a next study.
\\ \\
\indent Comparisons with precise theoretical methods, which include detailed nucleonic correlations, will provide guidance for future developments of the present approach. At the same time, upcoming experimental measurements of $\beta$-decay half-lives in the $N=126$ region by radioactive-beam facilities such as FAIR \cite{FAIR} will be crucial to test the reliability of the present approach in the heavy region.
In the future we plan to extend this method to deformed nuclei and to perform global calculations of $\beta$-decay rates for $r$-process simulations.


\begin{acknowledgments}
We would like to thank Diana Alvear Terrero and Ante Ravli\'c for useful discussions and for their help in cross-checking the original relativistic QRPA code (D3C*).
  We also thank Xavier Mougeot and Elena Litvinova for interesting
  discussions.  This work is supported by Bielefeld Universit\"at, by the European Research Council
  (ERC) under the European Union's Horizon 2020 research and
  innovation programme (ERC Advanced Grant KILONOVA No.~885281), and by
  the Deutsche Forschungsgemeinschaft (DFG, German Research
  Foundation) - Project-ID MA 4248/3-1.  
  This research used resources of the National Energy Research
Scientific Computing Center, a DOE Office of Science User Facility
supported by the Office of Science of the U.S. Department of Energy
under Contract No. DE-AC02-05CH11231 using NERSC award
NP-ERCAP0029601.
\end{acknowledgments}

\appendix

\section{Calculations of first-forbidden transitions}
\label{sec:appendix}

We first recall a few aspects of the general response formalism, which can also be found in the literature (see e.g. Ref.~\cite{ring2004nuclear}). Subsequently we expand on the calculation of FF transitions in this formalism.

\subsection{General response formalism and transition strength distribution}
The response formalism is based on the calculation of the transition strength distribution associated to a (one-body) given transition operator $\bm{O}$
\begin{eqnarray}
S(\bm{O},E) &=& \sum_f |\braket{f|\bm{O}|i}|^2 \delta (E-E_f) \, , \nonumber \\
&=& \frac{-1}{\pi} \lim_{\Delta \rightarrow 0} \mbox{Im} \sum_f  \frac{|\braket{i | \bm{O} | f}|^2}{\omega- {E}_f } \, \nonumber \\
&=& -\frac{1}{\pi} \lim_{\Delta\rightarrow 0^+} \mbox{Im } \, \Pi(\bm{O}, \omega) \; ,
\label{eq:strength_app1}
\end{eqnarray}
with $\omega = E + i \Delta$, and 
where we have introduced the polarizability $\Pi(\omega, \bm{O})$  defined as
\begin{eqnarray}
\Pi(\bm{O}, \omega) &=& \sum_f  \frac{|\braket{i | \bm{O} | f}|^2}{\omega- {E}_f } \nonumber \\
                     &=& \braket{i| \bm{O}^\dagger \bm{R}(\omega) \bm{O} |i} \; ,\nonumber \\
    \label{eq:polariz_app}
\end{eqnarray}
where $\bm{R}(\omega) = 1/ (\omega - \bm{H}(\omega))$, where $\bm{H}(\omega)$ denotes the Hamiltonian of the system (which can be energy-dependent depending on the adopted many-body approximation).

Expressing the polarizability in single a quasiparticle basis
$i = \{ \eta_i, k_i \}$, with  $k_i = \{ n_i, \pi_i,  j_i, \tau_i \}$
\begin{eqnarray}
\Pi(\bm{O}, \omega) &=& \sum_{1234} O_{12}^{*} R_{1423} (\omega) O_{34} \; ,
    \label{eq:polariz_app2}
\end{eqnarray}
where the response function $R_{1423} (\omega)$ is given by
\begin{eqnarray}
    R_{1423} (\omega) &=& \braket{i | a_2^\dagger a_1 \bm{R} (\omega) a_3^\dagger a_4  |i} \nonumber \\
    &=& \sum_f \frac{ \braket{i | a_2^\dagger a_1 |f} \braket{f | a_3^\dagger a_4 |i} }{\omega - E_f} 
\end{eqnarray}
As in the main text, in the case where $\bm{O}$ is a charge-exchange (GT or FF) transition operators, odd (resp. even) numbers will denote proton (resp. neutron) states.
\\ 

In an angular-momentum coupled form, the polarizability reads
\begin{eqnarray}
    \Pi(\bm{O}^J, \omega) &=& \braket{i| \bm{O}^{J\, \dagger} \bm{R}^J(\omega) \bm{O}^J |i} \; , \nonumber \\
                        &=& \sum_{(1234)} O_{(12)}^{J,*} R_{(1423)}^J (\omega) O_{(34)}^{J} \; .
                        \label{eq:pola_app_J}
\end{eqnarray}
where $(i) = \{ \eta_i, n_i, j_i, \pi_i, \tau_i  \}$, and
where the coupled response function $R_{(1423)}^J (\omega)$ is given by
\begin{eqnarray}
 &&  R_{(1423)}^J (\omega) =  (2J+1) \sum_{m_1,m_2,m_3,m_4} (-1)^{j_1 - m_1}  (-1)^{j_3 - m_3} \nonumber \\ 
 &&   \hspace{0.5cm} \times \begin{pmatrix}
        j_1 & j_2 & J \\
        m_1 & -m_2 & M
    \end{pmatrix}
    \begin{pmatrix}
        j_3 & j_4 & J \\
        m_3 & -m_4 & M
    \end{pmatrix}
    R_{1423} (\omega) \; . \nonumber \\ 
\end{eqnarray}

The equations above are general, and in the present framework, the response function is calculated by including QVC effects, as described in section~\ref{sec:formalism}.

\subsection{First-forbbiden transitions in the response formalism}
The FF transitions include interference terms between operators of the same rank. Such terms can be calculated directly if one has access to the transition density matrices $\rho({\bm{O}^J})_{k_1,k_2} = \braket{f || \bm{O}^J ||i}$ associated to the various transition operators $\bm{O}$.
In Ref.~\cite{Litvinova:2007gg} a formula is provided which relates the transition strength distribution to the transition density matrix, however this expression is defined only up to a phase, and therefore does not allow to compute the sign of the interferences.  
Thus, in the response formalism, one has to find an alternative way to compute the FF transitions. A procedure was described in Ref.~\cite{Mustonen:2014bya}, and involves the computation of mixed transition strength distributions.
Here we largely follow this procedure and adapt it to the present framework, which 
does not involve contour integrations as in Ref.~\cite{Mustonen:2014bya}. We provide more details below.

As discussed in section~\ref{sec:beta_decay_theory}, the beta-decay rate associated with FF transitions is given by
\begin{eqnarray}
\lambda_{\text{tot}} &=&  \frac{\mbox{ln 2} }{K}   \sum_{f , E_f < -m_e} \int_1^{W_f} dW W \sqrt{W^2-1} (W_f -W)^2 \nonumber \\
&& \hspace{2cm} \times F(Z+1, W) \, C_{\text{FF}}(W) \; ,
\label{eq:rateFF}
\end{eqnarray}
where $W = E_e/ m_e$, $W_f \equiv (M_i - M_f)/m_e = - (E_f/m_e) $, with $M_i$ and $M_f$ the initial and final nuclear masses, repectively,
and $F(Z+1, W) $ is the Fermi function \cite{BEHRENS1971}. 
The energy-dependent forbidden shape factor $C_{\text{FF}}(W)$ is a sum of four terms:
\begin{equation}
C_{\text{FF}}(W) = k + ka \, W + kb \, W^{-1} + kc \, W^2 \; .
\end{equation}
where $k, \; ka, \; kb, \; kc$ are given in Eq.~(\ref{eq:k})-(\ref{eq:kc}).

Below we consider separately the contributions of rank-0, rank-1 and rank-2 FF transition operators:
\begin{eqnarray}
    C_{\text{FF}}(W) = \sum_{J=0,1,2} C_{\text{FF}}^{(J)}(W) \; .
\end{eqnarray}
and detail the calculation for the $J=0$ term. Those with $J=1,2$ are done straightforwardly in the same way.

As detailed in section~\ref{sec:beta_decay_theory}, the forbidden shape factor $C_{\text{FF}}^{(J=0)}(W)$ has contribution from $k$ and $kb$ terms, as
\begin{eqnarray}
    C_{\text{FF}}^{(J=0)}(W) = k^{(0)} + kb^{(0)} \, W^{-1} \; .
\end{eqnarray}

For clarity, in the following we will make explicit the dependence of the quantities $k^{(0)}$ and $kb^{(0)}$ on the final-state energy $E_f$.
Thus we have
\begin{eqnarray}
    \lambda^{(J=0)} &=& \lambda^{(J=0)}_k + \lambda^{(J=0)}_{kb} \nonumber \\
    &=& \frac{\ln 2}{K} \sum_{f, E_f < - m_e} \Bigl( k^{(0)}(E_f) \, h_0(E_f) \nonumber \\
    && \hspace{2cm} + \;  kb^{(0)}(E_f) \, h_{-1}(E_f) \Bigr) \nonumber \\
    &=& \frac{\ln 2}{K}  \int^{-m_e} \mbox{d}E \sum_{f} \delta(E-E_f) \nonumber \\
    && \hspace{1cm} \times  \Bigl( k^{(0)}(E) \, h_0(E) + kb^{(0)}(E) \, h_{-1}(E) \Bigr) \nonumber \\
    &\equiv& \frac{\ln 2}{K}  \int^{-m_e} \mbox{d}E \left( S^{(0)}_k (E) \, h_0(E)  + S^{(0)}_{kb} (E) \, h_{-1}(E) \right) \; , \nonumber \\
    \label{eq:integ_1}
\end{eqnarray}
where the function $h_n(E_f)$ encompasses the lepton kinematics and is defined by
\begin{eqnarray}
    h_n (E_f) &=& \int_1^{W_f=- E_f /m_e} \mbox{d}W \, W^{n+1} \sqrt{W^2-1} (W_f -W)^2 \; , \nonumber \\
   && \hspace{1cm}  \times F(Z+1, W) \; ,  
\end{eqnarray}
and $S^{(0)}_{K} (E)$ are "strength" distributions
\begin{eqnarray}
    S^{(0)}_{K} (E) = \sum_{f} \delta(E-E_f) \, K^{(0)}(E) \; , \ \ \mbox{with } K \equiv k, \,  kb \; . \nonumber \\ \label{eq:str_app} 
\end{eqnarray}
Thus the contribution to the decay rate $\lambda^{(J=0)}$ can be obtained by integrating the distributions $S^{(0)}_{K} (E) $ multiplied by the corresponding leptonic function, up to the electron mass. 
Note that in principle the lower bound of the integral in Eq.~\eqref{eq:integ_1} is given by $\Delta B - (m_n -m_p)$ where $\Delta B = B_i - B_f$ is the binding energy difference between initial and final nuclear states, and $m_n$ and $m_p$ are neutron and proton masses, respectively. In practice, we simply include all states with excitation energy below $-m_e$.
\\
\\
\indent
Let us first consider the term $k^{(0)}(E_f)$. We have (see Eqs.~(\ref{eq:k}-\ref{eq:zeta0}))
\begin{eqnarray}
k^{(0)}(E_f) &=& \bigl[ \zeta_0^2 + \frac{1}{9} w^2 \bigr]_{(J=0)}  \nonumber \\
        &=& \left( \xi 'v + \xi w' +\frac{W_f}{3} w \right)^2 + \frac{1}{9} w^2 \nonumber \\
        &=& \frac{1}{9} \left( W_f^2 +1 \right) w^2 + \xi^2 w'^2 + (\xi 'v)^2 \nonumber \\
        && +\frac{2}{3} \xi \, W_f \, w \, w' + \frac{2}{3} W_f \, w \, (\xi 'v) + 2 \xi\,  w' \, (\xi 'v) \; . \nonumber \\
        \label{eq:k0_app}
\end{eqnarray}
As seen from Eq.~(\ref{eq:FF_w}, \ref{eq:FF_wp}, \ref{eq:gamma5}), each of the terms $X = w, w', \xi' v$ are $\propto \braket{f || \bm{O}^J || i}$ where $\bm{O}$ is the corresponding transition operator. 
Thus, the first three terms on the right-hand side of Eq.~(\ref{eq:k0_app}) are usual strength distributions, while the last three terms are "mixed strengths" corresponding to interferences between the FF rank-0 operators.

Indeed, the first three terms on the right-hand side of Eq.~(\ref{eq:k0_app}) will contribute to the "strength" $S^{(0)}_{k} (E)$ in Eq.~(\ref{eq:str_app}) as
\begin{eqnarray}
S^{(0)}_{k} (\bm{O}^J,E) &=&  \sum_{f} \delta(E-E_f) \, A(E_f) \, |X|^2 \nonumber \\
    &=& \sum_{f} \delta(E-E_f)  A'(E_f) \, |\braket{f || \bm{O}^J || i}|^2 \nonumber \\
    &=& A'(E) \times  \frac{-1}{\pi} \lim_{\Delta \rightarrow 0} \mbox{Im} \sum_{f} \frac{|\braket{i || \bm{O}^J || f}|^2}{\omega- E_f} \nonumber \\
    &= & A'(E) \times \frac{-1}{\pi} \lim_{\Delta \rightarrow 0} \mbox{Im} \, \Pi(\bm{O}^J, \omega)  \; , \nonumber \\
\end{eqnarray}
where $\omega = E+i\Delta$ and $A,A'$ are energy-dependent variables coming from the pre-factors in $X$ and $k^{(0)} (E_f)$.
We recognize the expression of the usual strength distribution as shown in Eq.~(\ref{eq:strength1}), which is obtained from the polarizability $\Pi^{(\bm{O})} (\omega)$ as in Eq.~(\ref{eq:pola_app_J})

On the other hand, the last three terms on the right-hand side of Eq.~(\ref{eq:k0_app}) contributes to the "strength" $S^{(0)}_{k} (E)$ in Eq.~(\ref{eq:str_app}) as
\begin{eqnarray}
&&S^{(0)}_{k} (\bm{O}^J,\bm{O'}^J,E)  \nonumber \\
&=& \sum_{f} \delta(E-E_f) \, A(E_f) \, X^* \, X' \nonumber \\
    &=&  \sum_{f} \delta(E-E_f)  \, A'(E_f) \braket{i || \bm{O}^{\dagger \, J} || f} \braket{f || \bm{O'}^J || i} \nonumber \\
     &=& A'(E) \times \frac{-1}{\pi} \lim_{\Delta \rightarrow 0} \mbox{Im} \sum_{f} \frac{ \braket{i || \bm{O}^{\dagger \, J} || f} \braket{f || \bm{O'}^J || i}}{\omega- E_f} \nonumber \\
    &= & A'(E) \times \frac{-1}{\pi} \lim_{\Delta \rightarrow 0} \mbox{Im} \, \Pi(\bm{O}^J,\bm{O'}^J, \omega) \; , \nonumber \\
\end{eqnarray}

where the "mixed" polarizability is given by 
\begin{eqnarray}
    \Pi(\bm{O}^J,\bm{O'}^J, \omega) &=& \sum_{(1423)} 
                          \bm{O}^{J\, *}_{(12)}  \, R^J_{(1423)} (\omega) \, \bm{O'}^J_{(34)} \; . \nonumber \\
\end{eqnarray}

Thus, all the FF contributions can be obtained by 
folding the response function with the same operator $\bm{O}^J$ on both sides to obtain the usual strength terms, or with different operators $\bm{O},\, \bm{O'}$ on each side to obtain the interferences between these transition operators.

%
%

\bibliography{biblio}

\begin{thebibliography}{73}%
\makeatletter
\providecommand \@ifxundefined [1]{%
 \@ifx{#1\undefined}
}%
\providecommand \@ifnum [1]{%
 \ifnum #1\expandafter \@firstoftwo
 \else \expandafter \@secondoftwo
 \fi
}%
\providecommand \@ifx [1]{%
 \ifx #1\expandafter \@firstoftwo
 \else \expandafter \@secondoftwo
 \fi
}%
\providecommand \natexlab [1]{#1}%
\providecommand \enquote  [1]{``#1''}%
\providecommand \bibnamefont  [1]{#1}%
\providecommand \bibfnamefont [1]{#1}%
\providecommand \citenamefont [1]{#1}%
\providecommand \href@noop [0]{\@secondoftwo}%
\providecommand \href [0]{\begingroup \@sanitize@url \@href}%
\providecommand \@href[1]{\@@startlink{#1}\@@href}%
\providecommand \@@href[1]{\endgroup#1\@@endlink}%
\providecommand \@sanitize@url [0]{\catcode `\\12\catcode `\$12\catcode
  `\&12\catcode `\#12\catcode `\^12\catcode `\_12\catcode `\%12\relax}%
\providecommand \@@startlink[1]{}%
\providecommand \@@endlink[0]{}%
\providecommand \url  [0]{\begingroup\@sanitize@url \@url }%
\providecommand \@url [1]{\endgroup\@href {#1}{\urlprefix }}%
\providecommand \urlprefix  [0]{URL }%
\providecommand \Eprint [0]{\href }%
\providecommand \doibase [0]{https://doi.org/}%
\providecommand \selectlanguage [0]{\@gobble}%
\providecommand \bibinfo  [0]{\@secondoftwo}%
\providecommand \bibfield  [0]{\@secondoftwo}%
\providecommand \translation [1]{[#1]}%
\providecommand \BibitemOpen [0]{}%
\providecommand \bibitemStop [0]{}%
\providecommand \bibitemNoStop [0]{.\EOS\space}%
\providecommand \EOS [0]{\spacefactor3000\relax}%
\providecommand \BibitemShut  [1]{\csname bibitem#1\endcsname}%
\let\auto@bib@innerbib\@empty
\bibitem [{\citenamefont {Burbidge}\ \emph {et~al.}(1957)\citenamefont
  {Burbidge}, \citenamefont {Burbidge}, \citenamefont {Fowler},\ and\
  \citenamefont {Hoyle}}]{RevModPhys.29.547}%
  \BibitemOpen
  \bibfield  {author} {\bibinfo {author} {\bibfnamefont {E.~M.}\ \bibnamefont
  {Burbidge}}, \bibinfo {author} {\bibfnamefont {G.~R.}\ \bibnamefont
  {Burbidge}}, \bibinfo {author} {\bibfnamefont {W.~A.}\ \bibnamefont
  {Fowler}},\ and\ \bibinfo {author} {\bibfnamefont {F.}~\bibnamefont
  {Hoyle}},\ }\bibfield  {title} {\bibinfo {title} {Synthesis of the elements
  in stars},\ }\href {https://doi.org/10.1103/RevModPhys.29.547} {\bibfield
  {journal} {\bibinfo  {journal} {Rev. Mod. Phys.}\ }\textbf {\bibinfo {volume}
  {29}},\ \bibinfo {pages} {547} (\bibinfo {year} {1957})}\BibitemShut
  {NoStop}%
\bibitem [{\citenamefont {Cameron}(1957)}]{Cameron:1957}%
  \BibitemOpen
  \bibfield  {author} {\bibinfo {author} {\bibfnamefont {A.~G.~W.}\
  \bibnamefont {Cameron}},\ }\href@noop {} {\emph {\bibinfo {title} {Stellar
  Evolution, Nuclear Astrophysics, and Nucleogenesis}}},\ \bibinfo {type}
  {Report}\ \bibinfo {number} {CRL-41}\ (\bibinfo  {institution} {Chalk
  River},\ \bibinfo {year} {1957})\ \bibinfo {note} {reprinted in D. M. Kahl,
  Ed., \emph{Stellar Evolution, Nuclear Astrophysics, and Nucleogenesis}
  (Dover, New York, 2013)}\BibitemShut {NoStop}%
\bibitem [{\citenamefont {Abbott}\ \emph {et~al.}(2017)\citenamefont {Abbott}
  \emph {et~al.}}]{LIGOScientific:2017vwq}%
  \BibitemOpen
  \bibfield  {author} {\bibinfo {author} {\bibfnamefont {B.~P.}\ \bibnamefont
  {Abbott}} \emph {et~al.} (\bibinfo {collaboration} {LIGO Scientific,
  Virgo}),\ }\bibfield  {title} {\bibinfo {title} {{GW170817: Observation of
  Gravitational Waves from a Binary Neutron Star Inspiral}},\ }\href
  {https://doi.org/10.1103/PhysRevLett.119.161101} {\bibfield  {journal}
  {\bibinfo  {journal} {Phys. Rev. Lett.}\ }\textbf {\bibinfo {volume} {119}},\
  \bibinfo {pages} {161101} (\bibinfo {year} {2017})},\ \Eprint
  {https://arxiv.org/abs/1710.05832} {arXiv:1710.05832 [gr-qc]} \BibitemShut
  {NoStop}%
\bibitem [{\citenamefont {{Cowan}}\ \emph {et~al.}(2021)\citenamefont
  {{Cowan}}, \citenamefont {{Sneden}}, \citenamefont {{Lawler}}, \citenamefont
  {{Aprahamian}}, \citenamefont {{Wiescher}}, \citenamefont {{Langanke}},
  \citenamefont {{Mart{\'\i}nez-Pinedo}},\ and\ \citenamefont
  {{Thielemann}}}]{Cowan.Sneden.ea:2021}%
  \BibitemOpen
  \bibfield  {author} {\bibinfo {author} {\bibfnamefont {J.~J.}\ \bibnamefont
  {{Cowan}}}, \bibinfo {author} {\bibfnamefont {C.}~\bibnamefont {{Sneden}}},
  \bibinfo {author} {\bibfnamefont {J.~E.}\ \bibnamefont {{Lawler}}}, \bibinfo
  {author} {\bibfnamefont {A.}~\bibnamefont {{Aprahamian}}}, \bibinfo {author}
  {\bibfnamefont {M.}~\bibnamefont {{Wiescher}}}, \bibinfo {author}
  {\bibfnamefont {K.}~\bibnamefont {{Langanke}}}, \bibinfo {author}
  {\bibfnamefont {G.}~\bibnamefont {{Mart{\'\i}nez-Pinedo}}},\ and\ \bibinfo
  {author} {\bibfnamefont {F.-K.}\ \bibnamefont {{Thielemann}}},\ }\bibfield
  {title} {\bibinfo {title} {{Origin of the heaviest elements: The rapid
  neutron-capture process}},\ }\href
  {https://doi.org/10.1103/RevModPhys.93.015002} {\bibfield  {journal}
  {\bibinfo  {journal} {Rev. Mod. Phys.}\ }\textbf {\bibinfo {volume} {93}},\
  \bibinfo {pages} {015002} (\bibinfo {year} {2021})}\BibitemShut {NoStop}%
\bibitem [{\citenamefont {Mendoza-Temis}\ \emph {et~al.}(2015)\citenamefont
  {Mendoza-Temis}, \citenamefont {Wu}, \citenamefont {Langanke}, \citenamefont
  {Mart\'{\i}nez-Pinedo}, \citenamefont {Bauswein},\ and\ \citenamefont
  {Janka}}]{Mendoza-Temis.Wu.ea:2015}%
  \BibitemOpen
  \bibfield  {author} {\bibinfo {author} {\bibfnamefont {J.~J.}\ \bibnamefont
  {Mendoza-Temis}}, \bibinfo {author} {\bibfnamefont {M.-R.}\ \bibnamefont
  {Wu}}, \bibinfo {author} {\bibfnamefont {K.}~\bibnamefont {Langanke}},
  \bibinfo {author} {\bibfnamefont {G.}~\bibnamefont {Mart\'{\i}nez-Pinedo}},
  \bibinfo {author} {\bibfnamefont {A.}~\bibnamefont {Bauswein}},\ and\
  \bibinfo {author} {\bibfnamefont {H.-T.}\ \bibnamefont {Janka}},\ }\bibfield
  {title} {\bibinfo {title} {{Nuclear robustness of the $r$ process in
  neutron-star mergers}},\ }\href {https://doi.org/10.1103/PhysRevC.92.055805}
  {\bibfield  {journal} {\bibinfo  {journal} {Phys. Rev. C}\ }\textbf {\bibinfo
  {volume} {92}},\ \bibinfo {pages} {055805} (\bibinfo {year}
  {2015})}\BibitemShut {NoStop}%
\bibitem [{\citenamefont {Eichler}\ \emph {et~al.}(2015)\citenamefont
  {Eichler}, \citenamefont {Arcones}, \citenamefont {Kelic}, \citenamefont
  {Korobkin}, \citenamefont {Langanke}, \citenamefont {Marketin}, \citenamefont
  {Martinez-Pinedo}, \citenamefont {Panov}, \citenamefont {Rauscher},
  \citenamefont {Rosswog}, \citenamefont {Winteler}, \citenamefont {Zinner},\
  and\ \citenamefont {Thielemann}}]{Eichler_2015}%
  \BibitemOpen
  \bibfield  {author} {\bibinfo {author} {\bibfnamefont {M.}~\bibnamefont
  {Eichler}}, \bibinfo {author} {\bibfnamefont {A.}~\bibnamefont {Arcones}},
  \bibinfo {author} {\bibfnamefont {A.}~\bibnamefont {Kelic}}, \bibinfo
  {author} {\bibfnamefont {O.}~\bibnamefont {Korobkin}}, \bibinfo {author}
  {\bibfnamefont {K.}~\bibnamefont {Langanke}}, \bibinfo {author}
  {\bibfnamefont {T.}~\bibnamefont {Marketin}}, \bibinfo {author}
  {\bibfnamefont {G.}~\bibnamefont {Martinez-Pinedo}}, \bibinfo {author}
  {\bibfnamefont {I.}~\bibnamefont {Panov}}, \bibinfo {author} {\bibfnamefont
  {T.}~\bibnamefont {Rauscher}}, \bibinfo {author} {\bibfnamefont
  {S.}~\bibnamefont {Rosswog}}, \bibinfo {author} {\bibfnamefont
  {C.}~\bibnamefont {Winteler}}, \bibinfo {author} {\bibfnamefont {N.~T.}\
  \bibnamefont {Zinner}},\ and\ \bibinfo {author} {\bibfnamefont {F.-K.}\
  \bibnamefont {Thielemann}},\ }\bibfield  {title} {\bibinfo {title} {The role
  of fission in neutron star mergers and its impact on ther-process peaks},\
  }\href {https://doi.org/10.1088/0004-637x/808/1/30} {\bibfield  {journal}
  {\bibinfo  {journal} {Astrophys. J.}\ }\textbf {\bibinfo {volume} {808}},\
  \bibinfo {pages} {30} (\bibinfo {year} {2015})}\BibitemShut {NoStop}%
\bibitem [{\citenamefont {Mumpower}\ \emph {et~al.}(2016)\citenamefont
  {Mumpower}, \citenamefont {Surman}, \citenamefont {McLaughlin},\ and\
  \citenamefont {Aprahamian}}]{MUMPOWER201686}%
  \BibitemOpen
  \bibfield  {author} {\bibinfo {author} {\bibfnamefont {M.}~\bibnamefont
  {Mumpower}}, \bibinfo {author} {\bibfnamefont {R.}~\bibnamefont {Surman}},
  \bibinfo {author} {\bibfnamefont {G.}~\bibnamefont {McLaughlin}},\ and\
  \bibinfo {author} {\bibfnamefont {A.}~\bibnamefont {Aprahamian}},\ }\bibfield
   {title} {\bibinfo {title} {The impact of individual nuclear properties on
  r-process nucleosynthesis},\ }\href
  {https://doi.org/https://doi.org/10.1016/j.ppnp.2015.09.001} {\bibfield
  {journal} {\bibinfo  {journal} {Progress in Particle and Nuclear Physics}\
  }\textbf {\bibinfo {volume} {86}},\ \bibinfo {pages} {86} (\bibinfo {year}
  {2016})}\BibitemShut {NoStop}%
\bibitem [{\citenamefont {Kajino}\ \emph {et~al.}(2019)\citenamefont {Kajino},
  \citenamefont {Aoki}, \citenamefont {Balantekin}, \citenamefont {Diehl},
  \citenamefont {Famiano},\ and\ \citenamefont {Mathews}}]{KAJINO2019109}%
  \BibitemOpen
  \bibfield  {author} {\bibinfo {author} {\bibfnamefont {T.}~\bibnamefont
  {Kajino}}, \bibinfo {author} {\bibfnamefont {W.}~\bibnamefont {Aoki}},
  \bibinfo {author} {\bibfnamefont {A.}~\bibnamefont {Balantekin}}, \bibinfo
  {author} {\bibfnamefont {R.}~\bibnamefont {Diehl}}, \bibinfo {author}
  {\bibfnamefont {M.}~\bibnamefont {Famiano}},\ and\ \bibinfo {author}
  {\bibfnamefont {G.}~\bibnamefont {Mathews}},\ }\bibfield  {title} {\bibinfo
  {title} {Current status of r-process nucleosynthesis},\ }\href
  {https://doi.org/https://doi.org/10.1016/j.ppnp.2019.02.008} {\bibfield
  {journal} {\bibinfo  {journal} {Progress in Particle and Nuclear Physics}\
  }\textbf {\bibinfo {volume} {107}},\ \bibinfo {pages} {109} (\bibinfo {year}
  {2019})}\BibitemShut {NoStop}%
\bibitem [{\citenamefont {Pfeiffer}\ \emph {et~al.}(2001)\citenamefont
  {Pfeiffer}, \citenamefont {Kratz}, \citenamefont {Thielemann},\ and\
  \citenamefont {Walters}}]{Pfeiffer.Kratz.ea:2001}%
  \BibitemOpen
  \bibfield  {author} {\bibinfo {author} {\bibfnamefont {B.}~\bibnamefont
  {Pfeiffer}}, \bibinfo {author} {\bibfnamefont {K.-L.}\ \bibnamefont {Kratz}},
  \bibinfo {author} {\bibfnamefont {F.-K.}\ \bibnamefont {Thielemann}},\ and\
  \bibinfo {author} {\bibfnamefont {W.~B.}\ \bibnamefont {Walters}},\
  }\bibfield  {title} {\bibinfo {title} {Nuclear structure studies for the
  astrophysical r-process},\ }\href@noop {} {\bibfield  {journal} {\bibinfo
  {journal} {Nucl. Phys. A}\ }\textbf {\bibinfo {volume} {693}},\ \bibinfo
  {pages} {282} (\bibinfo {year} {2001})}\BibitemShut {NoStop}%
\bibitem [{\citenamefont {Arcones}\ and\ \citenamefont
  {Mart{\'\i}nez-Pinedo}(2011)}]{Arcones.Martinez-Pinedo:2011}%
  \BibitemOpen
  \bibfield  {author} {\bibinfo {author} {\bibfnamefont {A.}~\bibnamefont
  {Arcones}}\ and\ \bibinfo {author} {\bibfnamefont {G.}~\bibnamefont
  {Mart{\'\i}nez-Pinedo}},\ }\bibfield  {title} {\bibinfo {title} {Dynamical
  $r$-process studies within the neutrino-driven wind scenario and its
  sensitivity to the nuclear physics input},\ }\href
  {https://doi.org/10.1103/PhysRevC.83.045809} {\bibfield  {journal} {\bibinfo
  {journal} {Phys. Rev. C}\ }\textbf {\bibinfo {volume} {83}},\ \bibinfo
  {pages} {045809} (\bibinfo {year} {2011})}\BibitemShut {NoStop}%
\bibitem [{\citenamefont {Mumpower}\ \emph {et~al.}(2014)\citenamefont
  {Mumpower}, \citenamefont {Cass}, \citenamefont {Passucci}, \citenamefont
  {Surman},\ and\ \citenamefont {Aprahamian}}]{Mumpower2014}%
  \BibitemOpen
  \bibfield  {author} {\bibinfo {author} {\bibfnamefont {M.}~\bibnamefont
  {Mumpower}}, \bibinfo {author} {\bibfnamefont {J.}~\bibnamefont {Cass}},
  \bibinfo {author} {\bibfnamefont {G.}~\bibnamefont {Passucci}}, \bibinfo
  {author} {\bibfnamefont {R.}~\bibnamefont {Surman}},\ and\ \bibinfo {author}
  {\bibfnamefont {A.}~\bibnamefont {Aprahamian}},\ }\bibfield  {title}
  {\bibinfo {title} {Sensitivity studies for the main r process: $\beta$-decay
  rates},\ }\href {https://doi.org/10.1063/1.4867192} {\bibfield  {journal}
  {\bibinfo  {journal} {AIP Advances}\ }\textbf {\bibinfo {volume} {4}},\
  \bibinfo {pages} {041009} (\bibinfo {year} {2014})},\ \Eprint
  {https://arxiv.org/abs/https://doi.org/10.1063/1.4867192}
  {https://doi.org/10.1063/1.4867192} \BibitemShut {NoStop}%
\bibitem [{\citenamefont {Shafer}\ \emph {et~al.}(2016)\citenamefont {Shafer},
  \citenamefont {Engel}, \citenamefont {Fr\"ohlich}, \citenamefont
  {McLaughlin}, \citenamefont {Mumpower},\ and\ \citenamefont
  {Surman}}]{Shafer2016}%
  \BibitemOpen
  \bibfield  {author} {\bibinfo {author} {\bibfnamefont {T.}~\bibnamefont
  {Shafer}}, \bibinfo {author} {\bibfnamefont {J.}~\bibnamefont {Engel}},
  \bibinfo {author} {\bibfnamefont {C.}~\bibnamefont {Fr\"ohlich}}, \bibinfo
  {author} {\bibfnamefont {G.~C.}\ \bibnamefont {McLaughlin}}, \bibinfo
  {author} {\bibfnamefont {M.}~\bibnamefont {Mumpower}},\ and\ \bibinfo
  {author} {\bibfnamefont {R.}~\bibnamefont {Surman}},\ }\bibfield  {title}
  {\bibinfo {title} {{$\ensuremath{\beta}$ decay of deformed $r$-process nuclei
  near $A=80$ and $A=160$, including odd-$A$ and odd-odd nuclei, with the
  Skyrme finite-amplitude method}},\ }\href
  {https://doi.org/10.1103/PhysRevC.94.055802} {\bibfield  {journal} {\bibinfo
  {journal} {Phys. Rev. C}\ }\textbf {\bibinfo {volume} {94}},\ \bibinfo
  {pages} {055802} (\bibinfo {year} {2016})}\BibitemShut {NoStop}%
\bibitem [{\citenamefont {Marketin}\ \emph {et~al.}(2016)\citenamefont
  {Marketin}, \citenamefont {Huther},\ and\ \citenamefont
  {Mart\'\i{}nez-Pinedo}}]{Marketin2016}%
  \BibitemOpen
  \bibfield  {author} {\bibinfo {author} {\bibfnamefont {T.}~\bibnamefont
  {Marketin}}, \bibinfo {author} {\bibfnamefont {L.}~\bibnamefont {Huther}},\
  and\ \bibinfo {author} {\bibfnamefont {G.}~\bibnamefont
  {Mart\'\i{}nez-Pinedo}},\ }\bibfield  {title} {\bibinfo {title} {{Large-scale
  evaluation of \ensuremath{\beta}-decay rates of r-process nuclei with the
  inclusion of first-forbidden transitions}},\ }\href
  {https://doi.org/10.1103/PhysRevC.93.025805} {\bibfield  {journal} {\bibinfo
  {journal} {Phys. Rev. C}\ }\textbf {\bibinfo {volume} {93}},\ \bibinfo
  {pages} {025805} (\bibinfo {year} {2016})}\BibitemShut {NoStop}%
\bibitem [{\citenamefont {Ring}\ and\ \citenamefont
  {Schuck}(2004)}]{ring2004nuclear}%
  \BibitemOpen
  \bibfield  {author} {\bibinfo {author} {\bibfnamefont {P.}~\bibnamefont
  {Ring}}\ and\ \bibinfo {author} {\bibfnamefont {P.}~\bibnamefont {Schuck}},\
  }\href {https://books.google.de/books?id=PTynSM-nMA8C} {\emph {\bibinfo
  {title} {The Nuclear Many-Body Problem}}},\ Physics and astronomy online
  library\ (\bibinfo  {publisher} {Springer},\ \bibinfo {year}
  {2004})\BibitemShut {NoStop}%
\bibitem [{\citenamefont {Engel}\ \emph {et~al.}(1999)\citenamefont {Engel},
  \citenamefont {Bender}, \citenamefont {Dobaczewski}, \citenamefont
  {Nazarewicz},\ and\ \citenamefont {Surman}}]{Engel1999}%
  \BibitemOpen
  \bibfield  {author} {\bibinfo {author} {\bibfnamefont {J.}~\bibnamefont
  {Engel}}, \bibinfo {author} {\bibfnamefont {M.}~\bibnamefont {Bender}},
  \bibinfo {author} {\bibfnamefont {J.}~\bibnamefont {Dobaczewski}}, \bibinfo
  {author} {\bibfnamefont {W.}~\bibnamefont {Nazarewicz}},\ and\ \bibinfo
  {author} {\bibfnamefont {R.}~\bibnamefont {Surman}},\ }\bibfield  {title}
  {\bibinfo {title} {$\ensuremath{\beta}$ decay rates of r-process
  waiting-point nuclei in a self-consistent approach},\ }\href
  {https://doi.org/10.1103/PhysRevC.60.014302} {\bibfield  {journal} {\bibinfo
  {journal} {Phys. Rev. C}\ }\textbf {\bibinfo {volume} {60}},\ \bibinfo
  {pages} {014302} (\bibinfo {year} {1999})}\BibitemShut {NoStop}%
\bibitem [{\citenamefont {Borzov}(2003)}]{Borzov2003}%
  \BibitemOpen
  \bibfield  {author} {\bibinfo {author} {\bibfnamefont {I.~N.}\ \bibnamefont
  {Borzov}},\ }\bibfield  {title} {\bibinfo {title} {Gamow-teller and
  first-forbidden decays near the $r$-process paths at $n=50,82,$ and $126$},\
  }\href {https://doi.org/10.1103/PhysRevC.67.025802} {\bibfield  {journal}
  {\bibinfo  {journal} {Phys. Rev. C}\ }\textbf {\bibinfo {volume} {67}},\
  \bibinfo {pages} {025802} (\bibinfo {year} {2003})}\BibitemShut {NoStop}%
\bibitem [{\citenamefont {M\"oller}\ \emph {et~al.}(2003)\citenamefont
  {M\"oller}, \citenamefont {Pfeiffer},\ and\ \citenamefont
  {Kratz}}]{Moeller2003}%
  \BibitemOpen
  \bibfield  {author} {\bibinfo {author} {\bibfnamefont {P.}~\bibnamefont
  {M\"oller}}, \bibinfo {author} {\bibfnamefont {B.}~\bibnamefont {Pfeiffer}},\
  and\ \bibinfo {author} {\bibfnamefont {K.-L.}\ \bibnamefont {Kratz}},\
  }\bibfield  {title} {\bibinfo {title} {New calculations of gross
  $\ensuremath{\beta}$-decay properties for astrophysical applications:
  Speeding-up the classical r process},\ }\href
  {https://doi.org/10.1103/PhysRevC.67.055802} {\bibfield  {journal} {\bibinfo
  {journal} {Phys. Rev. C}\ }\textbf {\bibinfo {volume} {67}},\ \bibinfo
  {pages} {055802} (\bibinfo {year} {2003})}\BibitemShut {NoStop}%
\bibitem [{\citenamefont {Nik\v{s}i\'c}\ \emph {et~al.}(2005)\citenamefont
  {Nik\v{s}i\'c}, \citenamefont {Marketin}, \citenamefont {Vretenar},
  \citenamefont {Paar},\ and\ \citenamefont {Ring}}]{Niksic2005}%
  \BibitemOpen
  \bibfield  {author} {\bibinfo {author} {\bibfnamefont {T.}~\bibnamefont
  {Nik\v{s}i\'c}}, \bibinfo {author} {\bibfnamefont {T.}~\bibnamefont
  {Marketin}}, \bibinfo {author} {\bibfnamefont {D.}~\bibnamefont {Vretenar}},
  \bibinfo {author} {\bibfnamefont {N.}~\bibnamefont {Paar}},\ and\ \bibinfo
  {author} {\bibfnamefont {P.}~\bibnamefont {Ring}},\ }\bibfield  {title}
  {\bibinfo {title} {\ensuremath{\beta}-decay rates of $r$-process nuclei in
  the relativistic quasiparticle random phase approximation},\ }\href
  {https://doi.org/10.1103/PhysRevC.71.014308} {\bibfield  {journal} {\bibinfo
  {journal} {Phys. Rev. C}\ }\textbf {\bibinfo {volume} {71}},\ \bibinfo
  {pages} {014308} (\bibinfo {year} {2005})}\BibitemShut {NoStop}%
\bibitem [{\citenamefont {Niu}\ \emph {et~al.}(2013)\citenamefont {Niu},
  \citenamefont {Niu}, \citenamefont {Liang}, \citenamefont {Long},
  \citenamefont {Nik\v{s}i\'c}, \citenamefont {Vretenar},\ and\ \citenamefont
  {Meng}}]{NIU2013172}%
  \BibitemOpen
  \bibfield  {author} {\bibinfo {author} {\bibfnamefont {Z.}~\bibnamefont
  {Niu}}, \bibinfo {author} {\bibfnamefont {Y.}~\bibnamefont {Niu}}, \bibinfo
  {author} {\bibfnamefont {H.}~\bibnamefont {Liang}}, \bibinfo {author}
  {\bibfnamefont {W.}~\bibnamefont {Long}}, \bibinfo {author} {\bibfnamefont
  {T.}~\bibnamefont {Nik\v{s}i\'c}}, \bibinfo {author} {\bibfnamefont
  {D.}~\bibnamefont {Vretenar}},\ and\ \bibinfo {author} {\bibfnamefont
  {J.}~\bibnamefont {Meng}},\ }\bibfield  {title} {\bibinfo {title}
  {$\beta$-decay half-lives of neutron-rich nuclei and matter flow in the
  r-process},\ }\href
  {https://doi.org/https://doi.org/10.1016/j.physletb.2013.04.048} {\bibfield
  {journal} {\bibinfo  {journal} {Physics Letters B}\ }\textbf {\bibinfo
  {volume} {723}},\ \bibinfo {pages} {172} (\bibinfo {year}
  {2013})}\BibitemShut {NoStop}%
\bibitem [{\citenamefont {Fang}\ \emph
  {et~al.}(2013{\natexlab{a}})\citenamefont {Fang}, \citenamefont {Brown},\
  and\ \citenamefont {Suzuki}}]{PhysRevC.88.024314}%
  \BibitemOpen
  \bibfield  {author} {\bibinfo {author} {\bibfnamefont {D.-L.}\ \bibnamefont
  {Fang}}, \bibinfo {author} {\bibfnamefont {B.~A.}\ \bibnamefont {Brown}},\
  and\ \bibinfo {author} {\bibfnamefont {T.}~\bibnamefont {Suzuki}},\
  }\bibfield  {title} {\bibinfo {title} {$\ensuremath{\beta}$-decay properties
  for neutron-rich kr--tc isotopes from deformed $pn$-quasiparticle
  random-phase approximation calculations with realistic forces},\ }\href
  {https://doi.org/10.1103/PhysRevC.88.024314} {\bibfield  {journal} {\bibinfo
  {journal} {Phys. Rev. C}\ }\textbf {\bibinfo {volume} {88}},\ \bibinfo
  {pages} {024314} (\bibinfo {year} {2013}{\natexlab{a}})}\BibitemShut
  {NoStop}%
\bibitem [{\citenamefont {Fang}\ \emph
  {et~al.}(2013{\natexlab{b}})\citenamefont {Fang}, \citenamefont {Brown},\
  and\ \citenamefont {Suzuki}}]{PhysRevC.88.034304}%
  \BibitemOpen
  \bibfield  {author} {\bibinfo {author} {\bibfnamefont {D.-L.}\ \bibnamefont
  {Fang}}, \bibinfo {author} {\bibfnamefont {B.~A.}\ \bibnamefont {Brown}},\
  and\ \bibinfo {author} {\bibfnamefont {T.}~\bibnamefont {Suzuki}},\
  }\bibfield  {title} {\bibinfo {title} {Investigating
  $\ensuremath{\beta}$-decay properties of spherical nuclei along the possible
  $r$-process path},\ }\href {https://doi.org/10.1103/PhysRevC.88.034304}
  {\bibfield  {journal} {\bibinfo  {journal} {Phys. Rev. C}\ }\textbf {\bibinfo
  {volume} {88}},\ \bibinfo {pages} {034304} (\bibinfo {year}
  {2013}{\natexlab{b}})}\BibitemShut {NoStop}%
\bibitem [{\citenamefont {Martini}\ \emph {et~al.}(2014)\citenamefont
  {Martini}, \citenamefont {P\'eru},\ and\ \citenamefont
  {Goriely}}]{PhysRevC.89.044306}%
  \BibitemOpen
  \bibfield  {author} {\bibinfo {author} {\bibfnamefont {M.}~\bibnamefont
  {Martini}}, \bibinfo {author} {\bibfnamefont {S.}~\bibnamefont {P\'eru}},\
  and\ \bibinfo {author} {\bibfnamefont {S.}~\bibnamefont {Goriely}},\
  }\bibfield  {title} {\bibinfo {title} {Gamow-teller strength in deformed
  nuclei within the self-consistent charge-exchange quasiparticle random-phase
  approximation with the gogny force},\ }\href
  {https://doi.org/10.1103/PhysRevC.89.044306} {\bibfield  {journal} {\bibinfo
  {journal} {Phys. Rev. C}\ }\textbf {\bibinfo {volume} {89}},\ \bibinfo
  {pages} {044306} (\bibinfo {year} {2014})}\BibitemShut {NoStop}%
\bibitem [{\citenamefont {Sarriguren}(2015)}]{PhysRevC.91.044304}%
  \BibitemOpen
  \bibfield  {author} {\bibinfo {author} {\bibfnamefont {P.}~\bibnamefont
  {Sarriguren}},\ }\bibfield  {title} {\bibinfo {title}
  {{$\ensuremath{\beta}$-decay properties of neutron-rich Ge, Se, Kr, Sr, Ru,
  and Pd isotopes from deformed quasiparticle random-phase approximation}},\
  }\href {https://doi.org/10.1103/PhysRevC.91.044304} {\bibfield  {journal}
  {\bibinfo  {journal} {Phys. Rev. C}\ }\textbf {\bibinfo {volume} {91}},\
  \bibinfo {pages} {044304} (\bibinfo {year} {2015})}\BibitemShut {NoStop}%
\bibitem [{\citenamefont {Ney}\ \emph {et~al.}(2020)\citenamefont {Ney},
  \citenamefont {Engel}, \citenamefont {Li},\ and\ \citenamefont
  {Schunck}}]{Ney2020}%
  \BibitemOpen
  \bibfield  {author} {\bibinfo {author} {\bibfnamefont {E.~M.}\ \bibnamefont
  {Ney}}, \bibinfo {author} {\bibfnamefont {J.}~\bibnamefont {Engel}}, \bibinfo
  {author} {\bibfnamefont {T.}~\bibnamefont {Li}},\ and\ \bibinfo {author}
  {\bibfnamefont {N.}~\bibnamefont {Schunck}},\ }\bibfield  {title} {\bibinfo
  {title} {Global description of ${\ensuremath{\beta}}^{\ensuremath{-}}$ decay
  with the axially deformed skyrme finite-amplitude method: Extension to
  odd-mass and odd-odd nuclei},\ }\href
  {https://doi.org/10.1103/PhysRevC.102.034326} {\bibfield  {journal} {\bibinfo
   {journal} {Phys. Rev. C}\ }\textbf {\bibinfo {volume} {102}},\ \bibinfo
  {pages} {034326} (\bibinfo {year} {2020})}\BibitemShut {NoStop}%
\bibitem [{\citenamefont {Hao}\ \emph {et~al.}(2023)\citenamefont {Hao},
  \citenamefont {Niu},\ and\ \citenamefont {Niu}}]{PhysRevC.108.L062802}%
  \BibitemOpen
  \bibfield  {author} {\bibinfo {author} {\bibfnamefont {Y.-W.}\ \bibnamefont
  {Hao}}, \bibinfo {author} {\bibfnamefont {Y.-F.}\ \bibnamefont {Niu}},\ and\
  \bibinfo {author} {\bibfnamefont {Z.-M.}\ \bibnamefont {Niu}},\ }\bibfield
  {title} {\bibinfo {title} {Impact of nuclear $\ensuremath{\beta}$-decay rates
  on the $r$-process rare-earth peak abundances},\ }\href
  {https://doi.org/10.1103/PhysRevC.108.L062802} {\bibfield  {journal}
  {\bibinfo  {journal} {Phys. Rev. C}\ }\textbf {\bibinfo {volume} {108}},\
  \bibinfo {pages} {L062802} (\bibinfo {year} {2023})}\BibitemShut {NoStop}%
\bibitem [{\citenamefont {Chen}\ \emph {et~al.}(2023)\citenamefont {Chen},
  \citenamefont {Fang}, \citenamefont {Hao}, \citenamefont {Niu},\ and\
  \citenamefont {Niu}}]{Chen:2023fpe}%
  \BibitemOpen
  \bibfield  {author} {\bibinfo {author} {\bibfnamefont {J.}~\bibnamefont
  {Chen}}, \bibinfo {author} {\bibfnamefont {J.~Y.}\ \bibnamefont {Fang}},
  \bibinfo {author} {\bibfnamefont {Y.~W.}\ \bibnamefont {Hao}}, \bibinfo
  {author} {\bibfnamefont {Z.~M.}\ \bibnamefont {Niu}},\ and\ \bibinfo {author}
  {\bibfnamefont {Y.~F.}\ \bibnamefont {Niu}},\ }\bibfield  {title} {\bibinfo
  {title} {{Impact of Nuclear \ensuremath{\beta}-decay Half-life Uncertainties
  on the r-process Simulations}},\ }\href
  {https://doi.org/10.3847/1538-4357/acaeab} {\bibfield  {journal} {\bibinfo
  {journal} {Astrophys. J.}\ }\textbf {\bibinfo {volume} {943}},\ \bibinfo
  {pages} {102} (\bibinfo {year} {2023})}\BibitemShut {NoStop}%
\bibitem [{\citenamefont {Gambacurta}\ \emph {et~al.}(2020)\citenamefont
  {Gambacurta}, \citenamefont {Grasso},\ and\ \citenamefont
  {Engel}}]{Gambacurta:2020dhb}%
  \BibitemOpen
  \bibfield  {author} {\bibinfo {author} {\bibfnamefont {D.}~\bibnamefont
  {Gambacurta}}, \bibinfo {author} {\bibfnamefont {M.}~\bibnamefont {Grasso}},\
  and\ \bibinfo {author} {\bibfnamefont {J.}~\bibnamefont {Engel}},\ }\bibfield
   {title} {\bibinfo {title} {{Gamow-Teller Strength in $^{48}$Ca and $^{78}$Ni
  with the Charge-Exchange Subtracted Second Random-Phase Approximation}},\
  }\href {https://doi.org/10.1103/PhysRevLett.125.212501} {\bibfield  {journal}
  {\bibinfo  {journal} {Phys. Rev. Lett.}\ }\textbf {\bibinfo {volume} {125}},\
  \bibinfo {pages} {212501} (\bibinfo {year} {2020})},\ \Eprint
  {https://arxiv.org/abs/2007.04957} {arXiv:2007.04957 [nucl-th]} \BibitemShut
  {NoStop}%
\bibitem [{\citenamefont {Gambacurta}\ and\ \citenamefont
  {Grasso}(2022)}]{Gambacurta:2021zlv}%
  \BibitemOpen
  \bibfield  {author} {\bibinfo {author} {\bibfnamefont {D.}~\bibnamefont
  {Gambacurta}}\ and\ \bibinfo {author} {\bibfnamefont {M.}~\bibnamefont
  {Grasso}},\ }\bibfield  {title} {\bibinfo {title} {{Quenching of Gamow-Teller
  strengths and two-particle\textendash{}two-hole configurations}},\ }\href
  {https://doi.org/10.1103/PhysRevC.105.014321} {\bibfield  {journal} {\bibinfo
   {journal} {Phys. Rev. C}\ }\textbf {\bibinfo {volume} {105}},\ \bibinfo
  {pages} {014321} (\bibinfo {year} {2022})},\ \Eprint
  {https://arxiv.org/abs/2109.06064} {arXiv:2109.06064 [nucl-th]} \BibitemShut
  {NoStop}%
\bibitem [{\citenamefont {Col\`o}\ \emph {et~al.}(2010)\citenamefont {Col\`o},
  \citenamefont {Sagawa},\ and\ \citenamefont
  {Bortignon}}]{PhysRevC.82.064307}%
  \BibitemOpen
  \bibfield  {author} {\bibinfo {author} {\bibfnamefont {G.}~\bibnamefont
  {Col\`o}}, \bibinfo {author} {\bibfnamefont {H.}~\bibnamefont {Sagawa}},\
  and\ \bibinfo {author} {\bibfnamefont {P.~F.}\ \bibnamefont {Bortignon}},\
  }\bibfield  {title} {\bibinfo {title} {Effect of particle-vibration coupling
  on single-particle states: A consistent study within the skyrme framework},\
  }\href {https://doi.org/10.1103/PhysRevC.82.064307} {\bibfield  {journal}
  {\bibinfo  {journal} {Phys. Rev. C}\ }\textbf {\bibinfo {volume} {82}},\
  \bibinfo {pages} {064307} (\bibinfo {year} {2010})}\BibitemShut {NoStop}%
\bibitem [{\citenamefont {Litvinova}\ and\ \citenamefont
  {Ring}(2006)}]{PhysRevC.73.044328}%
  \BibitemOpen
  \bibfield  {author} {\bibinfo {author} {\bibfnamefont {E.}~\bibnamefont
  {Litvinova}}\ and\ \bibinfo {author} {\bibfnamefont {P.}~\bibnamefont
  {Ring}},\ }\bibfield  {title} {\bibinfo {title} {Covariant theory of
  particle-vibrational coupling and its effect on the single-particle
  spectrum},\ }\href {https://doi.org/10.1103/PhysRevC.73.044328} {\bibfield
  {journal} {\bibinfo  {journal} {Phys. Rev. C}\ }\textbf {\bibinfo {volume}
  {73}},\ \bibinfo {pages} {044328} (\bibinfo {year} {2006})}\BibitemShut
  {NoStop}%
\bibitem [{\citenamefont {Niu}\ \emph {et~al.}(2012)\citenamefont {Niu},
  \citenamefont {Colo}, \citenamefont {Brenna}, \citenamefont {Bortignon},\
  and\ \citenamefont {Meng}}]{Niu:2012mi}%
  \BibitemOpen
  \bibfield  {author} {\bibinfo {author} {\bibfnamefont {Y.~F.}\ \bibnamefont
  {Niu}}, \bibinfo {author} {\bibfnamefont {G.}~\bibnamefont {Colo}}, \bibinfo
  {author} {\bibfnamefont {M.}~\bibnamefont {Brenna}}, \bibinfo {author}
  {\bibfnamefont {P.~F.}\ \bibnamefont {Bortignon}},\ and\ \bibinfo {author}
  {\bibfnamefont {J.}~\bibnamefont {Meng}},\ }\bibfield  {title} {\bibinfo
  {title} {{The Gamow-Teller response within Skyrme random-phase approximation
  plus particle-vibration coupling}},\ }\href
  {https://doi.org/10.1103/PhysRevC.85.034314} {\bibfield  {journal} {\bibinfo
  {journal} {Phys. Rev. C}\ }\textbf {\bibinfo {volume} {85}},\ \bibinfo
  {pages} {034314} (\bibinfo {year} {2012})},\ \Eprint
  {https://arxiv.org/abs/1203.6280} {arXiv:1203.6280 [nucl-th]} \BibitemShut
  {NoStop}%
\bibitem [{\citenamefont {Niu}\ \emph {et~al.}(2015)\citenamefont {Niu},
  \citenamefont {Niu}, \citenamefont {Col\`o},\ and\ \citenamefont
  {Vigezzi}}]{PhysRevLett.114.142501}%
  \BibitemOpen
  \bibfield  {author} {\bibinfo {author} {\bibfnamefont {Y.~F.}\ \bibnamefont
  {Niu}}, \bibinfo {author} {\bibfnamefont {Z.~M.}\ \bibnamefont {Niu}},
  \bibinfo {author} {\bibfnamefont {G.}~\bibnamefont {Col\`o}},\ and\ \bibinfo
  {author} {\bibfnamefont {E.}~\bibnamefont {Vigezzi}},\ }\bibfield  {title}
  {\bibinfo {title} {Particle-vibration coupling effect on the
  $\ensuremath{\beta}$ decay of magic nuclei},\ }\href
  {https://doi.org/10.1103/PhysRevLett.114.142501} {\bibfield  {journal}
  {\bibinfo  {journal} {Phys. Rev. Lett.}\ }\textbf {\bibinfo {volume} {114}},\
  \bibinfo {pages} {142501} (\bibinfo {year} {2015})}\BibitemShut {NoStop}%
\bibitem [{\citenamefont {Niu}\ \emph {et~al.}(2016)\citenamefont {Niu},
  \citenamefont {Colo}, \citenamefont {Vigezzi}, \citenamefont {Bai},\ and\
  \citenamefont {Sagawa}}]{Niu:2016kfj}%
  \BibitemOpen
  \bibfield  {author} {\bibinfo {author} {\bibfnamefont {Y.~F.}\ \bibnamefont
  {Niu}}, \bibinfo {author} {\bibfnamefont {G.}~\bibnamefont {Colo}}, \bibinfo
  {author} {\bibfnamefont {E.}~\bibnamefont {Vigezzi}}, \bibinfo {author}
  {\bibfnamefont {C.~L.}\ \bibnamefont {Bai}},\ and\ \bibinfo {author}
  {\bibfnamefont {H.}~\bibnamefont {Sagawa}},\ }\bibfield  {title} {\bibinfo
  {title} {{Quasiparticle random-phase approximation with
  quasiparticle-vibration coupling: Application to the Gamow-Teller response of
  the superfluid nucleus $^{120}$Sn}},\ }\href
  {https://doi.org/10.1103/PhysRevC.94.064328} {\bibfield  {journal} {\bibinfo
  {journal} {Phys. Rev. C}\ }\textbf {\bibinfo {volume} {94}},\ \bibinfo
  {pages} {064328} (\bibinfo {year} {2016})},\ \Eprint
  {https://arxiv.org/abs/1609.02341} {arXiv:1609.02341 [nucl-th]} \BibitemShut
  {NoStop}%
\bibitem [{\citenamefont {Niu}\ \emph {et~al.}(2018)\citenamefont {Niu},
  \citenamefont {Niu}, \citenamefont {Col\`o},\ and\ \citenamefont
  {Vigezzi}}]{Niu:2018art}%
  \BibitemOpen
  \bibfield  {author} {\bibinfo {author} {\bibfnamefont {Y.~F.}\ \bibnamefont
  {Niu}}, \bibinfo {author} {\bibfnamefont {Z.~M.}\ \bibnamefont {Niu}},
  \bibinfo {author} {\bibfnamefont {G.}~\bibnamefont {Col\`o}},\ and\ \bibinfo
  {author} {\bibfnamefont {E.}~\bibnamefont {Vigezzi}},\ }\bibfield  {title}
  {\bibinfo {title} {{Interplay of quasiparticle-vibration coupling and pairing
  correlations on \ensuremath{\beta} -decay half-lives}},\ }\href
  {https://doi.org/10.1016/j.physletb.2018.02.061} {\bibfield  {journal}
  {\bibinfo  {journal} {Phys. Lett. B}\ }\textbf {\bibinfo {volume} {780}},\
  \bibinfo {pages} {325} (\bibinfo {year} {2018})}\BibitemShut {NoStop}%
\bibitem [{\citenamefont {Robin}\ and\ \citenamefont
  {Litvinova}(2016)}]{Robin:2016wuh}%
  \BibitemOpen
  \bibfield  {author} {\bibinfo {author} {\bibfnamefont {C.}~\bibnamefont
  {Robin}}\ and\ \bibinfo {author} {\bibfnamefont {E.}~\bibnamefont
  {Litvinova}},\ }\bibfield  {title} {\bibinfo {title} {{Nuclear response
  theory for spin-isospin excitations in a relativistic quasiparticle-phonon
  coupling framework}},\ }\href {https://doi.org/10.1140/epja/i2016-16205-0}
  {\bibfield  {journal} {\bibinfo  {journal} {Eur. Phys. J. A}\ }\textbf
  {\bibinfo {volume} {52}},\ \bibinfo {pages} {205} (\bibinfo {year}
  {2016})}\BibitemShut {NoStop}%
\bibitem [{\citenamefont {Robin}\ and\ \citenamefont
  {Litvinova}(2018)}]{Robin:2018cjm}%
  \BibitemOpen
  \bibfield  {author} {\bibinfo {author} {\bibfnamefont {C.}~\bibnamefont
  {Robin}}\ and\ \bibinfo {author} {\bibfnamefont {E.}~\bibnamefont
  {Litvinova}},\ }\bibfield  {title} {\bibinfo {title} {{Coupling
  charge-exchange vibrations to nucleons in a relativistic framework: effect on
  Gamow-Teller transitions and $\beta$-decay half-lives}},\ }\href
  {https://doi.org/10.1103/PhysRevC.98.051301} {\bibfield  {journal} {\bibinfo
  {journal} {Phys. Rev. C}\ }\textbf {\bibinfo {volume} {98}},\ \bibinfo
  {pages} {051301} (\bibinfo {year} {2018})}\BibitemShut {NoStop}%
\bibitem [{\citenamefont {Litvinova}\ \emph {et~al.}(2020)\citenamefont
  {Litvinova}, \citenamefont {Robin},\ and\ \citenamefont
  {Wibowo}}]{Litvinova:2018pmr}%
  \BibitemOpen
  \bibfield  {author} {\bibinfo {author} {\bibfnamefont {E.}~\bibnamefont
  {Litvinova}}, \bibinfo {author} {\bibfnamefont {C.}~\bibnamefont {Robin}},\
  and\ \bibinfo {author} {\bibfnamefont {H.}~\bibnamefont {Wibowo}},\
  }\bibfield  {title} {\bibinfo {title} {{Temperature dependence of nuclear
  spin-isospin response and beta decay in hot astrophysical environments}},\
  }\href {https://doi.org/10.1016/j.physletb.2019.135134} {\bibfield  {journal}
  {\bibinfo  {journal} {Phys. Lett. B}\ }\textbf {\bibinfo {volume} {800}},\
  \bibinfo {pages} {135134} (\bibinfo {year} {2020})},\ \Eprint
  {https://arxiv.org/abs/1808.07223} {arXiv:1808.07223 [nucl-th]} \BibitemShut
  {NoStop}%
\bibitem [{\citenamefont {Robin}\ and\ \citenamefont
  {Litvinova}(2019)}]{Robin:2019jzm}%
  \BibitemOpen
  \bibfield  {author} {\bibinfo {author} {\bibfnamefont {C.}~\bibnamefont
  {Robin}}\ and\ \bibinfo {author} {\bibfnamefont {E.}~\bibnamefont
  {Litvinova}},\ }\bibfield  {title} {\bibinfo {title} {{Time-reversed
  particle-vibration loops and nuclear Gamow-Teller response}},\ }\href
  {https://doi.org/10.1103/PhysRevLett.123.202501} {\bibfield  {journal}
  {\bibinfo  {journal} {Phys. Rev. Lett.}\ }\textbf {\bibinfo {volume} {123}},\
  \bibinfo {pages} {202501} (\bibinfo {year} {2019})},\ \Eprint
  {https://arxiv.org/abs/1903.09182} {arXiv:1903.09182 [nucl-th]} \BibitemShut
  {NoStop}%
\bibitem [{\citenamefont {Liu}\ \emph {et~al.}(2023)\citenamefont {Liu},
  \citenamefont {Engel}, \citenamefont {Hinohara},\ and\ \citenamefont
  {Kortelainen}}]{Liu:2023xlv}%
  \BibitemOpen
  \bibfield  {author} {\bibinfo {author} {\bibfnamefont {Q.}~\bibnamefont
  {Liu}}, \bibinfo {author} {\bibfnamefont {J.}~\bibnamefont {Engel}}, \bibinfo
  {author} {\bibfnamefont {N.}~\bibnamefont {Hinohara}},\ and\ \bibinfo
  {author} {\bibfnamefont {M.}~\bibnamefont {Kortelainen}},\ }\bibfield
  {title} {\bibinfo {title} {{Effects of Quasiparticle-Vibration Coupling on
  Gamow-Teller Strength and $\beta$ Decay with the Skyrme Proton-Neutron
  Finite-Amplitude Method}},\ }\href@noop {} {\  (\bibinfo {year} {2023})},\
  \Eprint {https://arxiv.org/abs/2308.11802} {arXiv:2308.11802 [nucl-th]}
  \BibitemShut {NoStop}%
\bibitem [{\citenamefont {Suzuki}\ \emph {et~al.}(2012)\citenamefont {Suzuki},
  \citenamefont {Yoshida}, \citenamefont {Kajino},\ and\ \citenamefont
  {Otsuka}}]{SM1}%
  \BibitemOpen
  \bibfield  {author} {\bibinfo {author} {\bibfnamefont {T.}~\bibnamefont
  {Suzuki}}, \bibinfo {author} {\bibfnamefont {T.}~\bibnamefont {Yoshida}},
  \bibinfo {author} {\bibfnamefont {T.}~\bibnamefont {Kajino}},\ and\ \bibinfo
  {author} {\bibfnamefont {T.}~\bibnamefont {Otsuka}},\ }\bibfield  {title}
  {\bibinfo {title} {$\ensuremath{\beta}$ decays of isotones with neutron magic
  number of $n=126$ and $r$-process nucleosynthesis},\ }\href
  {https://doi.org/10.1103/PhysRevC.85.015802} {\bibfield  {journal} {\bibinfo
  {journal} {Phys. Rev. C}\ }\textbf {\bibinfo {volume} {85}},\ \bibinfo
  {pages} {015802} (\bibinfo {year} {2012})}\BibitemShut {NoStop}%
\bibitem [{\citenamefont {Zhi}\ \emph {et~al.}(2013)\citenamefont {Zhi},
  \citenamefont {Caurier}, \citenamefont {Cuenca-Garc\'{\i}a}, \citenamefont
  {Langanke}, \citenamefont {Mart\'{\i}nez-Pinedo},\ and\ \citenamefont
  {Sieja}}]{SM2}%
  \BibitemOpen
  \bibfield  {author} {\bibinfo {author} {\bibfnamefont {Q.}~\bibnamefont
  {Zhi}}, \bibinfo {author} {\bibfnamefont {E.}~\bibnamefont {Caurier}},
  \bibinfo {author} {\bibfnamefont {J.~J.}\ \bibnamefont {Cuenca-Garc\'{\i}a}},
  \bibinfo {author} {\bibfnamefont {K.}~\bibnamefont {Langanke}}, \bibinfo
  {author} {\bibfnamefont {G.}~\bibnamefont {Mart\'{\i}nez-Pinedo}},\ and\
  \bibinfo {author} {\bibfnamefont {K.}~\bibnamefont {Sieja}},\ }\bibfield
  {title} {\bibinfo {title} {Shell-model half-lives including first-forbidden
  contributions for $r$-process waiting-point nuclei},\ }\href
  {https://doi.org/10.1103/PhysRevC.87.025803} {\bibfield  {journal} {\bibinfo
  {journal} {Phys. Rev. C}\ }\textbf {\bibinfo {volume} {87}},\ \bibinfo
  {pages} {025803} (\bibinfo {year} {2013})}\BibitemShut {NoStop}%
\bibitem [{\citenamefont {Behrens}\ and\ \citenamefont
  {Buehring}(1971)}]{BEHRENS1971}%
  \BibitemOpen
  \bibfield  {author} {\bibinfo {author} {\bibfnamefont {H.}~\bibnamefont
  {Behrens}}\ and\ \bibinfo {author} {\bibfnamefont {W.}~\bibnamefont
  {Buehring}},\ }\bibfield  {title} {\bibinfo {title} {Nuclear beta decay},\
  }\href {https://doi.org/https://doi.org/10.1016/0375-9474(71)90489-1}
  {\bibfield  {journal} {\bibinfo  {journal} {Nuclear Physics A}\ }\textbf
  {\bibinfo {volume} {162}},\ \bibinfo {pages} {111} (\bibinfo {year}
  {1971})}\BibitemShut {NoStop}%
\bibitem [{\citenamefont {Hardy}\ and\ \citenamefont {Towner}(2009)}]{K}%
  \BibitemOpen
  \bibfield  {author} {\bibinfo {author} {\bibfnamefont {J.~C.}\ \bibnamefont
  {Hardy}}\ and\ \bibinfo {author} {\bibfnamefont {I.~S.}\ \bibnamefont
  {Towner}},\ }\bibfield  {title} {\bibinfo {title} {Superallowed
  ${0}^{+}\ensuremath{\rightarrow}{0}^{+}$ nuclear $\ensuremath{\beta}$ decays:
  A new survey with precision tests of the conserved vector current hypothesis
  and the standard model},\ }\href {https://doi.org/10.1103/PhysRevC.79.055502}
  {\bibfield  {journal} {\bibinfo  {journal} {Phys. Rev. C}\ }\textbf {\bibinfo
  {volume} {79}},\ \bibinfo {pages} {055502} (\bibinfo {year}
  {2009})}\BibitemShut {NoStop}%
\bibitem [{\citenamefont {Workman}\ and\ \citenamefont
  {Others}(2022)}]{Workman:2022ynf}%
  \BibitemOpen
  \bibfield  {author} {\bibinfo {author} {\bibfnamefont {R.~L.}\ \bibnamefont
  {Workman}}\ and\ \bibinfo {author} {\bibnamefont {Others}} (\bibinfo
  {collaboration} {Particle Data Group}),\ }\bibfield  {title} {\bibinfo
  {title} {{Review of Particle Physics}},\ }\href
  {https://doi.org/10.1093/ptep/ptac097} {\bibfield  {journal} {\bibinfo
  {journal} {PTEP}\ }\textbf {\bibinfo {volume} {2022}},\ \bibinfo {pages}
  {083C01} (\bibinfo {year} {2022})}\BibitemShut {NoStop}%
\bibitem [{\citenamefont {Behrens}\ and\ \citenamefont
  {B{\"u}hring}(1982)}]{behrens1982electron}%
  \BibitemOpen
  \bibfield  {author} {\bibinfo {author} {\bibfnamefont {H.}~\bibnamefont
  {Behrens}}\ and\ \bibinfo {author} {\bibfnamefont {W.}~\bibnamefont
  {B{\"u}hring}},\ }\href {https://books.google.de/books?id=b062AAAAIAAJ}
  {\emph {\bibinfo {title} {Electron Radial Wave Functions and Nuclear
  Betadecay}}},\ International series of monographs on physics\ (\bibinfo
  {publisher} {Clarendon Press},\ \bibinfo {year} {1982})\BibitemShut {NoStop}%
\bibitem [{\citenamefont {Warburton}(1991)}]{Warburton1991}%
  \BibitemOpen
  \bibfield  {author} {\bibinfo {author} {\bibfnamefont {E.~K.}\ \bibnamefont
  {Warburton}},\ }\bibfield  {title} {\bibinfo {title} {First-forbidden
  \ensuremath{\beta} decay in the lead region and mesonic enhancement of the
  weak axial current},\ }\href {https://doi.org/10.1103/PhysRevC.44.233}
  {\bibfield  {journal} {\bibinfo  {journal} {Phys. Rev. C}\ }\textbf {\bibinfo
  {volume} {44}},\ \bibinfo {pages} {233} (\bibinfo {year} {1991})}\BibitemShut
  {NoStop}%
\bibitem [{\citenamefont {Tselyaev}(1989)}]{TBA}%
  \BibitemOpen
  \bibfield  {author} {\bibinfo {author} {\bibfnamefont {V.~I.}\ \bibnamefont
  {Tselyaev}},\ }\href@noop {} {\bibfield  {journal} {\bibinfo  {journal} {Yad.
  Fiz}\ }\textbf {\bibinfo {volume} {50}},\ \bibinfo {pages} {1252} (\bibinfo
  {year} {1989})},\ \bibinfo {note} {[Sov. J. Nucl. Phys. 50, 780
  (1989)]}\BibitemShut {NoStop}%
\bibitem [{\citenamefont {Ring}(1996)}]{RING1996193}%
  \BibitemOpen
  \bibfield  {author} {\bibinfo {author} {\bibfnamefont {P.}~\bibnamefont
  {Ring}},\ }\bibfield  {title} {\bibinfo {title} {Relativistic mean field
  theory in finite nuclei},\ }\href
  {https://doi.org/https://doi.org/10.1016/0146-6410(96)00054-3} {\bibfield
  {journal} {\bibinfo  {journal} {Progress in Particle and Nuclear Physics}\
  }\textbf {\bibinfo {volume} {37}},\ \bibinfo {pages} {193} (\bibinfo {year}
  {1996})}\BibitemShut {NoStop}%
\bibitem [{\citenamefont {Lalazissis}\ \emph {et~al.}(1997)\citenamefont
  {Lalazissis}, \citenamefont {K\"onig},\ and\ \citenamefont {Ring}}]{NL3}%
  \BibitemOpen
  \bibfield  {author} {\bibinfo {author} {\bibfnamefont {G.~A.}\ \bibnamefont
  {Lalazissis}}, \bibinfo {author} {\bibfnamefont {J.}~\bibnamefont
  {K\"onig}},\ and\ \bibinfo {author} {\bibfnamefont {P.}~\bibnamefont
  {Ring}},\ }\bibfield  {title} {\bibinfo {title} {New parametrization for the
  lagrangian density of relativistic mean field theory},\ }\href
  {https://doi.org/10.1103/PhysRevC.55.540} {\bibfield  {journal} {\bibinfo
  {journal} {Phys. Rev. C}\ }\textbf {\bibinfo {volume} {55}},\ \bibinfo
  {pages} {540} (\bibinfo {year} {1997})}\BibitemShut {NoStop}%
\bibitem [{\citenamefont {Kucharek}\ and\ \citenamefont
  {Ring}(1991)}]{Kucharek1991}%
  \BibitemOpen
  \bibfield  {author} {\bibinfo {author} {\bibfnamefont {H.}~\bibnamefont
  {Kucharek}}\ and\ \bibinfo {author} {\bibfnamefont {P.}~\bibnamefont
  {Ring}},\ }\bibfield  {title} {\bibinfo {title} {Relativistic field theory of
  superfluidity in nuclei},\ }\href {https://doi.org/10.1007/BF01282930}
  {\bibfield  {journal} {\bibinfo  {journal} {Zeitschrift f{\"u}r Physik A
  Hadrons and Nuclei}\ }\textbf {\bibinfo {volume} {339}},\ \bibinfo {pages}
  {23} (\bibinfo {year} {1991})}\BibitemShut {NoStop}%
\bibitem [{\citenamefont {Serra}\ \emph {et~al.}(2001)\citenamefont {Serra},
  \citenamefont {Rummel},\ and\ \citenamefont {Ring}}]{Sierra2001}%
  \BibitemOpen
  \bibfield  {author} {\bibinfo {author} {\bibfnamefont {M.}~\bibnamefont
  {Serra}}, \bibinfo {author} {\bibfnamefont {A.}~\bibnamefont {Rummel}},\ and\
  \bibinfo {author} {\bibfnamefont {P.}~\bibnamefont {Ring}},\ }\bibfield
  {title} {\bibinfo {title} {Relativistic theory of pairing in infinite nuclear
  matter},\ }\href {https://doi.org/10.1103/PhysRevC.65.014304} {\bibfield
  {journal} {\bibinfo  {journal} {Phys. Rev. C}\ }\textbf {\bibinfo {volume}
  {65}},\ \bibinfo {pages} {014304} (\bibinfo {year} {2001})}\BibitemShut
  {NoStop}%
\bibitem [{\citenamefont {Bonche}\ \emph {et~al.}(1994)\citenamefont {Bonche},
  \citenamefont {Chabanat}, \citenamefont {Chen}, \citenamefont {Dobaczewski},
  \citenamefont {Flocard}, \citenamefont {Gall}, \citenamefont {Heenen},
  \citenamefont {Meyer}, \citenamefont {Tajima},\ and\ \citenamefont
  {Weiss}}]{BONCHE1994185}%
  \BibitemOpen
  \bibfield  {author} {\bibinfo {author} {\bibfnamefont {P.}~\bibnamefont
  {Bonche}}, \bibinfo {author} {\bibfnamefont {E.}~\bibnamefont {Chabanat}},
  \bibinfo {author} {\bibfnamefont {B.}~\bibnamefont {Chen}}, \bibinfo {author}
  {\bibfnamefont {J.}~\bibnamefont {Dobaczewski}}, \bibinfo {author}
  {\bibfnamefont {H.}~\bibnamefont {Flocard}}, \bibinfo {author} {\bibfnamefont
  {B.}~\bibnamefont {Gall}}, \bibinfo {author} {\bibfnamefont {P.}~\bibnamefont
  {Heenen}}, \bibinfo {author} {\bibfnamefont {J.}~\bibnamefont {Meyer}},
  \bibinfo {author} {\bibfnamefont {N.}~\bibnamefont {Tajima}},\ and\ \bibinfo
  {author} {\bibfnamefont {M.}~\bibnamefont {Weiss}},\ }\bibfield  {title}
  {\bibinfo {title} {Microscopic approach to collective motion},\ }\href
  {https://doi.org/https://doi.org/10.1016/0375-9474(94)90045-0} {\bibfield
  {journal} {\bibinfo  {journal} {Nuclear Physics A}\ }\textbf {\bibinfo
  {volume} {574}},\ \bibinfo {pages} {185} (\bibinfo {year}
  {1994})}\BibitemShut {NoStop}%
\bibitem [{\citenamefont {Bouyssy}\ \emph {et~al.}(1987)\citenamefont
  {Bouyssy}, \citenamefont {Mathiot}, \citenamefont {Van~Giai},\ and\
  \citenamefont {Marcos}}]{PhysRevC.36.380}%
  \BibitemOpen
  \bibfield  {author} {\bibinfo {author} {\bibfnamefont {A.}~\bibnamefont
  {Bouyssy}}, \bibinfo {author} {\bibfnamefont {J.-F.}\ \bibnamefont
  {Mathiot}}, \bibinfo {author} {\bibfnamefont {N.}~\bibnamefont {Van~Giai}},\
  and\ \bibinfo {author} {\bibfnamefont {S.}~\bibnamefont {Marcos}},\
  }\bibfield  {title} {\bibinfo {title} {{Relativistic description of nuclear
  systems in the Hartree-Fock approximation}},\ }\href
  {https://doi.org/10.1103/PhysRevC.36.380} {\bibfield  {journal} {\bibinfo
  {journal} {Phys. Rev. C}\ }\textbf {\bibinfo {volume} {36}},\ \bibinfo
  {pages} {380} (\bibinfo {year} {1987})}\BibitemShut {NoStop}%
\bibitem [{\citenamefont {Liang}\ \emph {et~al.}(2008)\citenamefont {Liang},
  \citenamefont {Van~Giai},\ and\ \citenamefont
  {Meng}}]{PhysRevLett.101.122502}%
  \BibitemOpen
  \bibfield  {author} {\bibinfo {author} {\bibfnamefont {H.}~\bibnamefont
  {Liang}}, \bibinfo {author} {\bibfnamefont {N.}~\bibnamefont {Van~Giai}},\
  and\ \bibinfo {author} {\bibfnamefont {J.}~\bibnamefont {Meng}},\ }\bibfield
  {title} {\bibinfo {title} {Spin-isospin resonances: A self-consistent
  covariant description},\ }\href
  {https://doi.org/10.1103/PhysRevLett.101.122502} {\bibfield  {journal}
  {\bibinfo  {journal} {Phys. Rev. Lett.}\ }\textbf {\bibinfo {volume} {101}},\
  \bibinfo {pages} {122502} (\bibinfo {year} {2008})}\BibitemShut {NoStop}%
\bibitem [{\citenamefont {Ravli\'c}\ \emph {et~al.}(2021)\citenamefont
  {Ravli\'c}, \citenamefont {Niu}, \citenamefont {Nik\v{s}i\'c}, \citenamefont
  {Paar},\ and\ \citenamefont {Ring}}]{Ravlic:2021uvo}%
  \BibitemOpen
  \bibfield  {author} {\bibinfo {author} {\bibfnamefont {A.}~\bibnamefont
  {Ravli\'c}}, \bibinfo {author} {\bibfnamefont {Y.~F.}\ \bibnamefont {Niu}},
  \bibinfo {author} {\bibfnamefont {T.}~\bibnamefont {Nik\v{s}i\'c}}, \bibinfo
  {author} {\bibfnamefont {N.}~\bibnamefont {Paar}},\ and\ \bibinfo {author}
  {\bibfnamefont {P.}~\bibnamefont {Ring}},\ }\bibfield  {title} {\bibinfo
  {title} {{Finite-temperature linear response theory based on relativistic
  Hartree Bogoliubov model with point-coupling interaction}},\ }\href
  {https://doi.org/10.1103/PhysRevC.104.064302} {\bibfield  {journal} {\bibinfo
   {journal} {Phys. Rev. C}\ }\textbf {\bibinfo {volume} {104}},\ \bibinfo
  {pages} {064302} (\bibinfo {year} {2021})}\BibitemShut {NoStop}%
\bibitem [{\citenamefont {Vale}\ \emph {et~al.}(2021)\citenamefont {Vale},
  \citenamefont {Niu},\ and\ \citenamefont {Paar}}]{PhysRevC.103.064307}%
  \BibitemOpen
  \bibfield  {author} {\bibinfo {author} {\bibfnamefont {D.}~\bibnamefont
  {Vale}}, \bibinfo {author} {\bibfnamefont {Y.~F.}\ \bibnamefont {Niu}},\ and\
  \bibinfo {author} {\bibfnamefont {N.}~\bibnamefont {Paar}},\ }\bibfield
  {title} {\bibinfo {title} {Nuclear charge-exchange excitations based on a
  relativistic density-dependent point-coupling model},\ }\href
  {https://doi.org/10.1103/PhysRevC.103.064307} {\bibfield  {journal} {\bibinfo
   {journal} {Phys. Rev. C}\ }\textbf {\bibinfo {volume} {103}},\ \bibinfo
  {pages} {064307} (\bibinfo {year} {2021})}\BibitemShut {NoStop}%
\bibitem [{\citenamefont {Tselyaev}(2013)}]{Tselyaev2013}%
  \BibitemOpen
  \bibfield  {author} {\bibinfo {author} {\bibfnamefont {V.~I.}\ \bibnamefont
  {Tselyaev}},\ }\href@noop {} {\bibfield  {journal} {\bibinfo  {journal}
  {Phys. Rev. C}\ }\textbf {\bibinfo {volume} {88}},\ \bibinfo {pages} {054301}
  (\bibinfo {year} {2013})}\BibitemShut {NoStop}%
\bibitem [{\citenamefont {Mustonen}\ \emph {et~al.}(2014)\citenamefont
  {Mustonen}, \citenamefont {Shafer}, \citenamefont {Zenginerler},\ and\
  \citenamefont {Engel}}]{Mustonen:2014bya}%
  \BibitemOpen
  \bibfield  {author} {\bibinfo {author} {\bibfnamefont {M.~T.}\ \bibnamefont
  {Mustonen}}, \bibinfo {author} {\bibfnamefont {T.}~\bibnamefont {Shafer}},
  \bibinfo {author} {\bibfnamefont {Z.}~\bibnamefont {Zenginerler}},\ and\
  \bibinfo {author} {\bibfnamefont {J.}~\bibnamefont {Engel}},\ }\bibfield
  {title} {\bibinfo {title} {{Finite Amplitude Method for Charge-Changing
  Transitions in Axially-Deformed Nuclei}},\ }\href
  {https://doi.org/10.1103/PhysRevC.90.024308} {\bibfield  {journal} {\bibinfo
  {journal} {Phys. Rev. C}\ }\textbf {\bibinfo {volume} {90}},\ \bibinfo
  {pages} {024308} (\bibinfo {year} {2014})},\ \Eprint
  {https://arxiv.org/abs/1405.0254} {arXiv:1405.0254 [nucl-th]} \BibitemShut
  {NoStop}%
\bibitem [{dat()}]{data}%
  \BibitemOpen
  \href@noop {} {}\bibinfo {howpublished} {IAEA LiveChart,
  \url{https://www-nds.iaea.org/relnsd/vcharthtml/VChartHTML.html}}\BibitemShut
  {NoStop}%
\bibitem [{\citenamefont {Litvinova}\ \emph {et~al.}(2007)\citenamefont
  {Litvinova}, \citenamefont {Ring},\ and\ \citenamefont
  {Tselyaev}}]{Litvinova:2007gg}%
  \BibitemOpen
  \bibfield  {author} {\bibinfo {author} {\bibfnamefont {E.}~\bibnamefont
  {Litvinova}}, \bibinfo {author} {\bibfnamefont {P.}~\bibnamefont {Ring}},\
  and\ \bibinfo {author} {\bibfnamefont {V.}~\bibnamefont {Tselyaev}},\
  }\bibfield  {title} {\bibinfo {title} {{Particle-vibration coupling within
  covariant density functional theory}},\ }\href
  {https://doi.org/10.1103/PhysRevC.75.064308} {\bibfield  {journal} {\bibinfo
  {journal} {Phys. Rev. C}\ }\textbf {\bibinfo {volume} {75}},\ \bibinfo
  {pages} {064308} (\bibinfo {year} {2007})},\ \Eprint
  {https://arxiv.org/abs/0705.1044} {arXiv:0705.1044 [nucl-th]} \BibitemShut
  {NoStop}%
\bibitem [{\citenamefont {Wang}\ \emph {et~al.}(2021)\citenamefont {Wang},
  \citenamefont {Huang}, \citenamefont {Kondev}, \citenamefont {Audi},\ and\
  \citenamefont {Naimi}}]{Wang.Huang.ea:2021}%
  \BibitemOpen
  \bibfield  {author} {\bibinfo {author} {\bibfnamefont {M.}~\bibnamefont
  {Wang}}, \bibinfo {author} {\bibfnamefont {W.}~\bibnamefont {Huang}},
  \bibinfo {author} {\bibfnamefont {F.}~\bibnamefont {Kondev}}, \bibinfo
  {author} {\bibfnamefont {G.}~\bibnamefont {Audi}},\ and\ \bibinfo {author}
  {\bibfnamefont {S.}~\bibnamefont {Naimi}},\ }\bibfield  {title} {\bibinfo
  {title} {The {AME} 2020 atomic mass evaluation ({II}). {Tables}, graphs and
  references*},\ }\href {https://doi.org/10.1088/1674-1137/abddaf} {\bibfield
  {journal} {\bibinfo  {journal} {Chinese Physics C}\ }\textbf {\bibinfo
  {volume} {45}},\ \bibinfo {pages} {030003} (\bibinfo {year}
  {2021})}\BibitemShut {NoStop}%
\bibitem [{\citenamefont {Tuli}\ and\ \citenamefont
  {Browne}(2019)}]{Tuli.Browne:2019}%
  \BibitemOpen
  \bibfield  {author} {\bibinfo {author} {\bibfnamefont {J.~K.}\ \bibnamefont
  {Tuli}}\ and\ \bibinfo {author} {\bibfnamefont {E.}~\bibnamefont {Browne}},\
  }\bibfield  {title} {\bibinfo {title} {Nuclear {Data} {Sheets} for {A}=82},\
  }\href {https://doi.org/10.1016/j.nds.2019.04.002} {\bibfield  {journal}
  {\bibinfo  {journal} {Nuclear Data Sheets}\ }\textbf {\bibinfo {volume}
  {157}},\ \bibinfo {pages} {260} (\bibinfo {year} {2019})}\BibitemShut
  {NoStop}%
\bibitem [{\citenamefont {Berger}\ \emph {et~al.}(1991)\citenamefont {Berger},
  \citenamefont {Girod},\ and\ \citenamefont {Gogny}}]{D1S}%
  \BibitemOpen
  \bibfield  {author} {\bibinfo {author} {\bibfnamefont {J.}~\bibnamefont
  {Berger}}, \bibinfo {author} {\bibfnamefont {M.}~\bibnamefont {Girod}},\ and\
  \bibinfo {author} {\bibfnamefont {D.}~\bibnamefont {Gogny}},\ }\bibfield
  {title} {\bibinfo {title} {Time-dependent quantum collective dynamics applied
  to nuclear fission},\ }\href
  {https://doi.org/https://doi.org/10.1016/0010-4655(91)90263-K} {\bibfield
  {journal} {\bibinfo  {journal} {Computer Physics Communications}\ }\textbf
  {\bibinfo {volume} {63}},\ \bibinfo {pages} {365} (\bibinfo {year}
  {1991})}\BibitemShut {NoStop}%
\bibitem [{\citenamefont {Kubodera}\ \emph {et~al.}(1978)\citenamefont
  {Kubodera}, \citenamefont {Delorme},\ and\ \citenamefont
  {Rho}}]{PhysRevLett.40.755}%
  \BibitemOpen
  \bibfield  {author} {\bibinfo {author} {\bibfnamefont {K.}~\bibnamefont
  {Kubodera}}, \bibinfo {author} {\bibfnamefont {J.}~\bibnamefont {Delorme}},\
  and\ \bibinfo {author} {\bibfnamefont {M.}~\bibnamefont {Rho}},\ }\bibfield
  {title} {\bibinfo {title} {Axial currents in nuclei},\ }\href
  {https://doi.org/10.1103/PhysRevLett.40.755} {\bibfield  {journal} {\bibinfo
  {journal} {Phys. Rev. Lett.}\ }\textbf {\bibinfo {volume} {40}},\ \bibinfo
  {pages} {755} (\bibinfo {year} {1978})}\BibitemShut {NoStop}%
\bibitem [{\citenamefont {Machleidt}(1989)}]{Machleidt1989}%
  \BibitemOpen
  \bibfield  {author} {\bibinfo {author} {\bibfnamefont {R.}~\bibnamefont
  {Machleidt}},\ }\bibinfo {title} {The meson theory of nuclear forces and
  nuclear structure},\ in\ \href {https://doi.org/10.1007/978-1-4613-9907-0_2}
  {\emph {\bibinfo {booktitle} {Advances in Nuclear Physics}}},\ \bibinfo
  {editor} {edited by\ \bibinfo {editor} {\bibfnamefont {J.~W.}\ \bibnamefont
  {Negele}}\ and\ \bibinfo {editor} {\bibfnamefont {E.}~\bibnamefont {Vogt}}}\
  (\bibinfo  {publisher} {Springer US},\ \bibinfo {address} {Boston, MA},\
  \bibinfo {year} {1989})\ pp.\ \bibinfo {pages} {189--376}\BibitemShut
  {NoStop}%
\bibitem [{\citenamefont {Goriely}\ \emph {et~al.}(2009)\citenamefont
  {Goriely}, \citenamefont {Chamel},\ and\ \citenamefont {Pearson}}]{HFB21}%
  \BibitemOpen
  \bibfield  {author} {\bibinfo {author} {\bibfnamefont {S.}~\bibnamefont
  {Goriely}}, \bibinfo {author} {\bibfnamefont {N.}~\bibnamefont {Chamel}},\
  and\ \bibinfo {author} {\bibfnamefont {J.~M.}\ \bibnamefont {Pearson}},\
  }\bibfield  {title} {\bibinfo {title} {Skyrme-hartree-fock-bogoliubov nuclear
  mass formulas: Crossing the 0.6 mev accuracy threshold with microscopically
  deduced pairing},\ }\href {https://doi.org/10.1103/PhysRevLett.102.152503}
  {\bibfield  {journal} {\bibinfo  {journal} {Phys. Rev. Lett.}\ }\textbf
  {\bibinfo {volume} {102}},\ \bibinfo {pages} {152503} (\bibinfo {year}
  {2009})}\BibitemShut {NoStop}%
\bibitem [{\citenamefont {Borzov}(2011)}]{Borzov2011}%
  \BibitemOpen
  \bibfield  {author} {\bibinfo {author} {\bibfnamefont {I.~N.}\ \bibnamefont
  {Borzov}},\ }\bibfield  {title} {\bibinfo {title} {Beta-decay of nuclei near
  the neutron shell n = 126},\ }\href
  {https://doi.org/10.1134/S1063778811100024} {\bibfield  {journal} {\bibinfo
  {journal} {Physics of Atomic Nuclei}\ }\textbf {\bibinfo {volume} {74}},\
  \bibinfo {pages} {1435} (\bibinfo {year} {2011})}\BibitemShut {NoStop}%
\bibitem [{\citenamefont {Fayans}\ \emph {et~al.}(2000)\citenamefont {Fayans},
  \citenamefont {Tolokonnikov}, \citenamefont {Trykov},\ and\ \citenamefont
  {Zawischa}}]{FAYANS200049}%
  \BibitemOpen
  \bibfield  {author} {\bibinfo {author} {\bibfnamefont {S.}~\bibnamefont
  {Fayans}}, \bibinfo {author} {\bibfnamefont {S.}~\bibnamefont
  {Tolokonnikov}}, \bibinfo {author} {\bibfnamefont {E.}~\bibnamefont
  {Trykov}},\ and\ \bibinfo {author} {\bibfnamefont {D.}~\bibnamefont
  {Zawischa}},\ }\bibfield  {title} {\bibinfo {title} {Nuclear isotope shifts
  within the local energy-density functional approach},\ }\href
  {https://doi.org/https://doi.org/10.1016/S0375-9474(00)00192-5} {\bibfield
  {journal} {\bibinfo  {journal} {Nuclear Physics A}\ }\textbf {\bibinfo
  {volume} {676}},\ \bibinfo {pages} {49} (\bibinfo {year} {2000})}\BibitemShut
  {NoStop}%
\bibitem [{\citenamefont {Reinhard}\ \emph {et~al.}(1999)\citenamefont
  {Reinhard}, \citenamefont {Dean}, \citenamefont {Nazarewicz}, \citenamefont
  {Dobaczewski}, \citenamefont {Maruhn},\ and\ \citenamefont {Strayer}}]{Sko}%
  \BibitemOpen
  \bibfield  {author} {\bibinfo {author} {\bibfnamefont {P.-G.}\ \bibnamefont
  {Reinhard}}, \bibinfo {author} {\bibfnamefont {D.~J.}\ \bibnamefont {Dean}},
  \bibinfo {author} {\bibfnamefont {W.}~\bibnamefont {Nazarewicz}}, \bibinfo
  {author} {\bibfnamefont {J.}~\bibnamefont {Dobaczewski}}, \bibinfo {author}
  {\bibfnamefont {J.~A.}\ \bibnamefont {Maruhn}},\ and\ \bibinfo {author}
  {\bibfnamefont {M.~R.}\ \bibnamefont {Strayer}},\ }\bibfield  {title}
  {\bibinfo {title} {Shape coexistence and the effective nucleon-nucleon
  interaction},\ }\href {https://doi.org/10.1103/PhysRevC.60.014316} {\bibfield
   {journal} {\bibinfo  {journal} {Phys. Rev. C}\ }\textbf {\bibinfo {volume}
  {60}},\ \bibinfo {pages} {014316} (\bibinfo {year} {1999})}\BibitemShut
  {NoStop}%
\bibitem [{\citenamefont {Marketin}\ \emph {et~al.}(2007)\citenamefont
  {Marketin}, \citenamefont {Vretenar},\ and\ \citenamefont {Ring}}]{D3Cs}%
  \BibitemOpen
  \bibfield  {author} {\bibinfo {author} {\bibfnamefont {T.}~\bibnamefont
  {Marketin}}, \bibinfo {author} {\bibfnamefont {D.}~\bibnamefont {Vretenar}},\
  and\ \bibinfo {author} {\bibfnamefont {P.}~\bibnamefont {Ring}},\ }\bibfield
  {title} {\bibinfo {title} {Calculation of \ensuremath{\beta}-decay rates in a
  relativistic model with momentum-dependent self-energies},\ }\href
  {https://doi.org/10.1103/PhysRevC.75.024304} {\bibfield  {journal} {\bibinfo
  {journal} {Phys. Rev. C}\ }\textbf {\bibinfo {volume} {75}},\ \bibinfo
  {pages} {024304} (\bibinfo {year} {2007})}\BibitemShut {NoStop}%
\bibitem [{\citenamefont {Mustonen}\ and\ \citenamefont
  {Engel}(2016)}]{PhysRevC.93.014304}%
  \BibitemOpen
  \bibfield  {author} {\bibinfo {author} {\bibfnamefont {M.~T.}\ \bibnamefont
  {Mustonen}}\ and\ \bibinfo {author} {\bibfnamefont {J.}~\bibnamefont
  {Engel}},\ }\bibfield  {title} {\bibinfo {title} {Global description of
  ${\ensuremath{\beta}}^{\ensuremath{-}}$ decay in even-even nuclei with the
  axially-deformed skyrme finite-amplitude method},\ }\href
  {https://doi.org/10.1103/PhysRevC.93.014304} {\bibfield  {journal} {\bibinfo
  {journal} {Phys. Rev. C}\ }\textbf {\bibinfo {volume} {93}},\ \bibinfo
  {pages} {014304} (\bibinfo {year} {2016})}\BibitemShut {NoStop}%
\bibitem [{\citenamefont {M\"arkisch}\ \emph {et~al.}(2019)\citenamefont
  {M\"arkisch}, \citenamefont {Mest}, \citenamefont {Saul}, \citenamefont
  {Wang}, \citenamefont {Abele}, \citenamefont {Dubbers}, \citenamefont
  {Klopf}, \citenamefont {Petoukhov}, \citenamefont {Roick}, \citenamefont
  {Soldner},\ and\ \citenamefont {Werder}}]{ga}%
  \BibitemOpen
  \bibfield  {author} {\bibinfo {author} {\bibfnamefont {B.}~\bibnamefont
  {M\"arkisch}}, \bibinfo {author} {\bibfnamefont {H.}~\bibnamefont {Mest}},
  \bibinfo {author} {\bibfnamefont {H.}~\bibnamefont {Saul}}, \bibinfo {author}
  {\bibfnamefont {X.}~\bibnamefont {Wang}}, \bibinfo {author} {\bibfnamefont
  {H.}~\bibnamefont {Abele}}, \bibinfo {author} {\bibfnamefont
  {D.}~\bibnamefont {Dubbers}}, \bibinfo {author} {\bibfnamefont
  {M.}~\bibnamefont {Klopf}}, \bibinfo {author} {\bibfnamefont
  {A.}~\bibnamefont {Petoukhov}}, \bibinfo {author} {\bibfnamefont
  {C.}~\bibnamefont {Roick}}, \bibinfo {author} {\bibfnamefont
  {T.}~\bibnamefont {Soldner}},\ and\ \bibinfo {author} {\bibfnamefont
  {D.}~\bibnamefont {Werder}},\ }\bibfield  {title} {\bibinfo {title}
  {Measurement of the weak axial-vector coupling constant in the decay of free
  neutrons using a pulsed cold neutron beam},\ }\href
  {https://doi.org/10.1103/PhysRevLett.122.242501} {\bibfield  {journal}
  {\bibinfo  {journal} {Phys. Rev. Lett.}\ }\textbf {\bibinfo {volume} {122}},\
  \bibinfo {pages} {242501} (\bibinfo {year} {2019})}\BibitemShut {NoStop}%
\bibitem [{FAI()}]{FAIR}%
  \BibitemOpen
  \href@noop {} {\bibinfo {title} {Facility for antiprotons and ion research in
  europe ({FAIR}), \url{https://fair-center.org}}}\BibitemShut {NoStop}%
\end{thebibliography}%

\end{document}